\documentclass[twocolumn]{aa}
\usepackage{lscape}
\usepackage{graphicx}
\usepackage{natbib}
\bibpunct{(}{)}{;}{a}{}{,}
\usepackage{amsmath}
\usepackage{txfonts}
\usepackage{longtable}
\usepackage{supertabular}
\begin{document}
\bibliographystyle{astron}
%
%
%


\def\jnl@style{\it}
\def\aaref@jnl#1{{\jnl@style#1}}

\def\aaref@jnl#1{{\jnl@style#1}}

\def\aj{\aaref@jnl{AJ}}                   
\def\araa{\aaref@jnl{ARA$\&$A}}           
\def\apj{\aaref@jnl{ApJ}}                 
\def\apjl{\aaref@jnl{ApJ}}                
\def\apjs{\aaref@jnl{ApJS}}               
\def\ao{\aaref@jnl{Appl.~Opt.}}           
\def\apss{\aaref@jnl{Ap$\&$SS}}           
\def\aap{\aaref@jnl{A$\&$A}}              
\def\aapr{\aaref@jnl{A$\&$A~Rev.}}        
\def\aaps{\aaref@jnl{A$\&$AS}}            
\def\azh{\aaref@jnl{AZh}}                 
\def\baas{\aaref@jnl{BAAS}}               
\def\jrasc{\aaref@jnl{JRASC}}             
\def\memras{\aaref@jnl{MmRAS}}            
\def\mnras{\aaref@jnl{MNRAS}}             
\def\pra{\aaref@jnl{Phys.~Rev.~A}}        
\def\prb{\aaref@jnl{Phys.~Rev.~B}}        
\def\prc{\aaref@jnl{Phys.~Rev.~C}}        
\def\prd{\aaref@jnl{Phys.~Rev.~D}}        
\def\pre{\aaref@jnl{Phys.~Rev.~E}}        
\def\prl{\aaref@jnl{Phys.~Rev.~Lett.}}    
\def\pasp{\aaref@jnl{PASP}}               
\def\pasj{\aaref@jnl{PASJ}}               
\def\qjras{\aaref@jnl{QJRAS}}             
\def\skytel{\aaref@jnl{S\&T}}             
\def\solphys{\aaref@jnl{Sol.~Phys.}}      
\def\sovast{\aaref@jnl{Soviet~Ast.}}      
\def\ssr{\aaref@jnl{Space~Sci.~Rev.}}     
\def\zap{\aaref@jnl{ZAp}}                 
\def\nat{\aaref@jnl{Nature}}              
\def\iaucirc{\aaref@jnl{IAU~Circ.}}       
\def\aplett{\aaref@jnl{Astrophys.~Lett.}} 
\def\apspr{\aaref@jnl{Astrophys.~Space~Phys.~Res.}}
\def\bain{\aaref@jnl{Bull.~Astron.~Inst.~Netherlands}} 
\def\fcp{\aaref@jnl{Fund.~Cosmic~Phys.}}  
\def\gca{\aaref@jnl{Geochim.~Cosmochim.~Acta}}   
\def\grl{\aaref@jnl{Geophys.~Res.~Lett.}} 
\def\jcp{\aaref@jnl{J.~Chem.~Phys.}}      
\def\jgr{\aaref@jnl{J.~Geophys.~Res.}}    
\def\jqsrt{\aaref@jnl{J.~Quant.~Spec.~Radiat.~Transf.}}
\def\memsai{\aaref@jnl{Mem.~Soc.~Astron.~Italiana}}
\def\nphysa{\aaref@jnl{Nucl.~Phys.~A}}   
\def\physrep{\aaref@jnl{Phys.~Rep.}}   
\def\physscr{\aaref@jnl{Phys.~Scr}}   
\def\planss{\aaref@jnl{Planet.~Space~Sci.}}   
\def\procspie{\aaref@jnl{Proc.~SPIE}}   

\let\astap=\aap
\let\apjlett=\apjl
\let\apjsupp=\apjs
\let\applopt=\ao

   \title{The surface brightness of the Galaxy at the Solar Neighbourhood}

   \author{A.-L. Melchior\inst{1}
     \and 
     F. Combes\inst{2}
     \and
     A. Gould\inst{3}\fnmsep\thanks{Visiting Astronomer, LPCC,
   Coll\`ege de France, 11, place Marcelin Berthelot, F-75231 Paris,
   France}
 } 

\offprints{}

   \institute{LERMA, Universit\'e Pierre et Marie Curie and Observatoire
   de Paris, 61, avenue de l'Observatoire, F-75\,014 Paris, France\\
   \email{Anne-Laure.Melchior@obspm.fr}
\and
   LERMA, Observatoire de Paris, 61, avenue de l 'Observatoire, F-75\,014
   Paris, France\\ \email{Francoise.Combes@obspm.fr}
\and
Department of Astronomy, Ohio State University, Columbus Ohio 43210,
   USA\\ \email{gould@astronomy.ohio-state.edu}
  }

   \date{}

   \abstract{We present a new determination of the surface brightness
of our Galaxy at the Solar Neighbourhood as observed from outside the
Galaxy. We rely on various existing optical and infra-red surveys to
obtain a multiwavelength estimate. On the one hand, scattered light
does not contribute significantly to the surface brightness. On the
other hand, optical and infrared integrated all-sky surveys (Pioneer 10/11
and COBE/DIRBE) show a systematically larger value than our synthetic
local estimate based on Hipparcos data. This local estimate is also
compatible with our Galactic simulations normalised at the Solar
Neighbourhood and assuming an homogeneous stellar distribution. We
interpret this disagreement as a signature of the presence of a local
minimum of the stellar density compatible with Gould's belt. 
According to this result, the global luminosity of the Milky Way
should follow the Tully-Fisher relation established for external
galaxies.
 }

   \maketitle 
%

\section{Introduction} 
\label{sec:intro}
Modelling of the multiwavelength continuum emission of high-z galaxies
is essential to simulate realistically the formation and evolution of
galaxies as will be observed with the next generation of instruments.
The data samples of galaxies are currently limited especially in the
infrared and millimeter wavelengths, but will be significantly
enlarged in the coming years. On the one hand, theoretical modelling
based on physical principles actually relies on a large number of
variables and requires well-sampled galaxy spectra to be constrained.
On the other hand, empirical approaches
\citep[e.g.][]{Guiderdoni:1998,Dale:2001} minimise the number of
parameters to reproduce the current observations. Both approaches need
to be validated with well-sampled continuum spectra of galaxies. It is
customary to use the ``Solar Neighbourhood'' as a zero-point for these
models. Even though various multiwavelength observations are available
for the Galaxy and the Solar Neighbourhood, most authors quote Mathis
et al. (1981), who were investigating the Local Interstellar Radiation
Field. Although this work is robust to understand the absorption of
photons by dust grains in the Local Neighbourhood, it did not intend
to provide a spectral measurement of the surface brightness of our
Galaxy at the Solar Neighbourhood as observed from outside the Galaxy.

In this paper, we compile various data sets to measure this quantity,
which could be used as a zero-point for spectral modelling of external
galaxies. We perform a cosecant averaging based on a plane parallel
approximation for this measurement. In Sect. \ref{sec:methods}, we
explain the basis for this approximation and its expected
behaviour. We compare this approximation (integration in a cone) with
the method usually used for external galaxies (integration in a
cylinder) through simulations. In Sect. \ref{sec:synth}, we present a
synthetic approach based on the Hipparcos data set and a stellar
spectral flux library to estimate an optical spectrum of surface
brightness of our Galaxy. We also present the two different types of
integration (cone/cylinder), described here as corrected direct and
indirect methods. In Sect. \ref{sec:osb} and \ref{sec:isb}, we present
a compilation of various all-sky surveys with the cosecant averaging
method.  In the optical, we use the stellar catalogues Tycho-2 and
USNO-A2 as well as the Pioneer 10/11 integrated maps. In the
infra-red, we project in $\mathrm{cosec} b$ the COBE/DIRBE survey.
Last, we discuss all these estimates.

\section{Methods and simulations}
\label{sec:methods}
The purpose of this paper is to determine at different wavelengths the
surface brightness of our Galaxy at the Solar Neighbourhood as
observed from outside the Galaxy. (Cf. Appendix \ref{app:sb} for a
formal definition.)  However, such an observational measurement
imposes strong constraints: (1) we want to compare different
wavelengths (which could be sampled differently); (2) we would prefer
to avoid the Galactic disc area, which is affected by crowding
(stellar catalogues); (3) we would like to sample a significant volume
in order to reduce local fluctuations.

The available all-sky surveys consist in stellar catalogues or
integrated maps, at different wavelengths. In order to minimise the
systematics due to this diversity, we adopt the same averaging
procedure for each data set. The idea is to apply a cosecant
averaging on the data to estimate the surface brightness of the
Galaxy. This method, also called integration in a ``cone'', satisfies
the previous constraints, while approximating the integration in a
cylinder, usually used to compute surface brightness of external
galaxies. In this section, we present the different aspects of this
approximation with respect to the standard method. Firstly, we present
the modelling performed for our Galaxy and the methods used to compute
the surface brightness. We describe the ``integration in a cylinder'',
which corresponds to the standard method to compute a surface
brightness. We explain the difficulties due to our position near the
plane of the Galaxy, and the corrections required to estimate the
surface brightness corresponding to a face-on observation.  In
parallel, we present the alternate method used throughout this paper,
called here ``integration in a cone''. We discuss and compare the
trends of both. Secondly, we simulate the $\mathrm{cosec} b$
projection that will be applied to stellar catalogues.

\subsection{Definitions}
We consider two extreme geometries for our Galaxy: (1) an exponential
disc with scale length $h_R=2.5$\,kpc \citep{Robin:1992} and
(variable) scale height\footnote{Similarly to
\citet{Miller:1979}, we take $h_Z=80$\,pc for $M_V\le 0$, $h_Z=250$\,pc for
$M_V\ge 8$, and a linear interpolation in between. We also consider a
fixed scale height ($h_Z=120$\,pc), compatible with the variable scale
height modelling in Fig. \ref{fig:ppexpo}. See also Appendix
\ref{sect:LF}.}  $h_Z$($M_V$), with a density normalised at the Solar
neighbourhood; (2) a plane parallel model with a constant number density and
a total thickness of $2 h_Z=240$\,pc. In both cases, we assume that
the Sun is located at a cylindrical Galactocentric radius $r_q =
R_\odot = 8$\,kpc, at $z_\odot=15$\,pc above the Galactic plane
\citep{Humphreys:1995} and that the total extension of the Galaxy with
respect to the centre is $R_{gal}=14\,kpc$.  The numbers
$n(r,b,l,M_V)$ of $M_V$ stars per pc$^3$ at the position $(r,b,l)$ for
the 2 geometries are as follows:
\begin{eqnarray} 
 n^{D. Expon.}&&(r,b,l,M_V) = n_\odot(M_V)\exp(z_\odot/h_Z(M_V)) 
\nonumber\\ &&\exp({R_\odot/h_R}) \exp({-r_q/h_R}) \exp(-|z+z_\odot|/h_Z(M_V)) 
\label{eq:dexp} 
\end{eqnarray} 
with the cylindrical Galactocentric radius $r_q = {({r_\odot}^2 +
r^2 {\cos{b}}^2 - 2 r_\odot r \cos{b}
\cos{l})}^{1/2}$; the number density of
$M_V$ stars per pc$^3$ $n_\odot(M_V) = \phi(M_V)$ is the luminosity
function $\phi(M_V)$ at the Solar Neighbourhood, as defined in
Appendix \ref{sect:LF}; $z=r \sin{b}$ is the vertical position with
respect to the Sun;
\begin{equation} 
n^{PP}(r,b,l,M_V) = 
\begin{cases} 
n_\odot(M_V) \exp(z_\odot/h_Z(M_V))  & \text{if}\, |z+z_\odot| < h_Z, \\ 0 & \text{if}\, 
|z+z_\odot| \geq h_Z, 
\end{cases} 
\label{eq:pp} 
\end{equation} 
We account for extinction with a uniform dust layer following the
prescription provided and discussed in Appendix \ref{sec:ext}.
\begin{figure}
\resizebox{0.5\textwidth}{!}{\includegraphics{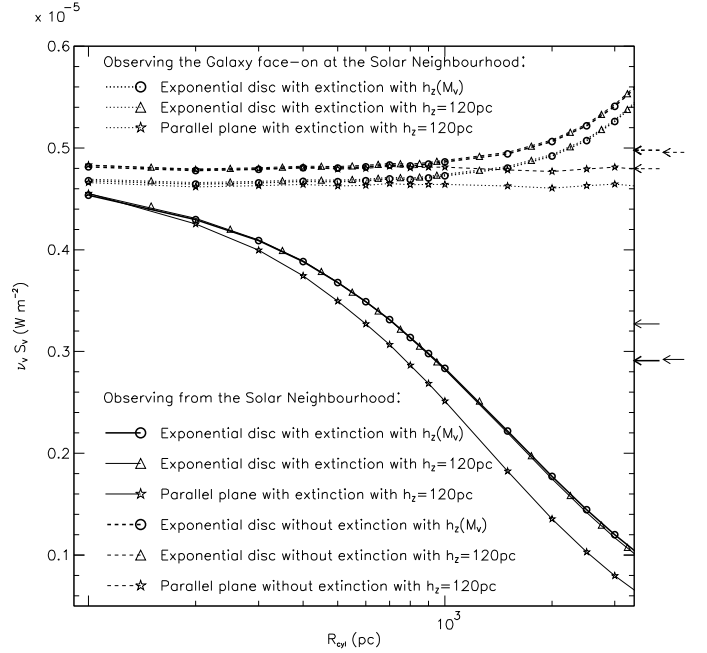}}
\caption{Simulations. Surface Brightness in V band of the Milky Way as
a function of the cylinder radius $R_{cyl}$ for the models of
Sect. \ref{sec:methods}. The cylinders corresponding to $R_{cyl}$ are
centered on the Sun and perpendicular to the Galactic plane. The
curves with symbols correspond to the surface brightness computed with
a cylinder for the exponential disc or plane parallel geometry. We
consider two different observers: one observing from the Solar
Neighbourhood and one observing the Galaxy face-on. We compare an
ideal galaxy without extinction with the real configuration with
extinction. The arrows on the right-hand side correspond to the
integration in a cone with the cosecant law approximation, fitted for
$1/\sin |b|\le 6$. The caption for the line sizes (and types) of these
horizontal lines are defined in Fig. \ref{fig:ppexpo}.}\label{fig:cyl}
\end{figure}
\begin{figure}
\resizebox{0.5\textwidth}{!}{\includegraphics{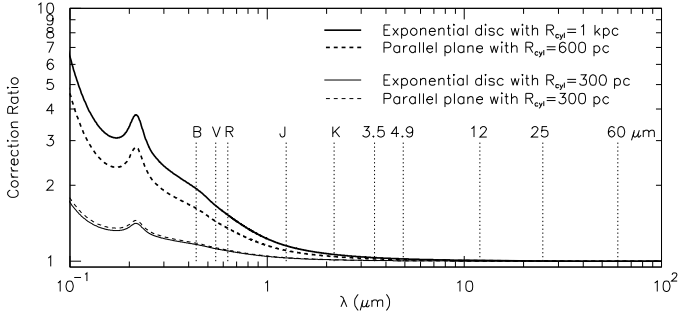}}
\caption{Correction ratio as a function of wavelength necessary to derive the
surface brightness of the Galaxy (observed face-on) from observations
at the Solar Neighbourhood. We display the corrections (thick lines)
for a cylinder of radius 1\,kpc and 600\,pc (corresponding to the
$1/\sin b < 6$ cut for the cosecant law approximation). We also
provide for comparison purposes the corresponding curves (thin lines)
for a cylinder radius of 300\,pc. For the wavelength dependence of the
extinction, we assume $R_V=3.1$ and use the prescription of
\citet{Fitzpatrick:1999}.  The different wavelengths for which we
project in $\mathrm{cosec} b$ all-sky data are indicated (and
numerical values provided in Table
\protect\ref{tab:extval}).}\label{fig:ratio}
\end{figure}

\subsubsection{Integration in a cylinder}
We define the average surface brightness $\nu S_\nu$ of our Galaxy at
the Solar neighbourhood in the V band.
\begin{eqnarray}
\label{eq:cyl}
\nu_V S_V = && [ \int_{r,b,l,{M_V}} n(r,b,l,M_V)\, 
f(A_V(r,b)) L(M_V) \nonumber\\&&r^2\cos b \, d r \, d b \, d l \, d M_V
]_{r_q<R_{cyl}} / {\pi R_{cyl}^2}
\end{eqnarray}
where $L(M_V)$ is the luminosity\footnote{We derive these luminosities
in V with the formula $L(M_V)= C_V 10^{-0.4 M_V}\times 4\pi
(D_{(=10pc)})^2$, where $C_V=1.97\times 10^{-8}$ W m$^{-2}$.} of a
star of $M_V$ magnitude, $f(A_V(r,b))=10^{-0.4 A_V(r,b)}$ is the
extinction factor and $R_{cyl}$ is the radius of the cylinder used to
compute the surface brightness (see Fig. \ref{fig:cyl}).  In the
absence of extinction and for $R_{cyl}<2$kpc, the two geometries are
equivalent and the surface brightness is nearly independent of the
cylinder radius. With extinction, the surface brightness measured from
the Solar Neighbourhood decreases with increasing cylinder
radius. This is due to the extinction affecting the $|b| < 90\deg$
lines of sight crossing the dust plane. Of course, this effect is not
observed when one computes the surface brightness for an observer of
the face-on Galaxy, as only the lines of sight\footnote{For our
Galaxy, the extinction corresponding to the total disc thickness is
0.12 Vmag.}  ($|b| = 90\deg$) perpendicular to the Galactic plane are
integrated. In Fig. \ref{fig:ratio}, we display the correction factor
necessary to convert our local estimates (measured from the Solar
Neighbourhood) into surface brightness of the face-on Galaxy (see also
Tab. \ref{tab:extval}).
\begin{figure*}
\begin{tabular}{cc}
\resizebox{0.5\textwidth}{!}{\includegraphics{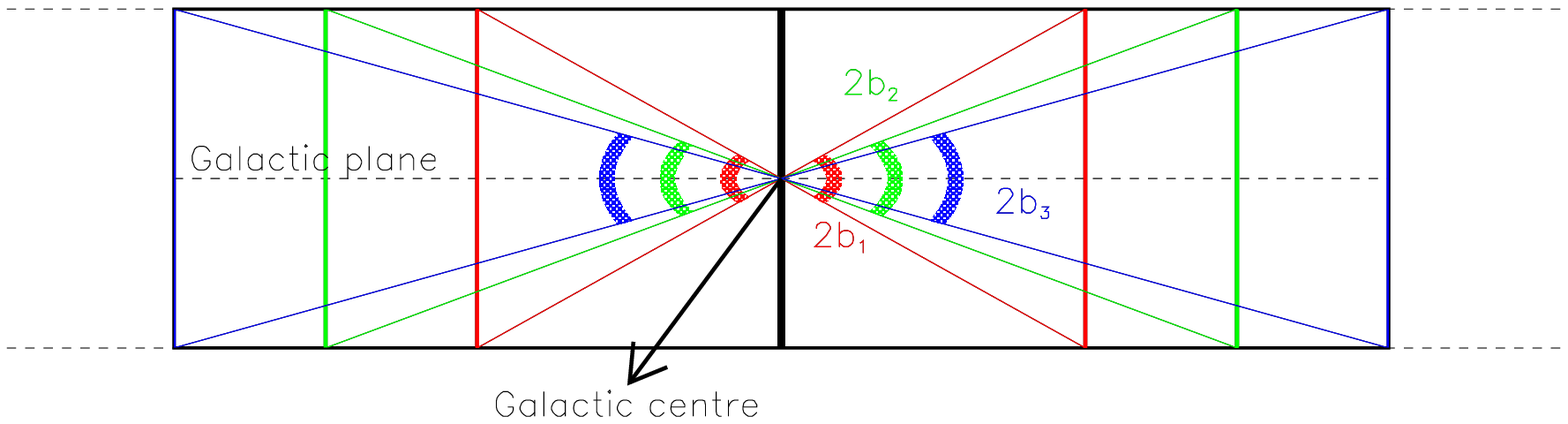}}
&
\resizebox{0.5\textwidth}{!}{\includegraphics{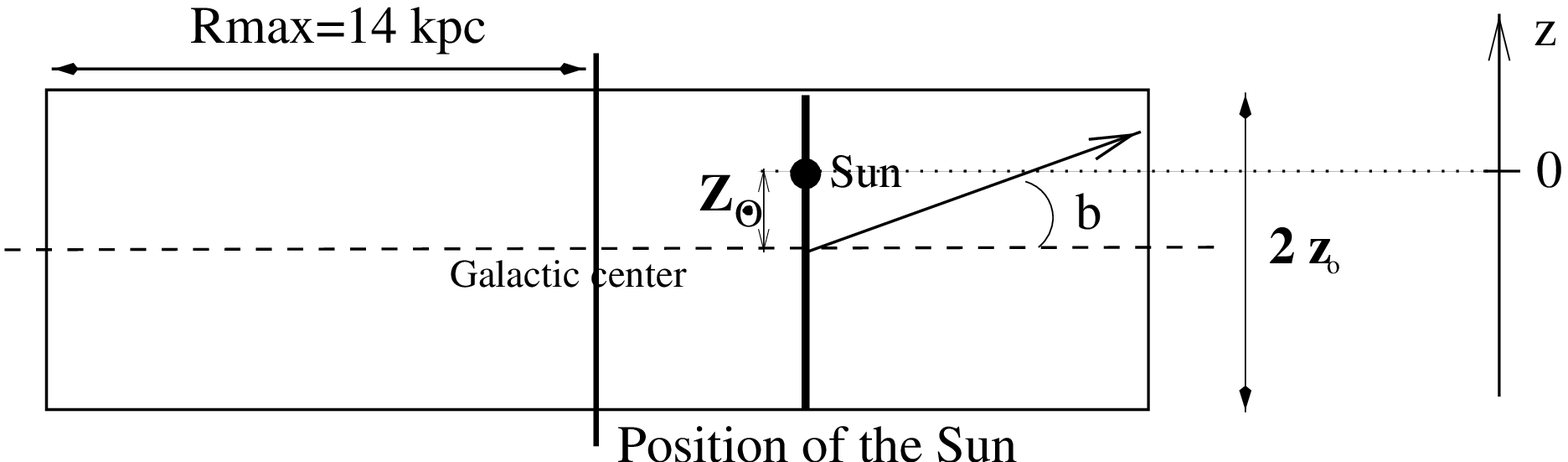}}\\
(a) & (b)
\end{tabular}
\caption{Plane parallel approximation and cosecant method. The matter
(HI gas or stars for instance) is assumed to be organised in planes
parallel to the Galactic plane. {\bf (a):} We assume an infinite thick
sheet of uniform density centred on the Galactic plane. In this
configuration, the column density perpendicular to the plane can be
derived from the integration of the column density along various
directions $b=b_1$, $b_2$ or $b_3$... as defined in
Eq. \ref{eq:def}. {\bf (b):} In this paper, we consider a truncated
sheet for the plane parallel configuration, and then study this
approximation for an exponential disc.}
\label{fig:pppeda}
\end{figure*}

\subsubsection{Integration in a ``cone'' - cosecant-law approximation}
We integrate the flux received at the Sun in a cone, according to the
cosecant-law approximation {(see Fig. \ref{fig:pppeda})}. The slope of the
corresponding function $\nu_\lambda I_\lambda(|b|)$ is the surface
brightness of our Galaxy defined above, assuming an homogeneous
stellar density galaxy.
\begin{eqnarray}  
\label{eq:con}
\nu_V I_V(|b|) = && 2 \times \int_{r,l,{M_V}} 
n(r,b,l,M_V) \nonumber\\ 
&&\frac{f(A_V(r,b)) L(M_V)}{4\pi} \cos b db \, dr \, d l \, d M_V 
\end{eqnarray}
\begin{figure}
\resizebox{0.5\textwidth}{!}{\includegraphics{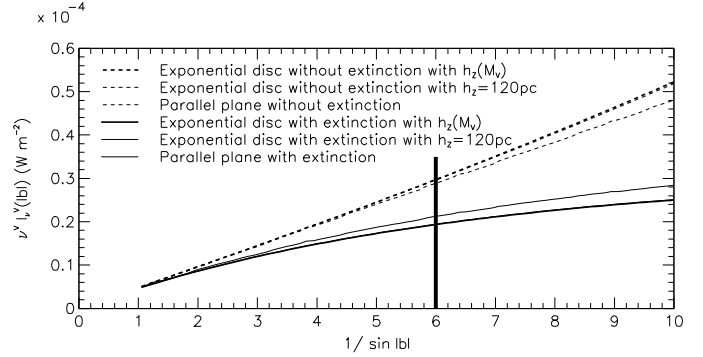}} 
\caption{Simulations. We present the cosecant laws computed
for exponential disc and plane parallel geometries.  We also consider
the effect of extinction. When not indicated otherwise, the slopes are
measured for $1/\sin |b| < 6$.  The fixed and variable $h_z$
geometries are barely distinguishable. }\label{fig:ppexpo}
\end{figure}
\begin{figure}
\resizebox{0.5\textwidth}{!}{\includegraphics{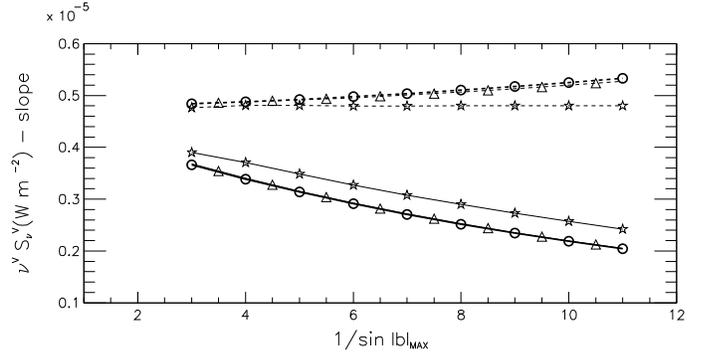}}
\caption{Simulations. V Surface Brightness derived from the cosecant
laws as a function of the chosen cut in $1/\sin |b|$. The symbols and
lines follow the same convention as Fig. \ref{fig:cyl}. The line
widths and types are also defined in
Fig. \ref{fig:ppexpo}.}\label{fig:cut}
\end{figure}
The normalisation factor in front of Eq. \ref{eq:con} is due to the
normalisation of this method with respect to the integration in a
cylinder. As explained in Appendix \ref{sect:norm}, it has been
determined for a uniform plane parallel geometry.  Figure
\ref{fig:ppexpo} displays the $\nu_V I_V(|b|)$ function, used to compute the
surface brightness. To a good approximation, the curves are linear.

In the case of a plane parallel approximation without extinction, the
slope of the $I_V(b)$ function with respect to $1/\sin b$ is equal to
the value at $b=90\deg$, namely:
\begin{equation}
\frac{I_V(b)-I_V(b=90\deg)}{1/{\sin b}-1} \sim I_V(b=90\deg) 
\label{eq:def}
\end{equation}
For a linear function, Eq. \ref{eq:def} is an equality and is
equivalent to $\nu_V S_V=I_V(b=90\deg)=I_V(b)\sin b$, where $\nu_V
I_V(b=90\deg)$ corresponds to the surface brightness. Figure
\ref{fig:ppexpo} shows that, without extinction, the exponential disc
geometry is very close to this approximation. In the presence of
extinction, the two geometries (plane parallel and exponential disc)
are also very close.  Figure \ref{fig:cut} displays the sensitivity of
the cosecant-law approximation to the cut in $(1/\sin |b|)_{MAX}$.

\subsubsection{Comparison}
\label{sect:comparison}
Plane parallel  and exponential disc geometries provide very close
results. In Figure \ref{fig:cyl}, we compare the two modes of
integration in V.

In the absence of extinction, the two modes of integration (in a cone
and in a cylinder) are equivalent for $R_{cyl} < 2$kpc. With
extinction, a flattening is observed for both methods. For integration
in a cone, it is sensitive to the cut in $1/\sin b$ used to compute
the slope (see Fig. \ref{fig:cut}). For integration in a cylinder, the
surface brightness decreases as the radius of the cylinder considered
increases (see Fig. \ref{fig:cyl}).

As displayed in Fig. \ref{fig:cyl}, we estimate that $ 1/\sin |b|\le
6$ is the best compromise to sample the surface brightness of the
Solar Neighbourhood over the whole spectral range.  The cosecant-law
approximation (computed with $1/\sin |b|\le 6$) is compatible with the
surface brightness computed in a cylinder of radius $1$\,kpc. The
plane parallel model provides a similar equivalence at $600$\,pc. For
this simple geometry, this radius can be compared to the maximum
radius probed by the integration in a cone: $h_Z/\tan b_{max}$. For
the parameter values considered, stars up to a radius of $710$\,pc are
probed. For the exponential disc, we expect a similar difference (but
asymmetric).
\begin{table}
\caption{Simulated V surface brightness expressed in different
units. The first and second columns provide the surface brightness
estimates derived from the Solar Neighbourhood (S.N.) and for the
face-on Galaxy observed at the Solar Neighbourhood. The third column
provides the estimate in the absence of extinction. The last column
gives the unit of value provided in the different lines.}
\label{tab:sb} 
\begin{center}
\begin{tabular}{rrr|l}
\hline\hline
from S.N. & Face-on $@$ S.N. & without extinction& units\\ \hline
$2.91 \times 10^{-6}$ & $4.72 \times 10^{-6}$ & $4.98 \times 10^{-6}$
& W\,m$^{-2}$\\
$9.7$ & $15.6$ & $16.5$ & L$_\odot$\,pc$^{-2}$\\
$23.90$ & $23.37$ & $23.31$ & mag\,arcsec$^{-2}$\\\hline
\end{tabular}
\end{center}
\end{table}

Table \ref{tab:sb} provides the values thus obtained for the V surface
brightness of our Galaxy at the Solar Neighbourhood for the different
cases. It is important to note that the difference between the surface
brightness estimated without extinction and the face-on Galaxy surface
brightness (with extinction) is so small that scattered light can be
neglected (less than 6$\%$ of the surface brightness).

\begin{figure}
\resizebox{0.5\textwidth}{!}  {  \includegraphics{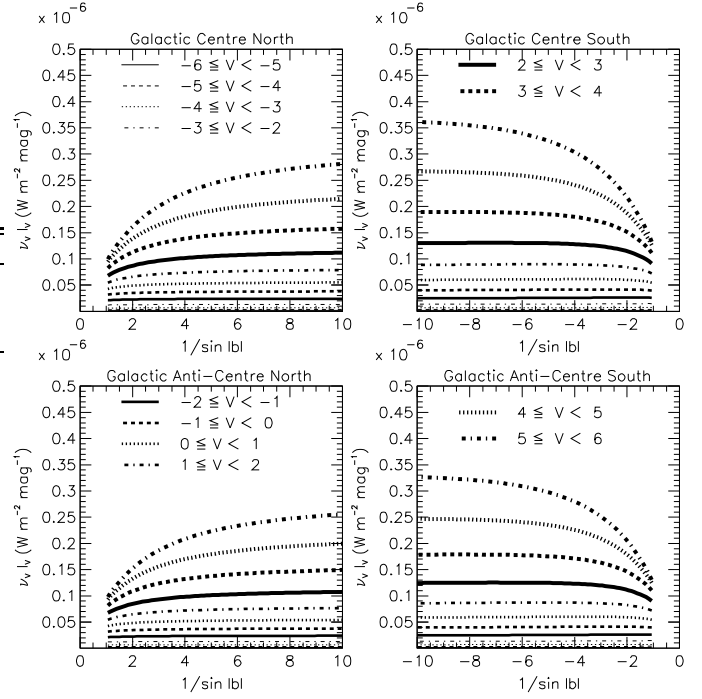}}
\caption{Simulations in V with extinction. Integrated $V$ flux in 1 mag
bins as a function of $1/\sin |b|$ for the bright stars ($V<6$).}
\label{fig:myconlisrf3_1}
\end{figure}
\begin{figure}
 \resizebox{0.5\textwidth}{!}  {\includegraphics{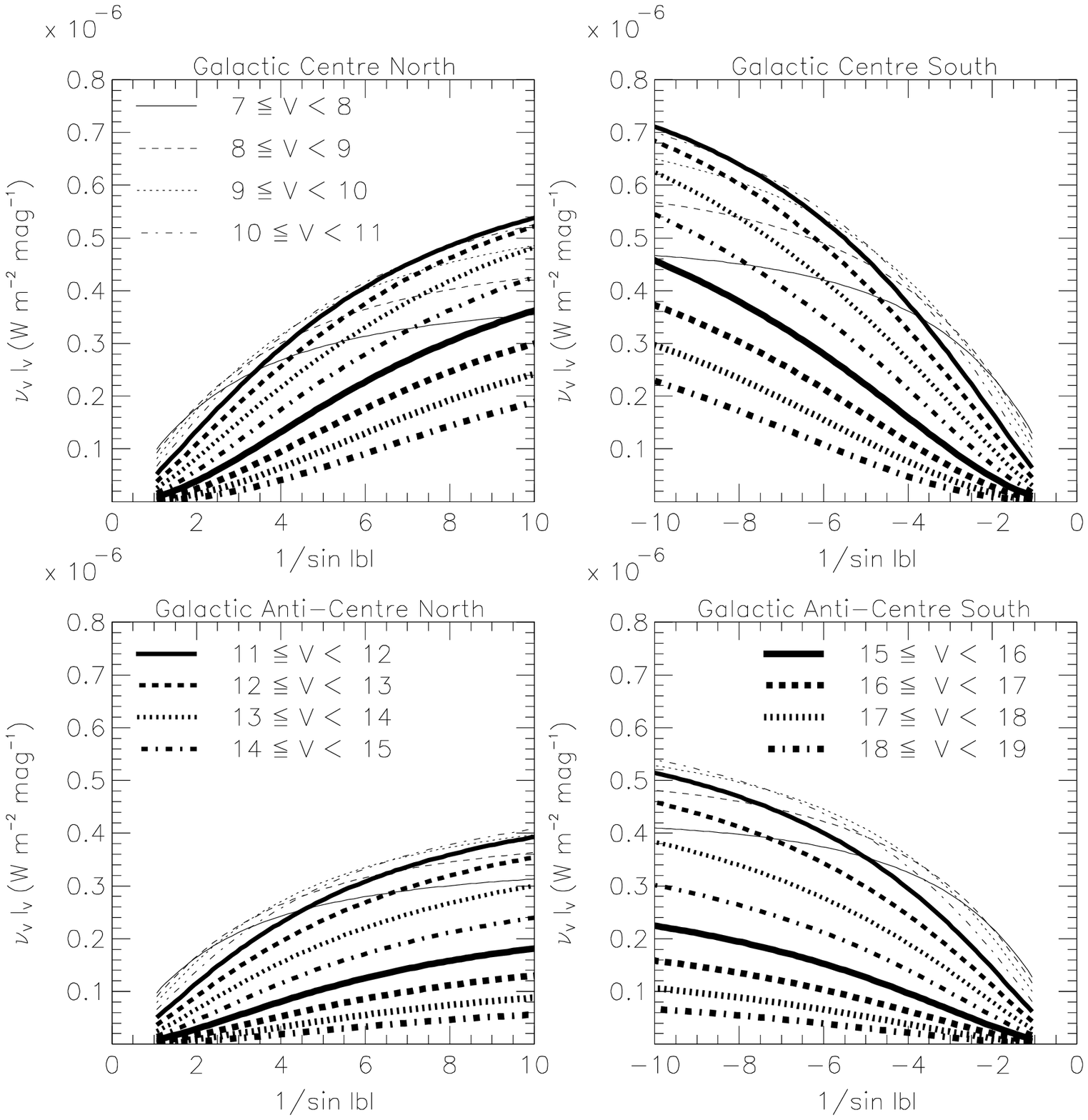}} 
\caption{Simulations in V with extinction. Integrated $V$
flux in 1 mag bins as a function of $1/\sin |b| $ for fainter
stars ($V\ge 7$).}
\label{fig:myconlisrf3_2} 
\end{figure}
\begin{figure}
 \resizebox{0.5\textwidth}{!}  {\includegraphics{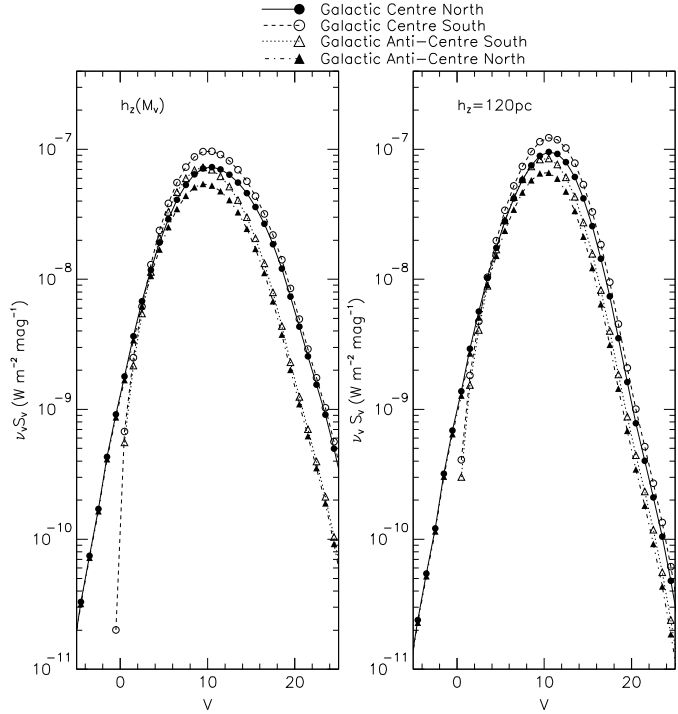}} 
\caption{Simulations in V with extinction. Contribution of each apparent
V magnitude bin to the surface brightness for 4 sky areas. On the left
(resp. right) panel, the stellar density (Eq. \ref{eq:dexp}) is
computed with a variable (resp. fixed) scale height $h_Z(M_V)$
(resp. $h_Z=120$\,pc).}
\label{fig:newconsl3}
\end{figure}

\subsection{Simulation of a catalogue reduction (observational procedure)}
\label{ssec:toy}
In the following, we will compile several whole sky stellar catalogues
to measure the surface brightness of the Solar Neighbourhood.  In
order to understand the possible trends that could affect this
observational procedure, we simulate this reduction for a V catalogue
based on the cosecant-law approximation. We take our exponential
geometry model with extinction and a variable scale height $h_Z(M_V)$
and decompose the $I(|b|)$ function into apparent V magnitude bins. We
also divide the sky into 4 areas, in order to study asymmetries
between the North and South Galactic hemispheres and the Galactic
Centre and Anti-Centre directions.  The sum of the slope of each bin
provides the surface brightness at the Solar Neighbourhood.

Figures~\ref{fig:myconlisrf3_1} and \ref{fig:myconlisrf3_2} display
this simulation for $V<6$ and $V\ge 7$ stars.  The flux is higher in
the direction of Galactic Centre. We observe a significant flattening
for large $1/\sin|b|$, due to extinction. We also clearly detect the
asymmetric position of the Sun above the Galactic plane
($z_\odot=15$pc). The left (resp. right) panel of
Fig. \ref{fig:newconsl3} summarises the contribution of each apparent
V magnitude bin to the surface brightness for $h_Z(M_V)$
(resp. $h_Z=120$\, pc). Table \ref{tab:simu} summarises the
corresponding contribution per apparent magnitude bins. It also
provides the fitted value at $b=90^\circ$. 

Each bin with $V<2$ and $V>21$ stars contributes to no more than about
$1\%$ of the surface brightness, while the main contributors (86.5$\%$
of the total surface brightness) are $5 < V < 16$ stars. Uncertainties
on the very bright stars, which are often missed in stellar catalogue
(due to their saturation, extinction in the stellar disc, etc.), and
dim stars (close to the detection limit) will not bias the surface
brightness estimate.

Note that this method, uncorrected for extinction, provides an
estimate of the surface brightness computed from the Solar
Neighbourhood, and has to be corrected as indicated in
Fig. \ref{fig:ratio} and Table \ref{tab:extval} to obtain a face-on
value.

\section{Stellar Surface Brightness: a synthetic approach with the Hipparcos data}
\label{sec:synth}
The Hipparcos Catalogue \citep{Perryman:1997,vanLeeuwen:1997} with the
Hipparcos Input Catalogue (HIC) \citep{Turon:1992,Turon:1995}
constitute a unique database to explore the Solar
Neighbourhood. Spectral type, parallax, and V magnitude are
provided for each $V<7.3$ star. We restrict the sample to $13\,374$
stars with a $3\sigma$ parallax and with the constraints on the
completion limits described in Appendix \ref{sect:LF}. In the
following, we use the stellar spectral flux library from
\cite{Pickles:1998} to produce synthetic spectra: a spectrum is
associated with each star and normalised to the observed V
flux. Hipparcos stars are then combined in two different ways to get
estimates of the surface brightness. In Sect. \ref{ssec:corrdirect},
we directly sum the flux of the observed stars with a $\sin |b|$
weighting. However, this direct measurement requires a volume
correction due to the apparent magnitude limit of the sample
used. This corrected direct measurement corresponds to the cosecant
averaging (integration in a cone).  In Sect. \ref{ssec:indirect}, we
compute the surface brightness with the luminosity of the stars with
the integration in the cylinder approach. Appropriate volume
corrections are also required.
\begin{figure}
 \resizebox{0.5\textwidth}{!}  {\includegraphics{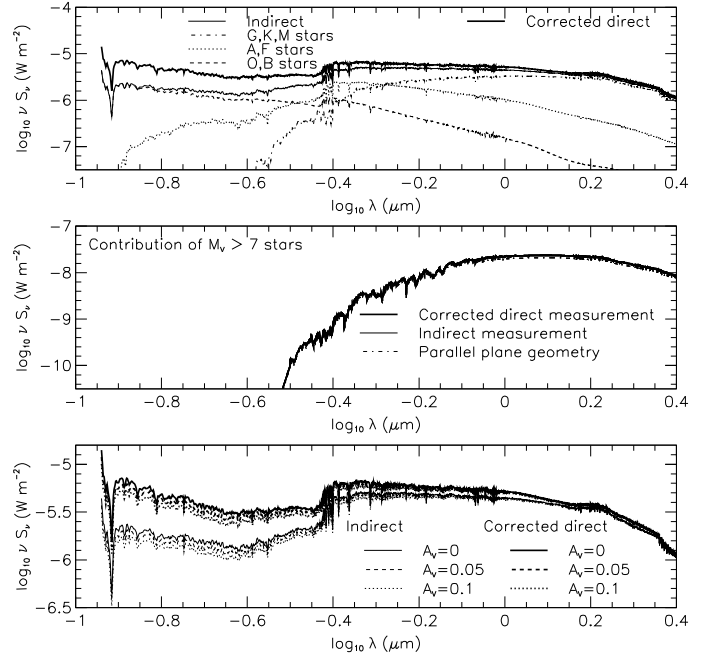}} 
\caption{Corrected direct and indirect synthetic estimates of the
surface brightness of our Galaxy at the Solar Neighbourhood. In the
top panel, we display the two estimates obtained with the compilation
of the $V<7.3$ Hipparcos Catalogue stars. For the indirect estimate,
we indicate the contribution of the different types of stars. In the
middle panel, we display the contribution of $M_V > 7$ stars with our
simulation tool for different configurations. We also consider $\pm 1
\sigma$ luminosity functions to estimate the scatter due to
uncertainties in the luminosity function. These dim stars mainly
contribute to the near-infrared part of the spectrum, but this
integrated surface brightness remains an order of magnitude below
their bright counter-parts. In the bottom panel, we study the
influence of the extinction, perpendicular to the Galactic plane, which
might have a marginal influence in the UV. We assume a fixed
extinction $A_V$ applies to the whole population.}
\label{fig:synth}
\end{figure}

\subsection{Corrected direct measurement of starlight: a synthetic
spectrum}
\label{ssec:corrdirect}
The cosecant-law approximation (see Eq. \ref{eq:def}) enables to
compute directly the surface brightness as the sum of the stellar
fluxes weighted by $\sin |b|$.  Even though the amount of dust is
small at the Solar neighbourhood, it significantly affects the surface
brightness computation (see Fig. \ref{fig:cyl}). We have shown in
Sect. \ref{sec:methods} that the surface brightness of our Galaxy
observed face-on is very close to the surface brightness computed for
a Galaxy without dust. Hence, we subtract the extinction to each star
relying on our dust modelling (see Appendix \ref{sec:ext}), and obtain
the lower bound:
\begin{eqnarray}
\nu S_\nu  &>&
2 \times \sum_i^{\mathrm{all\, stars}} \nu F_\nu(M_V,SpT)/f(A_V)
\times \sin |b| \nonumber \\ &>& 2 \times \sum_i^{\mathrm{all\,
stars}} \frac{L_\lambda(M_V,SpT)}{4\pi r_i^2} \times \sin |b|
\label{eq:direct}
\end{eqnarray}
where $F_\nu(M_V,SpT)$ (resp.\ $L_\lambda(M_V,SpT)$) is the flux
(resp.\ luminosity) spectrum in W~m$^{-2}$~Hz$^{-1}$ (resp.\ W)
corresponding to a star with a magnitude $M_V$ and a spectral type
SpT, and $r_i$ the actual distance of this star.

According to our simulations (Table \ref{tab:simu}), the cut in
apparent magnitude ($V<7.3$) limits our sensitivity to 17\% of the
total $V$ stellar surface brightness. We apply a volume correction
relying on the knowledge of the Hipparcos parallaxes together with the
well-defined completeness limit ($V<7.3$): each absolute magnitude bin
($M_V-\frac{1}{2} \le m < M_V+\frac{1}{2}$) is complete within a
radius $R(M_V+\frac{1}{2})$ (see Appendix \ref{ssect:LF2}).  We compute
\begin{eqnarray} 
\nu S_\nu = 2 \times \sum_i^{\mathrm{M_V<7}} \sum^{x_i \in}_{{ \left[M_V-\frac{1}{2},
M_V+\frac{1}{2}\left[\right.\right.}}
\frac{L_\lambda({x_i},{SpT})\sin |b|}{4\pi r^2} \times   f^{LF}_i
\nonumber \\
\times \frac{W^{sph}_*(R_{max})}{W^{sph}_*(R(x_i))}&  
\label{eq:corrdirect} 
\end{eqnarray}
here $L_\lambda({x_i},SpT)$ is the spectrum (in W) corresponding to a
star with a magnitude $M_V-\frac{1}{2}\le {x_i} < M_V+\frac{1}{2}$ and
a spectral type SpT, $ f^{LF}_i$ is the correction of the luminosity
function provided in Fig. \ref{fig:lfcorr} applied for stars with $M_V
\le -1.5$, and $R_{max}=1$kpc is the cylinder radius discussed in
Sect. \ref{sec:methods}. The last factor corresponds to the volume
correction with: $W^{sph}_*(y) = \int_{r<y} n(r,b,l) \sin|b|/r^2 dv $,
$dv=\cos(b)\,db\,dl\,dr $. $W^{sph}_*(y)$ corresponds to the stellar
surface density weighted by $\sin|b|/r^2$ computed within a sphere of
radius $y$. $W^{sph}_*R(x_i))$ is computed for each magnitude bin,
while $W^{sph}_*(R_{max})$ is obtained with a weighted sphere with the
radius of the cylinder radius $R_{max}$. The corresponding spectrum is
displayed in Figure \ref{fig:synth}. We check that if we take
$W^{sph}_*(5 R_{max})$ instead of $W^{sph}_*(R_{max})$, the estimate
of the surface brightness increases by less than $14.5\%$.

This measurement takes into account the real distribution of stars but
assumes an smooth and homogeneous distribution of stars of the LF
along the exponential profile of the Milky Way for the corrections.

\subsection{Indirect measurement of starlight: a synthetic spectrum}
\label{ssec:indirect}
The second possibility is to compute the surface brightness as the
mean stellar luminosity spectrum per unit area at the Solar
Neighbourhood. We rely on Hipparcos data for intrinsically bright stars
($M_V<7$) and perform the appropriate corrections, as follows:
\begin{gather}
 \nu I_\nu = \sum_{i}^{M_V<7}  
 \sum^{x_i\in}_{\left[M_V-\frac{1}{2}, M_V+\frac{1}{2}\left[\right.\right.} \frac{L_\lambda(x_i,SpT)}
 {\pi R(x_i)^2} \times f^{LF}_i \times \frac{\Sigma^{cyl}_i}{\Sigma^{sph}_i}
\end{gather} 
where $L_\lambda(M_V,SpT)$ is the spectrum (in W) corresponding to a
star with a magnitude $M_V$ and a spectral type SpT, $ f^{LF}_i$ is
the correction of the luminosity function.  The last factor
corresponds to the volume correction for each magnitude
bin. $\Sigma^{cyl}_i$ is the stellar surface density computed for the
cylinder of radius $R_{max}$, while $\Sigma^{sph}_i$ is the stellar
surface density integrated over a sphere of radius $R(x_i)$ normalised
to the surface $\pi R(x_i)^2$.

This measurement assumes a smooth homogeneous distribution of the
stars and a uniform LF along the exponential profile of the Milky Way.

\subsection{Discussion}
\label{ssect:dis}
Figure~\ref{fig:synth} presents the corrected direct and indirect
estimates.  The two curves are very close in the optical and  		 
near-infrared, while they differ significantly in the UV. This
difference, which cannot be explained with the extinction (see bottom
panel of Fig. \ref{fig:synth}), is understood as due to the
inhomogeneous distribution of the OB associations
 (\citealt{deZeeuw:1999}, see also Appendix \ref{ssect:3D}). If we remove
the very bright stars $(M_V<-4.5)$, the corrected direct estimate is
reduced in the UV. If we remove the bright stars $(M_V<-1.5)$, all the
discrepancies are removed. 	

The corrected direct estimate, based on the fluxes, is very sensitive
to the anisotropic distribution of OB associations, while the indirect
estimate is more robust, as it is based on luminosities, but does not
account for the real spatial distribution of stars. The uncertainties
on the bright end of the LF (see Fig. \ref{fig:lf}) are also a source
of error.  By chance, our indirect estimate is in good agreement with
the UV measurements of \citet{Gondhalekar:1980} (see
Fig. \ref{fig:last}). However, it is a different quantity that has
been measured: the sum of the UV light over the whole sky (S2/68
Sky-survey TD1-satellite data) without any extinction correction.
This might mean that, due to an obvious selection bias, we detect only
stars that are moderately extinguished, but this might also be a
coincidence as the measured quantities are different.

The significant difference of our two estimates in the UV range is due
to the irregular distribution of OB stars (considered homogeneous in
the indirect estimate).

The surface brightness obtained in the optical can be compared with
the study of \citet{Flynn:2006}. These authors use accurate data on
the local luminosity function and the disc's vertical structure to
measure the following surface brightnesses (see their Table 5):
$\mu_B=23.51$ mag.arcsec$^{-2}$, $\mu_V=22.93$ mag.arcsec$^{-2}$, and
$\mu_I=22.03$ mag.arcsec$^{-2}$. These values are in good agreement
with our corrected direct estimates (see Tables \ref{tab:opt} and
\ref{tab:IR}). Our indirect estimates are slightly weaker for the
reasons discussed above.

\section{Optical surface brightness}
\label{sec:osb}
\subsection{USNO-A2 and TYCHO-2 data}
\begin{figure}[Hb]
 \resizebox{0.5\textwidth}{!}{\includegraphics{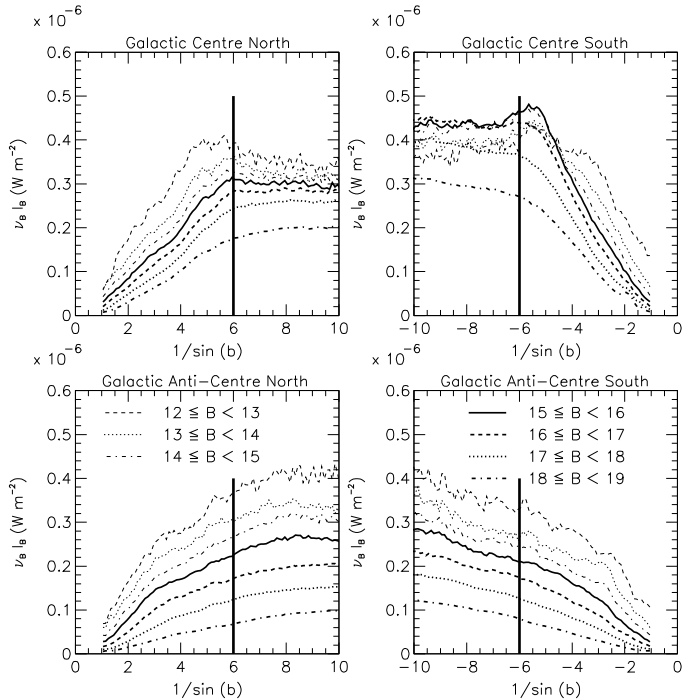}}
 \caption{Cosecant laws for $12\le B<19$ stars from the USNO-A2
 catalogue. Irregularities, significant towards the Galactic
 centre, are minimised by our cut $1/\sin |b| < 6$. (See Table \ref{tab:usnob4}.)}\label{fig:usnob4}
\end{figure}
\begin{figure}
\resizebox{0.5\textwidth}{!}{ \includegraphics{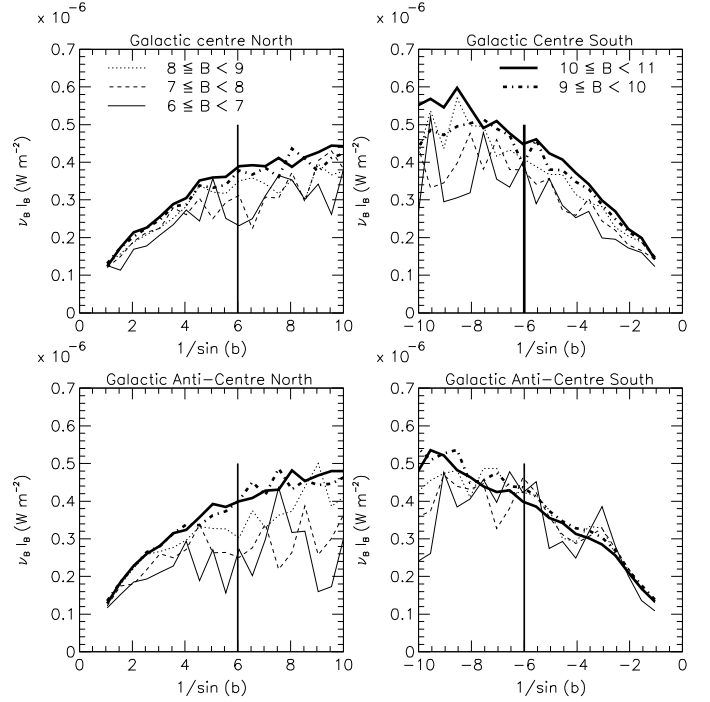}}
\caption{Similar to Fig.~\ref{fig:usnob4} for B 
 cosecant laws computed for $6\le B<11$ TYCHO-2 stars. (See Table
\ref{tab:usnob4}.)}\label{fig:tychob4}
\end{figure}
\begin{table}
\begin{center}
\caption{Contribution of the $B<6.$ stars to the surface
brightness (see Fig. \ref{fig:bright}). $2\,783$ Hipparcos stars are
thus been used.  Values with the observed
extinction are provided together with the values corrected for
extinction.}
\label{tab:bhipp6} 
\begin{tabular}{rrr}\hline\hline
$\lambda$ ($\mu$m) & With extinction & Without extinction \\\hline
0.440 ($B_J$) & 5.640 $\times 10^{-7}$ & 6.186 $\times 10^{-7}$\\\hline
\end{tabular}
\end{center}
\end{table}
The USNO-A2 B and R magnitudes have been calibrated as explained in
Appendix \ref{sec:calib}.
\subsubsection{B measurements}
\label{sssec:bmeas}
Figures \ref{fig:usnob4} and \ref{fig:tychob4} display the cosecant
laws obtained for 4 sky areas with the USNO-A2 and TYCHO-2 stellar
catalogues and for each magnitude bin.  Table \ref{tab:usnob4}
summarises the slopes measured with the fit to the cosecant laws and
provides a comparison with simulations. These simulations are based on
the method described in Sect. \ref{ssec:toy}. A B luminosity function
has been derived from the V luminosity function displayed in
Fig. \ref{fig:lf}, assuming that 50$\%$ of the $V<2$ stars are red
giants. We relying on \citet[p:137-145]{Lang:1992} to relate the V
absolute magnitudes to spectral types from \citet{Pickles:1998}, and
to perform this transformation. We thus find excellent agreement
between our simulations and the TYCHO-2 data. The USNO-A2 data provide
estimates systematically larger than our simulations. The emission in
the Galactic Centre South direction is perturbed as displayed in
Fig. \ref{fig:usnob4}.  Note that simulations show that $B>19$ stars
do not contribute more than a few percent to the total surface
brightness.

\begin{figure}
\resizebox{0.5\textwidth}{!}  {\includegraphics{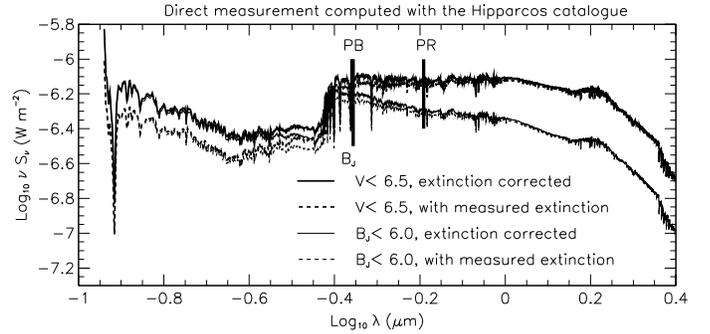}}
\caption{Direct contribution of bright stars to the surface
brightness. These synthetic estimates have been computed with
Eq. \ref{eq:direct} for $V<6.5$ stars and for $B<6$ stars from the
Hipparcos catalogue. $8\,013$ and $2\,783$ stars have thus been used,
while the few neglected stars do not contribute to more than 10$\%$ in
V (see Table \ref{tab:dir65}). The vertical lines indicate the
wavelengths discussed in the text (B$_J$ and Pioneer B/R). Note that
the extinction considered here is the extinction affecting each
individual star not the global extinction perpendicular to the
Galactic plane.}
\label{fig:bright} 
\end{figure}

We combine the B USNO-A2 and TYCHO-2 estimates for $B>6$ stars with
the synthetic uncorrected estimate based on ($B<6$) Hipparcos stars
(see Table \ref{tab:bhipp6} and Fig. \ref{fig:bright}), and correct
the extinction with the global factor provided in
Fig. \ref{fig:ratio} and Table \ref{tab:extval}, as follows:
\begin{eqnarray}
\label{eq:bsb}
\nu_B S_{B} = && (5.640 \times 10^{-7} + 3.004 \times 10^{-6}) \times
1.72 \nonumber \\ = && 6.13 \times 10^{-6} {\mathrm W}\, {\mathrm
m}^{-2} =23.51\, {\mathrm {mag}} \, {\mathrm {arcsec}}^{-2} \\
\end{eqnarray}

\begin{figure}
 \resizebox{0.5\textwidth}{!}{\includegraphics{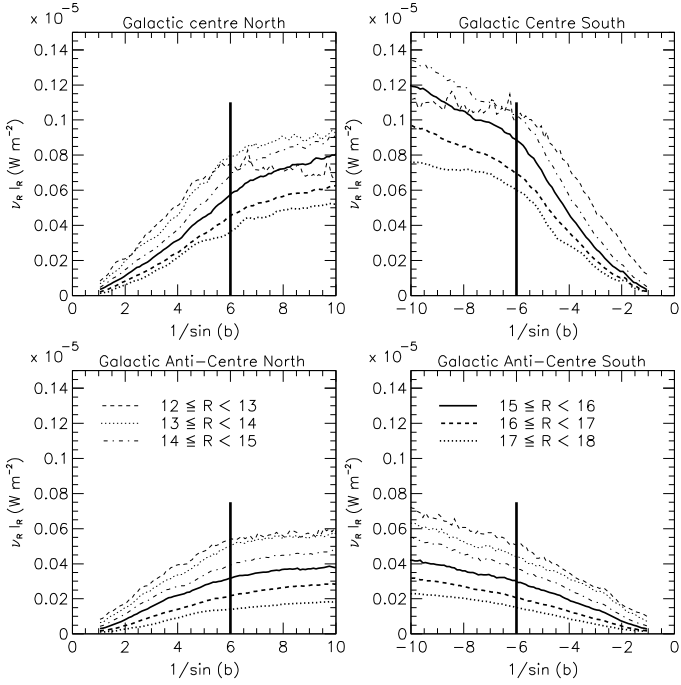}}
 \caption{Integrated R fluxes as a function of
\protect$1/\sin (b)$ computed with the USNO-A2 data for $12\le R<18$
stars. We decompose the contributions originating from different
sky areas. No correction has been applied. (See Table
\ref{tab:usnor4}.) }\label{fig:usnor4}
\end{figure}

\begin{table}
\begin{center}
\caption{Contribution of the $R<6.$ stars to the surface
brightness computed the same way as those displayed in
Fig. \ref{fig:bright}. $7\,010$ Hipparcos stars are thus considered. Values
with the observed extinction are provided together with the values
corrected for extinction.}
\label{tab:dirr6} 
\begin{tabular}{rrr}\hline\hline
$\lambda$ ($\mu$m) & With extinction & Without extinction \\\hline
0.640 (R) &  $ 6.98\times 10^{-7}$ &  $7.41 \times 10^{-7}$\\\hline
\end{tabular}
\end{center}
\end{table}

\subsubsection{R measurements}
\label{sssec:rmeas}
The data reduction of the R measurements of the calibrated USNO-A2
data is presented in Table \ref{tab:usnor4} and Figure
\ref{fig:usnor4}. We clearly observe an excess of the stellar data
with $12<R<18$ with respect to our simulation\footnote{{\bf To account
for red giants,} we apply in R the modification of the luminosity
function described in Sect. \ref{sssec:bmeas} to perform this
simulation.}. This cannot not be accounted for the thick disc
dominated by dim stars ($M_V>4$), which do not contribute significant
to the surface brightness (see Fig. \ref{fig:newconsl2}). We carefully
check with the USNO-B1
\citep{Monet:2003} catalogue that our calibration is satisfactory. In
addition, this excess has been confirmed with the CCD UCAC-1 data (see
Appendix \ref{sec:compucac}).

We complete the $12\le R<18$ USNO-A2 estimate with simulations
presented in Table \ref{tab:usnor4}, and correct the extinction with
the global factor provided in Fig. \ref{fig:ratio} and Table
\ref{tab:extval}, as follows:
\begin{eqnarray}
\label{eq:rsb}
\nu_R S_{R} = && (2.205 \times 10^{-6} + 2.555 \times 10^{-6}) \times
1.41 \nonumber \\ = && 6.71 \times 10^{-6} {\mathrm W}\, {\mathrm
m}^{-2} =22.6\, {\mathrm {mag}} \, {\mathrm {arcsec}}^{-2} \\
\end{eqnarray}

\begin{table*}
\begin{center}
\caption{$B$(0.437$\mu$m) and $R$(0.644$\mu$m) (Pioneer 10/11
derived) surface brightness, given in $10^{-9}\, \rm W\,m^{-2}$. We
give the values fitted for $1/\sin (b) <6$ (see Fig. \protect
\ref{fig:pion4}).  The surface brightness thus
obtained accounts for the $V>6.5$ stars as well as for a possible
diffuse component.}
\label{tab:pion} 
\begin{tabular}{rrrrrrrrr}
\hline\hline
&\multicolumn{4}{c}{Galactic Centre}
&\multicolumn{4}{c}{Galactic Anti-Centre}\\
$\lambda$ ($\mu$m)&\multicolumn{2}{c}{North} &
\multicolumn{2}{c}{South} & \multicolumn{2}{c}{North} &
\multicolumn{2}{c}{South}\\ &
{$|b|=90\degr$} & {slope}&
{$|b|=90\degr$} & {slope}&
{$|b|=90\degr$} & {slope}&
{$|b|=90\degr$} & {slope}\\\hline
0.437 (B)  & $1050.$ & $\mathbf{1148.}$ & $1048.$ & $\mathbf{1444.}$ & $1046.$ & $\mathbf{886.}$ & $996.$ & $\mathbf{1040.}$\\
0.644 (R)  & $1368.$ & $\mathbf{1684.}$ & $1542.$ & $\mathbf{1958.}$ & $1320.$ & $\mathbf{1164.}$ & $1500.$ & $\mathbf{1234.}$\\
\multicolumn{9}{c}{}\\
\multicolumn{9}{c}{Sum}\\ \hline
0.437 (B) & \multicolumn{4}{c}{$2098.$ ($|b|=90\degr$), $2592.$ (slope)}
&\multicolumn{4}{c}{$2042.$ ($|b|=90\degr$), $1926.$ (slope)}\\ 0.644 (R)  &
\multicolumn{4}{c}{$2910.$ ($|b|=90\degr$), $3642.$ (slope)}
&\multicolumn{4}{c}{$2820.$ ($|b|=90\degr$), $2398.$ (slope)}\\ \hline
\multicolumn{9}{c}{}\\
\multicolumn{9}{c}{\bf Total B (0.437$\mu$m) value: $4140.$
($|b|=90\degr$), $4518.$ (slope)}\\
\multicolumn{9}{c}{\bf Total R (0.644$\mu$m) value: $5730.$
($|b|=90\degr$), $6040$ (slope)}\\ \hline \hline
\end{tabular}
\end{center}
\end{table*}
\begin{figure}
\resizebox{0.5\textwidth}{!}{ \includegraphics{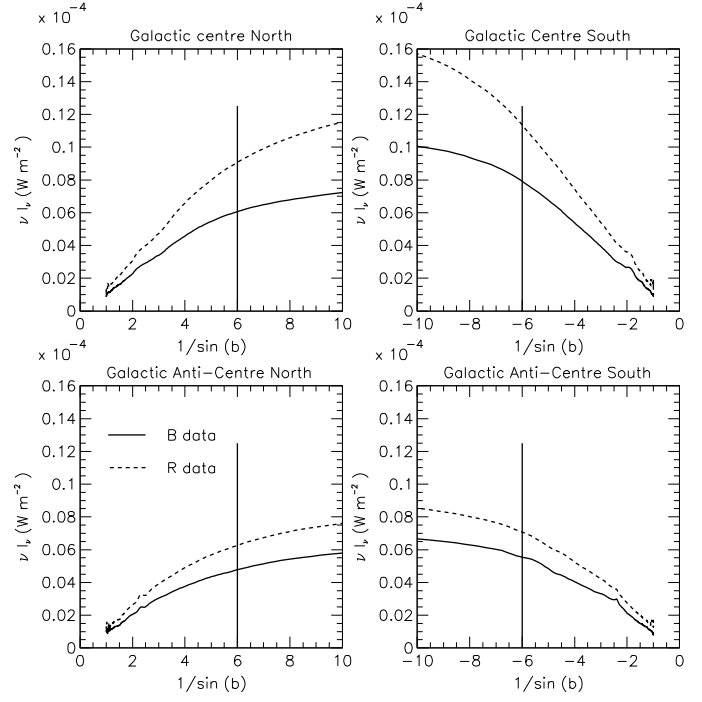}}
\caption{Cosecant laws for Pioneer 10/11 B (full-line) and R
(dashed-line) data \protect \citep{Gordon:1998}. This corresponds to
$V>6.5$ stars and diffuse emission. There is a hole in the Galactic
Anti-Centre North area (bottom left) that has been removed. No
correction has been applied. (See Table \protect
\ref{tab:pion}.)}\label{fig:pion4}
\end{figure}
\begin{table}
\begin{center}
\caption{Contribution of the $V<6.5$ stars to the surface
brightness (see Fig. \ref{fig:bright}). Values with the observed
extinction are provided together with the values corrected for
extinction. We estimate that we do not miss$^1$ more than 3$\%$ of the
brightness in V.}
\label{tab:dir65} 
\begin{tabular}{rrr}\hline\hline
$\lambda$ ($\mu$m) & With extinction & Without extinction \\\hline
0.437 (B) & 7.020 $\times 10^{-7}$ & 7.751 $\times 10^{-7}$\\
0.644 (R) & 7.261 $\times 10^{-7}$ & 7.699 $\times 10^{-7}$\\
V bad stars$^1$ & 2.465$\times 10^{-8}$& \\ \hline\hline
\end{tabular}
\end{center}
$^1$ This corresponds to stars with uncertain parallax or unspecified
spectral type neglected for the previous direct estimates.
\end{table}

\subsection{Pioneer 10/11 data}
\label{ssec:pion}
As discussed in \cite{Gordon:1998} (see also \citealt{Leinert:1998}),
the Pioneer 10/11 data provide an excellent all-sky survey in B
(0.437$\mu$m) and R (0.644$\mu$m) for the study of integrated flux.
In contrast to the Tycho star mapper background analysis of
\cite{Wicenec:1995}, the Pioneer 10/11 data taken beyond 3.26 AU
provide sky-maps devoid of detectable zodiacal light. We use here the
Pioneer 10/11 maps kindly provided by K. Gordon (see
\citealt{Gordon:1998} and Witt, Gordon \& Cohen, in preparation): they
correspond to the first iteration maps discussed in
\cite{Gordon:1998}. These maps contain all relevant flux (starlight
and possible diffuse components), except for stars brighter than
$V=6.5$ (see \citealt{Toller:1987}).  Figure \ref{fig:pion4} displays
the $1/\sin(b)$ laws obtained in B and R for the different sky areas,
while Table \ref{tab:pion} provides the corresponding slopes.  The R
fluxes are systematically larger than the B fluxes: as discussed in
the following, this is a consequence of the extinction.

Relying on Eq. \ref{tab:dir65}, we compute the contribution of the
bright ($V<6.5$) stars, independently with a synthetic direct estimate
based on the Hipparcos Catalogue, which is complete for these
stars. We nevertheless neglect stars with uncertain parallax or
undefined spectral type, but estimate that they do not contribute more
than 10$\%$ in V. 

We combine the Pioneer B and R estimates with the ($V<6.5$) stars
synthetic uncorrected estimates, and correct the extinction with the
global factor provided in Fig. \ref{fig:ratio} and Table
\ref{tab:extval} (see also Appendix \ref{sec:ext}), as follows:
\begin{eqnarray}
\label{eq:pion}
S_{B} = && (7.02 \times 10^{-7} + 4.52 \times 10^{-6}) \times 1.72
\nonumber \\ = && 8.98 \times 10^{-6} {\mathrm W}\, {\mathrm m}^{-2}
=23.1\, {\mathrm {mag}} \, {\mathrm {arcsec}}^{-2} \\ S_{R} = && (7.26
\times 10^{-7} + 6.04 \times 10^{-6}) \times 1.41
\nonumber \\ = && 9.54 \times 10^{-6} {\mathrm W}\, {\mathrm m}^{-2}
=22.3\, {\mathrm {mag} } \, {\mathrm {arcsec}}^{-2}
\end{eqnarray}
Our estimate is brighter than the estimate of the surface brightness
obtained in B by \citet{vanderKruit:1986} with the Pioneer data
($\mu_B = 23.8\pm 0.1$mag \, arcsec$^{-2}$), who did not correct for
extinction.

\begin{figure}
\resizebox{0.5\textwidth}{!}  { 
\includegraphics{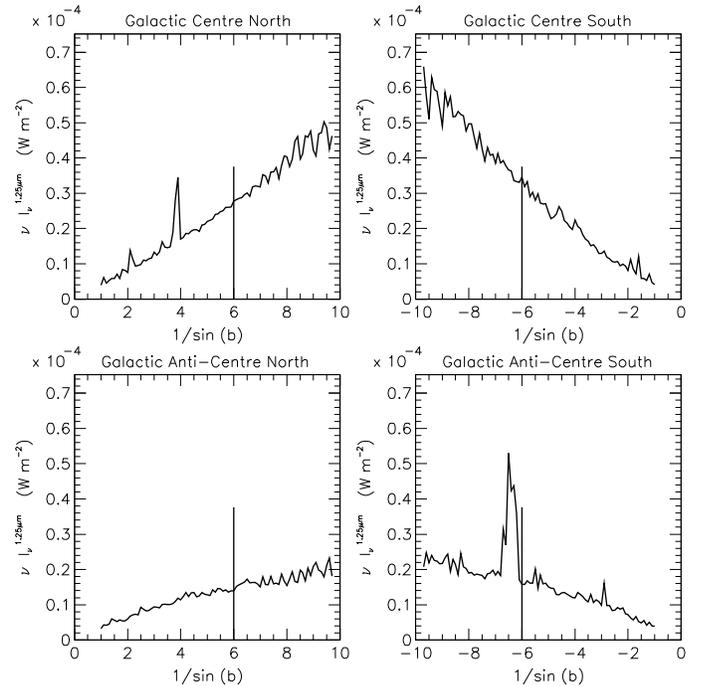}} 
\caption{Cosecant laws for the 1.25 $\mu$m COBE/DIRBE data averaged
over Galactic latitude. These values are derived from Zodi-Subtracted
Mission Average (ZSMA) data distribution. The sharp peaks correspond
to bright stars ($0<V<1$ and M type). We performed similar plots for
the other 9 COBE/DIRBE available wavelengths. Wiggles observed for
$1/\sin| b|>6$ are due to the COBE/DIRBE sampling (pixels of
0.32$\degr\times$0.32$\degr$).}\label{fig:cobe}
\end{figure}

\vspace{1cm}

The values obtained here are systematic brighter than our synthetic
estimates and than the local values provided by \citet{Flynn:2006}. As
later discussed in Sect. \ref{sec:dis}, we suggest that this disagreement is
due to a larger volume probed with the USNO-A2 and Pioneer 10/11
all-sky surveys.

\section{Infrared surface brightness: COBE/DIRBE data}
\label{sec:isb}
We consider the Zodi-Subtracted Mission Average (ZSMA) maps
\citep{Kelsall:1998} from the COBE/DIRBE project. We plot the cosecant
laws for the wavelengths 1.25, 2.2, 3.5, 4.9, 12, 25, 60, 100, 140 and
240$\mu$m as shown in Figure \ref{fig:cobe} (for 1.25$\mu$m). We then
perform a linear fit (slope and $|b|=90{\degr}$ values) with the
parameters provided in Table \ref{tab:cobe}.  We are sensitive to
bright point sources (like Betelgeuse) up to 4.9$\mu$m, that we
eliminate with a $5\sigma$ clip when necessary. Unlike
\cite{Boulanger:1988}, we do not subtract nearby dust complexes from
the study as we are interested in the light of all the components.
Otherwise, we observe quite a linear behaviour except at 12 and 25
$\mu$m where the light emission is very irregular, as observed by
\cite{Boulanger:1988} with IRAS data. These data exhibit a large value
of the $|b|=90{\degr}$ intensity and a small value of the slope.  This
large $|b|=90\degr$ intensity is characteristic of a ``non-Galactic''
or at least irregular behaviour, and also seems to affect marginally
the 4.90$\mu$m data. At 12 and 25 $\mu$m, we observe an unexplained
excess for $1/\sin|b|<2$ ($b>30\degr$), which affects the fit. A
strange behaviour was also noted by \cite{Boulanger:1988} in this area
and interpreted as a consequence of imperfect Zodiacal light
subtraction.  Hence, we perform the fit for $2<1/\sin|b|<6$ at these
wavelengths in order to minimise the effect, but these measurements
are probably still affected by noise and systematic effects. Note that
\cite{Arendt:1998}, who studied the Galactic foreground from the same
ZSMA data, have identified similar spurious effects and explained them
with residuals of the Zodiacal light subtraction.

\begin{figure*}
 \includegraphics{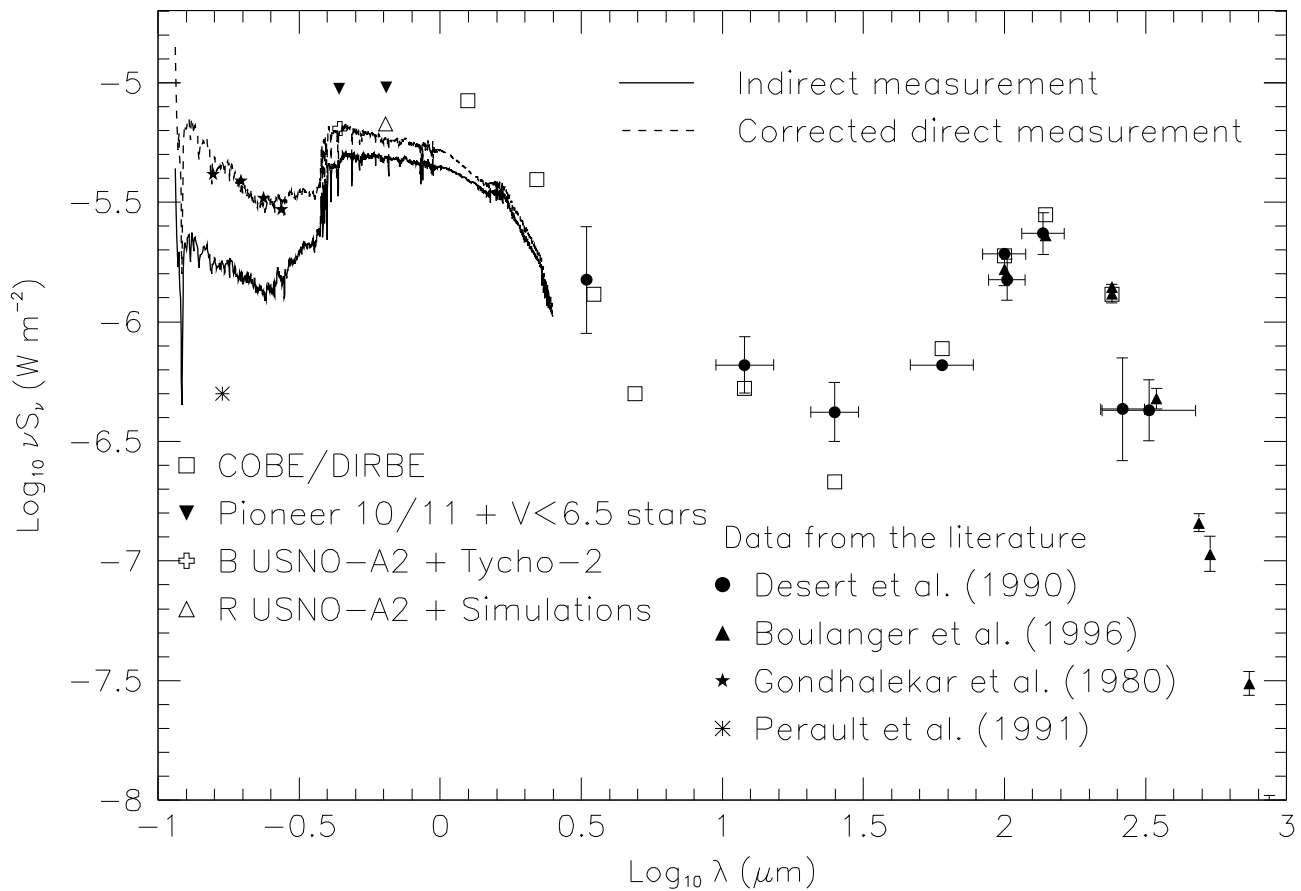} \caption{Summary of the measurements
 discussed throughout this paper.  The data points with no error bars
 have been computed with the cosecant law extrapolation method applied
 at given wavelengths (COBE/DIRBE, Pioneer 10/11, USNO-A2 \&
 Tycho-2). The Pioneer 10/11 and COBE/DIRBE $1.25\mu$m and $2.2\mu$m
 are corrected for extinction (cf. Table \ref{tab:extval}) by the
 amount indicated by the vertical lines attached to the Pioneer 10/11
 points.  In parallel, the {\em dashed} and {\em full line}
 curves correspond to the stellar synthetic estimates based on
 Hipparcos data.  In the IR, our estimates based on COBE/DIRBE are
 compatible with \protect \cite{Desert:1990}, \protect
 \cite{Boulanger:1996} (symbols with error bars). As discussed in \S\
 \protect \ref{sec:isb}, the 12$\mu$m and 25$\mu$m data points are
 probably affected by uncertainties due to Zodiacal light residuals.
 In the UV, the uncertainties are large due to the sensitivity to
 inhomogeneities (OB associations). Our indirect estimate is
 compatible with the all-sky average performed by
 \citet{Gondhalekar:1980}.}
\label{fig:last}
\end{figure*}

For $\lambda < 5 \mu$m, we observe the expected asymmetry between the
North and South hemispheres due to the position of the Sun above the
Galactic plane: the integrated radiation intensity measured in the
Galactic Center (resp. Anti-Centre) South panel is larger than the one
measured in the Galactic Center (resp. Anti-Centre) North panel. At
longer wavelengths, we observe an inverted behaviour: the radiation
intensity measured in Galactic North panels is larger than the one
measured in the South panels.  This is explained by the presence of a
warp of the Galactic disc (beyond 12 kpc) towards the North and
Galactic Centre \citep[see e.g.][]{Burton:1986,Wouterloot:1990}. Last,
for $\lambda\ge 60\mu$m, we are affected by nearby complexes (that we
choose not to remove): this tends to enlarge the uncertainties but
accounts for local fluxes. This probably explains in part the
relatively low values of the intercepts ($|b|=90\degr$) found for
those wavelengths.  However, as further discussed in the next section,
and displayed in Figure
\ref{fig:last}, the values of the slopes thus obtained are fully
compatible with those used by \cite{Desert:1990} based in part on IRAS
data, and those computed by \cite{Boulanger:1996} based on COBE/DIRBE
and FIRAS data .

{
\begin{table*}
\begin{minipage}[t]{2\columnwidth}
\caption{Summary for surface brightness measurements for the usual
optical bands and comparison with other estimates. The various
estimates discussed in this paper have been gathered and provided in
the different usual units. As discussed in the text, the values
obtained for Pioneer 10/11, probing a larger volume than the
(corrected estimate) synthetic estimate based on the Hipparcos data,
are systematically larger (brighter). Our (corrected direct) synthetic
estimate is compatible with the \protect\citet{Flynn:2006} data in B
and V. In the I band, the \citet{Flynn:2006} estiomate is between our
synthetic and Pioneer 10/11 estimates. This could be a biais of our
incompleteness correction, used for synthetic estimates, based on V
data, which assumes homogeneity of the stellar distribution. However,
the Pioneer data (supported by the COBE/DIRBE near-infrared) estimates
suggest that the surface brightness of the Milky Way is brighter. This
could explain why \citet{Flynn:2006} observed that the Milky Way
appears to be underluminous with respect to the main locus of the
Tully-Fisher relation for external galaxies.}\label{tab:opt}
\centering
\renewcommand{\footnoterule}{}  
\begin{tabular}{l|c|ccccccc|} 
\hline\hline             
&Unit & U & B & V & R$_C$ & R$_J$ & I$_C$ & I$_J$ \\
$\lambda  $& $\dot{A}$ & 3660 & 4400  & 5530  & 6400 & 6930  & 7900  &
8785  \\ 
$c_\lambda$  &W m$^{-2}$ & $1.48 \times 10^{-8}$  & $2.90 \times
10^{-8}$ & $1.97 \times 10^{-8}$ &$1.44 \times 10^{-8}$  & $1.25
\times 10^{-8}$  & $9.66 \times 10^{-9}$  & $7.77 \times 10^{-9}$   \\ 
 L$^\lambda_\odot = \lambda F^\lambda_\odot$ & L$_\odot$& 0.282 &
0.558 & 0.745 & 0.753 & 0.737 & 0.661 & 0.622\\ 
\hline\hline 
 \multicolumn{8}{c}{\bf This paper}&\\
\multicolumn{8}{l}{Synthetic estimates} &          \\
- corrected direct & $ 10^{-6}$ W m$^{-2}$& $3.662 $  &$6.249$ &$6.081
$  &$5.848 $ &$5.758 $ &$ 5.46$ &$ 5.347$ \\  
& L$_\odot$ pc$^{-2}$ & 9.08 & 15.49 & 15.07 & 14.50 & 14.27 & 13.56 &
13.25 \\
& L$^\lambda_\odot$ pc$^{-2}$ & 32.20 & 27.76 & 20.23 & 19.26 & 19.36
& 20.51 & 21.30\\ 	 
& {\bf mag/''} & {\bf 23.34} & {\bf 23.49} & {\bf 23.10} & {\bf
22.80} & {\bf 22.66} & {\bf 22.44} & {\bf 22.23}\\\hline
- indirect &$10^{-6}$W m$^{-2}$& $2.433 $  & $4.618 $ & $4.924 $ &
$4.864 $ &  $4.817 $ &  $ 4.597 $ & $ 4.495 $ \\
& L$_\odot$ pc$^{-2}$ & 6.03 & 11.45 & 12.21 & 12.06 & 11.94 & 11.40 &
11.14 \\
& L$^\lambda_\odot$ pc$^{-2}$ & 21.38 & 20.52 & 16.39 & 16.36 & 16.20
& 17.25 & 17.91\\
& {\bf mag/''} & {\bf 23.78} & {\bf 23.81} & {\bf 23.33} & {\bf
23.00} & {\bf 22.86} & {\bf 22.63} & {\bf 22.41}\\ \hline
{USNO-A2 } &$ 10^{-6}$W m$^{-2}$ & & $6.13 $ & & $6.71$&
&&\\ 
  & L$_\odot$ pc$^{-2}$ & & 15.19 & & 16.63 &&& \\ 
  & L$^\lambda_\odot$ pc$^{-2}$ & & 27.23 & & 22.08 &&&\\
  &  {\bf mag/''} & & {\bf 23.51} & & {\bf 22.6} &&& \\\hline
{Pioneer 10/11}\footnote{Values provided in parenthesis have been
obtained as an interpolation based on the corrected direct template
adjusted on the Pioneer 10/11 and COBE/DIRBE measurements (factor $1.7$).}& $10^{-6}$ W m$^{-2}$ &  ($6.23$) & $8.98 $ &
($10.34$) & $9.54$& ($9.79$) & ($9.30$) & ($ 9.51$) \\
  & L$_\odot$ pc$^{-2}$ & (15.44) & 22.26  & (25.63) & 23.65 & (24.27)
& (23.05)& (23.58)\\  
  & L$^\lambda_\odot$ pc$^{-2}$ & (54.76) & 39.89 & (34.41) & (31.41)
& (32.93) & (34.87) & (37.90) \\
  & {\bf  mag/''} & ({\bf 22.76}) & {\bf 23.09} & ({\bf 22.52}) & {\bf 22.27}
&({\bf 22.09}) & ({\bf 21.86}) & ({\bf 21.60})\\ \hline 
{Basic simulations}&$10^{-6}$W m$^{-2}$ & & $5.13 $ & $4.23 $ &
$4.76$& & &  \\  
  & L$_\odot$ pc$^{-2}$ & & 12.72 & 10.51 & 11.81 &\multicolumn{3}{l|}{}\\ 
  & L$^\lambda_\odot$ pc$^{-2}$ & &22.79 & 14.11 & 15.68
&\multicolumn{3}{l|}{}\\  
  & {\bf  mag/''} & & {\bf 23.67} & {\bf 23.49} &{\bf 23.02}
&\multicolumn{3}{l|}{}\\  \hline\hline 
\multicolumn{8}{c}{\bf Comparison with other works\footnote{The
surface brightness measurements in Wm$^{-2}$, L$_\odot$ pc$^{-2}$ and
L$^\lambda_\odot$  pc$^{-2}$ provided for the estimates of
\citet{vanderKruit:1986} and \citet{Flynn:2006} have been derived with
the calibration constants $c_\lambda$ and L$^\lambda_\odot$
provided on the top of this table and computed with
Eq. \protect\ref{eq:mu}. These L$^\lambda_\odot$ pc$^{-2}$ values are
compatible with the original values within 10$\%$ uncertainties, due to
the difference of L$^\lambda_\odot$ solar luminosity considered.}}&\\
v. der Kruit (1986)\nocite{vanderKruit:1986} &$10^{-6}$ W m$^{-2}$ & &
{\em 4.68} & & &\multicolumn{3}{l|}{}\\ & L$_\odot$ pc$^{-2}$ & &{\em
11.61} & & &\multicolumn{3}{l|}{}\\ & L$^\lambda_\odot$ pc$^{-2}$ & &
{\em 20.81} & & &\multicolumn{3}{l|}{}\\ & {\bf mag/''} & & {\bf 23.8}
& &&\multicolumn{3}{l|}{}\\ \hline
\citet{Flynn:2006} & $10^{-6}$ W m$^{-2}$ & & {\em 6.12}  & {\em 7.09}
&  & & {\em 7.96} & \\ 
  & L$_\odot$ pc$^{-2}$ & & {\em 15.16}  & {\em 17.57}
&  & & {\em 19.74} & {}\\ 
  & L$^\lambda_\odot$ pc$^{-2}$ & & {\em 27.70}  & {\em 23.59} &  & &
{\em 29.86}&\multicolumn{1}{l|}{}\\ 
  & {\bf  mag/''} & & {\bf 23.51} &  {\bf 22.93} && & {\bf 22.03 }& \multicolumn{1}{l|}{}\\  \hline\hline
\end{tabular}
\end{minipage}
\end{table*}
}
%
\begin{table*}
\begin{minipage}[t]{2\columnwidth}
\caption{Summary for surface brightness measurement in the
near-infrared and in the infrared comparison with previous
measurements. Our measurements are compatible with those of
\protect\citet{Desert:1990}. Both are systematically larger than our
synthetic estimates based on the V luminosity function, corrected for
incompleteness assuming a homogeneous stellar density.}\label{tab:IR}
\centering
\renewcommand{\footnoterule}{}  
\begin{tabular}{ll|cccccccccc|} 
\hline\hline             
& Unit & J & K & L &&&&&&& \\
$\lambda$  &$\mu$m& 1.25 & 2.20 & 3.50 & 4.90 & 12 & 25 & 60 & 100 &
140 & 240\\\hline \hline 
 \multicolumn{8}{c}{\bf This paper}&&&&\\
DIRBE &$ 10^{-6}$ W m$^{-2}$ & $7.50 $& $3.74 $& $1.31  $& $0.50 $ &
$0.53 $ & $0.21 $ & $0.77$ & $1.88 $ & $2.81$ & $1.30$\\ \hline
\multicolumn{8}{l}{Synthetic estimates} & &&&         \\
- corrected direct & $ 10^{-6}$ W m$^{-2}$& $4.30$  &$2.00$ &&&&&&&&\\
- indirect & $ 10^{-6}$ W m$^{-2}$& $3.92$  &$1.85$ &&&&&&&&\\
\hline\hline             
 \multicolumn{8}{c}{\bf Previous works}&&&&\\
 & & & \multicolumn{2}{r}{3.3$\mu$m} &&
 &    &  & &&\\ \hline\hline  
\citet{Desert:1990}\footnote{We assume a column density of hydrogen
$N_H=3\times 10^{20}$ cm$^{-2}$.} &  $10^{-6}$ W m$^{-2}$& &
\multicolumn{2}{r}{$1.5\pm 0.8$} && 
$0.66\pm 0.18$ &   $0.42\pm 0.12$ & $0.66\pm 0.04$ & $1.92\pm 0.07$ && \\ \hline\hline  
\end{tabular}
\end{minipage}
\end{table*}

\section{Discussion}
\label{sec:dis}
On the one hand, Figure \ref{fig:last} provides a summary of all
the measurements performed in this paper. We also superpose data from
the literature as indicated. \cite{Perault:1991} provides a cosecant
measurement in the UV of the diffuse component. Interestingly, it
confirms that it does not have a substantial contribution.  On the
other hand, Table \ref{tab:opt} gives numerical values of the surface
brightness for UBVRI filters and a comparison with other works. Table
\ref{tab:IR} summarises similarly the values for near-infrared and infrared.

We have shown with simulations that the surface brightness of the
Galaxy at the Solar Neighbourhood as observed from outside the Galaxy
can be computed with a cosecant averaging of all-sky surveys. While
face-on estimates performed on external galaxies are not very
sensitive to extinction, our measurements severely suffer from
extinction due to our internal view point. We have computed the
extinction corrections necessary to apply to the direct cosecant
averaging estimates.  Relying on these results, we compute synthetic
spectral estimates of the optical surface brightness. We use the
Hipparcos catalogue and the Hipparcos Input Catalogue together with a
stellar spectral flux library to estimate the expected contribution of
stars in the optical. We use the knowledge of the stellar parallaxes
and of the completeness limit to perform this estimate. We attempt to
correct the incompleteness (uncertain parallaxes and stars below the
completion limits) with corrections based on our simulations.  We then
compile optical and infra-red sky-surveys to obtain direct
observational estimates.

We observe a systematic difference between our synthetic estimates and
the optical and infra-red sky-surveys. The estimate based on USNO-A2 R
flux in the magnitude range $12 \le R <18$ exhibits a large excess
with respect to our simulations. This excess is compatible with the
disagreement observed between Pioneer 10/11 and our synthetic
estimates.  On the other hand, in the B band the TYCHO and USNO-A2
data are in relatively good agreement with the simulations and with
our synthetic estimate. The Pioneer 10/11 estimate nevertheless
exhibits an excess (similar to what is observed in R). Various
arguments could be evoked to provide tentative explanations for these
differences:
\begin{itemize}
\item The USNO-A2 data exhibit a very irregular pattern in B, which is not
observed in R. This might be an indication of patchy absorption
affecting solely the B data. Patchy extinction is not included in our
simulations nor in the extinction correction. However, we would expect
the Pioneer 10/11 data to be affected the same way. There might be
some other not understood problem with the B data.
\item The synthetic estimates are based on the Hipparcos data only
(distance limited sample with $V<7.3$), while the all-sky surveys
integrate the whole sky. In practice, the all-sky surveys probe an
equivalent cylinder radius of 1\,kpc, while the synthetic estimates are
based on the Hipparcos data complete within $\sim$100-200\,pc depending on
the magnitude range. In order to provide a surface brightness estimate
within 1\,kpc, we have applied appropriate corrections. However, these
corrections are normalised to the local stellar density measured by
Hipparcos and assume a uniform stellar density following the
exponential disc geometry.

We suggest that this disagreement could be a signature of the presence
of the local minimum of the stellar density linked with Gould's
belt. As displayed in Fig. \ref{fig:meanmag1}, the magnitude
interval $10\le V < 18$ is roughly composed of $\langle
M_V\rangle=4-5$ stars. In addition, we know that the strongest
contribution comes from $-3<M_V<6$ (see Fig. \ref{fig:newconsl2}). The
irregular distribution of OB stars (not accounted for in our
correction for synthetic estimates, nor in our modelling) plays an
important role.
\item The all-sky surveys estimates follow the same profile as the
synthetic estimates. This is also compatible with a minimum of the
stellar density (the bright stars are the main contributors to the
surface brightness).
\item  The Pioneer 10/11 data points, when uncorrected for extinction,
are in good agreement with our synthetic estimate. However, they seem
at odds with respect to the DIRBE/COBE J and K data. This is difficult
to understand as the synthetic estimates are corrected for extinction
(star by star).
\end{itemize}

{The recent work of \citet{Flynn:2006} based on (partly) different
data sets and a careful modelling exhibits similar trends. Their
Tuorla study (which results are quoted in Tab. \ref{tab:opt}) probes
stars out to circa 200\,pc. Their B and V estimates are consistent
with our corrected direct synthetic estimates, since the volume probed
with the Hipparcos data is similar. In parallel, their I-band estimate
is systematically larger than our synthetic estimate and closer to our
Pioneer 10/11 estimates. This is in agreement with our Gould's belt
interpretation as I-band surface brightness should be less sensitive
to inhomogeneities, and suggests that our incompleteness correction,
assuming homogeneity of the stellar distribution (see
Eq. \ref{eq:corrdirect}), probably misses some stars in I. Our results
could explain why \citet{Flynn:2006} find that the global luminosity
of the Milky Way appears to be under-luminous with respect to the main
locus of the Tully-Fisher relation as observed for external galaxies.}

We thus conclude that the presence of Gould's belt affects our
measurements of the surface brightness of the Galaxy at the Solar
Neighbourdhood. All-sky surveys (Pioneer 10/11 and DIRBE/COBE) do not
suffer from magnitude limits and enable to probe the whole volume
expected. Our synthetic estimates based on Hipparcos data
underestimate the surface brightness by a factor about 2. We
tentatively interpret this difference as a signature of Gould's
belt. The stellar catalogues tend partially to support this
interpretation. In the R band, we observed an excess in the USNO-A2
data compatible with the Pioneer 10/11 estimate. 

\begin{acknowledgements}
We thank G. Lagache for her efficient help for the COBE/DIRBE data
handling, R.\ Drimmel for providing an adapted version of his
published J,K model, and A.\ Robin for providing us simulations prior
to their publication.  We are most grateful to K.\ Gordon and his
collaborators for providing their Pioneer 10/11 data, which have been
very valuable for this study.  We thank Guy Simon for preparing an
adapted catalogue from the DENIS database. We also acknowledge useful
discussions with J.\ Guibert and M.\ P\'erault at various stages of
this work.  Work by AG was supported by grant AST 0452758 from the US
NSF. ALM has been supported during part of this work by a grant from
the Soci\'et\'e de Secours des Amis des Sciences. This research has
made use of the NASA/IPAC Extragalactic Database (NED), which is
operated by the Jet Propulsion Laboratory, California Institute of
Technology, under contract with the National Aeronautics and Space
Administration.
\end{acknowledgements}


\begin{thebibliography}{38}
\expandafter\ifx\csname natexlab\endcsname\relax\def\natexlab#1{#1}\fi
\bibitem[{Allen(1973)}]{Allen:1973}
Allen, C.~W. 1973, {A}strophysical {Q}uantities (London: Athlone
Press)
\bibitem[Arendt et al.(1998)]{Arendt:1998} Arendt, R.~G.~et al.\ 
1998, \apj, 508, 74 astro-ph/9805323
\bibitem[{Arenou {et~al.}(1992)Arenou, Grenon, \&
G\'omez}]{Arenou:1992} Arenou, F., Grenon, M., \&  G\'omez, A. 1992,
\aap, 258, 104   
\bibitem[Bergh{\" o}fer \& Breitschwerdt(2002)]{Berghofer:2002} 
Bergh{\" o}fer, T.~W.~\& Breitschwerdt, D.\ 2002, \aap, 390, 299 
\bibitem[{Boulanger {et~al.}(1996)Boulanger, Abergel, Bernard, Burton,
D\'esert, Hartmann, Lagache, \& Puget}]{Boulanger:1996} Boulanger, F.,
Abergel, A., Bernard, J.-P., Burton, W. B., D\'esert, F.-X., Hartmann,
D., Lagache, G., \& Puget, J.-L. 1996, \aap, 312, 256
\bibitem[{Boulanger \& P\'erault(1988)}]{Boulanger:1988}
Boulanger, F. \& P\'erault, M. 1988, \aap, 330, 964
\bibitem[{{Bucciarelli} {et~al.}(2001){Bucciarelli}, {Garc{\' i}a Yus},
  {Casalegno}, {Postman}, {Lasker}, {Sturch}, {Lattanzi}, {McLean},
  {Costa}, {Falasca}, {Le Poole}, {Massone}, {Potter}, {Rosenberg},
  {Borgman}, {Doggett}, {Morrison}, {Pizzuti}, {Pompei}, {Rehner},
  {Siciliano}, \& {Wolfe}}]{Bucciarelli:2001} {Bucciarelli}, B., et
  al.  2001, \aap, 368, 335
\bibitem[Burton \& Te Lintel Hekkert(1986)]{Burton:1986} Burton, 
W.~B.~\& Te Lintel Hekkert, P.\ 1986, \aaps, 65, 427 
\bibitem[Cambr{\' e}sy, Reach, Beichman, \& 
Jarrett(2001)]{Cambresy:2001} Cambr{\' e}sy, L., Reach, W.~T., 
Beichman, C.~A., \& Jarrett, T.~H.\ 2001, \apj, 555, 563 
\bibitem[{Dale {et~al.}(2001)}]{Dale:2001} Dale, D.~A., Helou, G., Contursi, A.,
  Silbermann, N.~A., \& Kolhatkar, S. 2001, \apj, 549, 215
\bibitem[{D\'esert {et~al.}(1990)}]{Desert:1990}
D\'esert, F., Boulanger, F., \& Puget, J.-L. 1990, \aap, 237, 215
\bibitem[de Zeeuw et al.(1999)]{deZeeuw:1999} de Zeeuw, P.~T., 
Hoogerwerf, R., de Bruijne, J.~H.~J., Brown, A.~G.~A., \& Blaauw, A.\ 1999, 
\aj, 117, 354 
\bibitem[{Drimmel \& Spergel(2001)}]{Drimmel:2001}
Drimmel, R. \& Spergel, D.~N. 2001, \apj, 556, 181
\bibitem[Drimmel et al.(2003)]{Drimmel:2003} Drimmel, R., 
Cabrera-Lavers, A., \& L{\' o}pez-Corredoira, M.\ 2003, \aap, 409, 205 
\bibitem[Dwek et al.(1997)]{Dwek:1997} Dwek, E.~et al.\ 1997, 
\apj, 475, 565 
\bibitem[{Fioc \& Rocca-Volmerange(1997)}]{Fioc:1997}
Fioc, M. \& Rocca-Volmerange, B. 1997, \aap, 326, 950
\bibitem[Fitzpatrick(1999)]{Fitzpatrick:1999} Fitzpatrick, E.~L.\ 1999, 
\pasp, 111, 63 
 \bibitem[Fouqu{\' e} et al.(2000)]{Fouque:2000} Fouqu{\' e}, P.~et 
al.\ 2000, \aaps, 141, 313 
\bibitem[Flynn et al. (2006)]{Flynn:2006} Flynn, C., \& Holmberg, J.\ 2003, \mnras,  accepted (astro-ph/0608193)
\bibitem[Gispert, Lagache, \& Puget(2000)]{Gispert:2000} Gispert, 
R., Lagache, G., \& Puget, J.~L.\ 2000, \aap, 360, 1 
\bibitem[{Gondhalekar {et~al.}(1980)Gondhalekar, Philips, \&
  Wilson}]{Gondhalekar:1980} Gondhalekar, P.~M., Philips, A., \&
  Wilson, R. 1980, \aap, 85, 272
\bibitem[{Gordon {et~al.}(1998)Gordon, Witt, \& Friedmann}]{Gordon:1998}
Gordon, K., Witt, A.~N., \& Friedmann, B.~C. 1998, \apj, 498, 540
\bibitem[Guiderdoni et al.(1998)]{Guiderdoni:1998} Guiderdoni, B., 
Hivon, E., Bouchet, F.~R., \& Maffei, B.\ 1998, \mnras, 295, 877 
\bibitem[Hakkila et al.(1997)]{Hakkila:1997} Hakkila, J., Myers, 
J.~M., Stidham, B.~J., \& Hartmann, D.~H.\ 1997, \aj, 114, 2043 
\bibitem[Hauser \& Dwek(2001)]{Hauser:2001} Hauser, M.~G.~\& Dwek, 
E.\ 2001, \araa, 39, 249 
\bibitem[{Humphreys \& Larsen(1995)}]{Humphreys:1995}
Humphreys, R. \& Larsen, J.~Q. 1995, \aj, 110, 2183
\bibitem[{Hayakawa {et~al.}(1981)}]{Hayakawa:1981}
{Hayakawa}, S., et al. 1981, \aap, 100, 116
\bibitem[{Henry(2002)}]{Henry:2002}
{Henry}, R. C. 2002, \aap, in press
\bibitem[H{\o}g et al.(2000)]{Hog:2000} H{\o}g, E.~et al.\ 2000, 
\aap, 355, L27 
\bibitem[{Jahreiss \& Wielen(1997)}]{Jahreiss:1997}
Jahreiss, H. \& Wielen, R. 1997, in HIPPARCOS '97. Presentation of the
Hipparcos and Tycho catalogs and first astrophysical results of the
Hipparcos space astrometry mission, ed. B. Battrick, M.A.C. Perryman
\& P.L. Bernacca, 675
\bibitem[{Kovalevsky(1998)}]{Kovalevsky:1998}
Kovalevsky, J. 1998, \aap, 340, L35
\bibitem[Kelsall et al.(1998)]{Kelsall:1998} Kelsall, T.~et al.\ 
1998, \apj, 508, 44 
\bibitem[Lallement et al.(2003)]{Lallement:2003} Lallement, R., Welsh, 
B.~Y., Vergely, J.~L., Crifo, F., \& Sfeir, D.\ 2003, \aap, 411, 447 
\bibitem[Lang (1992)]{Lang:1992} Lang, K., \ 1992, Astrophysical Data:
Planets and Stars. (New-York: Springer-Verlag) 
\bibitem[{{Leinert} {et~al.}(1998){Leinert}, {Bowyer}, {Haikala},
{Hanner}, {Hauser}, {Levasseur-Regourd}, {Mann}, {Mattila}, {Reach},
{Schlosser}, {Staude}, {Toller}, {Weiland}, {Weinberg}, \&
{Witt}}]{Leinert:1998} {Leinert}, C., et al. 1998, \aaps, 127, 1
\bibitem[{Li \& Draine(2001)}]{Li:2001}
Li, A. \& Draine, B.~T. 2001, \apj, 554, 778
\bibitem[{Lutz \& Kelker(1973)}]{Lutz:1973}
Lutz, T. \& Kelker, D. 1973, \pasp, 85, 573
\bibitem[Maihara et al.(2001)]{Maihara:2001} Maihara, T.~et al.\ 
2001, \pasj, 53, 25 
\bibitem[{Ma\'iz-Apell\'aniz(2001)}]{Maiz:2001}
Ma\'iz-Apell\'aniz, J. 2001, \aj, 121, 2742
\bibitem[{Mathis {et~al.}(1983)Mathis, Mezger, \& Panagia}]{Mathis:1983}
Mathis, J.~S., Mezger, P.~G., \& Panagia, N. 1983, \aap, 128, 212
\bibitem[{Mattila(1980{\natexlab{b}})}]{Mattila:1980a}
Mattila, K. 1980{\natexlab{b}}, \aaps, 39, 53
\bibitem[{Mattila(1980{\natexlab{a}})}]{Mattila:1980b}
---. 1980{\natexlab{a}}, \aap, 82, 373
\bibitem[{Mezger {et~al.}(1982)}]{Mezger:1982}
Mezger, P.~G., Mathis, J.~S., \& Panagia, N. 1982, \aap, 105, 372
\bibitem[{Miller \& Scalo(1979)}]{Miller:1979}
Miller, G.~E. \& Scalo, J.~M. 1979, \apjs, 41, 513
\bibitem[Monet et al.(2003)]{Monet:2003} Monet, D.~G., et al.\ 
2003, \aj, 125, 984 
\bibitem[{Monet et al.}(1998)]{Monet:1998} Monet D. et al. 1998,
USNO-A V2.0, A Catalog of Astrometric Standards (U.S. Naval
Observatory Flagstaff Station (USNOFS) and Universities Space Research
Association (USRA) stationed at USNOFS.)
\bibitem[{Murthy \& Henry(1995)}]{Murthy:1995}
Murthy J., \& Henry, R. C. 1995, \apj, 448, 848
\bibitem[{P\'erault {et~al.}(1991)}]{Perault:1991} P\'erault, M., Lequeux, J., Hanus, M., \&
  Joubert, M. 1991, \aap, 246, 243
\bibitem[{Perryman {et~al.}(1997)}]{Perryman:1997} Perryman, M., et
  al. 1997, \aap, 323, L49
\bibitem[{Pickles(1998)}]{Pickles:1998}
Pickles, A.~J. 1998, \pasp, 110, 863
\bibitem[{Poppel(1997)}]{Poppel:1997}
Poppel, W. 1997, Fundamentals of Cosmic Physics, 18, 1
\bibitem[{Reyl\'e \& Robin(2001{\natexlab{a}})}]{Reyle:2001a}
Reyl\'e, C. \& Robin, A.~C. 2001{\natexlab{a}}, \aap, 373, 886
\bibitem[{Robin {et~al.}(1992)Robin, Cr\'ez\'e, \& Mohan}]{Robin:1992}
Robin, A.~C., Cr\'ez\'e, M., \& Mohan, V. 1992, \aap, 265, 32
\bibitem[{Robin {et~al.}(2000)}]{Robin:2000}
Robin, A.~C., Reyle, \& Cr\'ez\'e. 2000, \aap, 359, 103
\bibitem[Schlegel et al.(1998)]{Schlegel:1998} Schlegel, D.~J., 
Finkbeiner, D.~P., \& Davis, M.\ 1998, \apj, 500, 525 \bibitem[Sfeir, Lallement, Crifo, \& Welsh(1999)]{Sfeir:1999} 
Sfeir, D.~M., Lallement, R., Crifo, F., \& Welsh, B.~Y.\ 1999, \aap, 346, 
785 
\bibitem[{Smith(1987)}]{Smith:1987}
Smith, H. 1987, \aap, 188, 233
\bibitem[{Toller {et~al.}(1987)}]{Toller:1987}
Toller, G., Tanabe, H., \& Weinberg, J.-L. 1987, \aap, 188, 24
\bibitem[{Turon {et~al.}(1992)}]{Turon:1992} Turon, C., et
  al. 1992, \aap, 258, 74
\bibitem[{Turon {et~al.}(1995)}]{Turon:1995}
  Turon, C., et al. 1995, \aap, 304, 82
\bibitem[van der Kruit(1986)]{vanderKruit:1986} van der Kruit, P.~C.\ 
1986, \aap, 157, 230 
 
\bibitem[{{van Leeuwen} {et~al.}(1997){van Leeuwen}, Evans, Grenon,
Grossmann,
  Mignard, \& Perryman}]{vanLeeuwen:1997} {van Leeuwen}, F., Evans,
  D.~W., Grenon, M., Grossmann, V., Mignard, F., \& Perryman,
  M. A.~C. 1997, \aap, 323, 61
\bibitem[{Vergely {et~al.}(1998)}]{Vergely:1998}
Vergely, J.-L., Freire Ferrero, R., Egret, D., \& K\"opper, J. 1998,
\aap, 340, 543  
\bibitem[Welsh, Sfeir, Sirk, \& Lallement(1999)]{Welsh:1999} 
Welsh, B.~Y., Sfeir, D.~M., Sirk, M.~M., \& Lallement, R.\ 1999, \aap, 352, 
308 
\bibitem[Welsh et al.(2002)]{Welsh:2002} Welsh, B.~Y., Sallmen, 
S., Sfeir, D., Shelton, R.~L., \& Lallement, R.\ 2002, \aap, 394, 691 
\bibitem[{Werner \& Salpeter(1969)}]{Werner:1969}
Werner, M.~W. \& Salpeter, W.~E. 1969, \mnras, 145, 249
\bibitem[{Wicenec \& {van Leeuwen}(1995)}]{Wicenec:1995}
Wicenec, A. \& {van Leeuwen}, F. 1995, \aap, 304, 160
\bibitem[Wouterloot, Brand, Burton, \& Kwee(1990)]{Wouterloot:1990} 
Wouterloot, J.~G.~A., Brand, J., Burton, W.~B., \& Kwee, K.~K.\ 1990, \aap, 
230, 21 
\bibitem[{Zacharias {et~al.}(2000)}]{Zacharias:2000} Zacharias, N., et al. 2000, \aj, 120, 2131
\end{thebibliography}

\clearpage


\section*{List of the Appendices}
\contentsline {section}{\numberline {A}Tables}{15}
\contentsline {section}{\numberline {B}Surface brightness definition}{20}
\contentsline {section}{\numberline {C}Normalisation factor of the cosecant int
egration method}{20}
\contentsline {section}{\numberline {D}Luminosity function (LF)}{21}
\contentsline {subsection}{\numberline {D.1}Uncertainties}{21}
\contentsline {subsection}{\numberline {D.2}Computation of the LF for direct/in
direct estimates}{21}
\contentsline {subsection}{\numberline {D.3}3D visualisation of the Hipparcos d
ata}{22}
\contentsline {section}{\numberline {E}Extinction Modelling}{22}
\contentsline {section}{\numberline {F}Comparison of the USNO-A2 with UCAC-1 da
ta}{23}
\contentsline {section}{\numberline {G}Calibration of the USNO-A2 data}{24}

\vspace{1cm}

\hrule

\vspace{1cm}

\appendix
\section{Tables}
Table \ref{tab:simu} is discussed in Sect. \ref{sect:comparison}, and
corresponds to Fig. \ref{fig:myconlisrf3_1} and
\ref{fig:myconlisrf3_2}. Table \ref{tab:usnob4} is discussed in
Sect. \ref{sssec:bmeas} and corresponds to Fig. \ref{fig:usnob4} and
\ref{fig:tychob4}.Table \ref{tab:usnor4} is discussed in
Sect. \ref{sssec:rmeas} and corresponds to
Fig. \ref{fig:usnor4}. Table \ref{tab:cobe} is discussed in
Sect. \ref{sec:isb}.

\begin{table*}
\caption{Simulated V surface brightness (in $10^{-9}\, \rm W\,m^{-2}$)
contributions obtained for each magnitude bin and sky area derived
from the cosecant laws (see Fig. \ref{fig:myconlisrf3_1} and
\ref{fig:myconlisrf3_2}). We give the values fitted for $1/\sin |b|
<6$.  The last column provides the percentages of the total surface
brightness (PTSB) corresponding to each apparent magnitude bin. }
\label{tab:simu} 
\begin{center}
\begin{tabular}{rrrrrrrrrr}
\hline\hline
&\multicolumn{4}{c}{Galactic Centre}
&\multicolumn{4}{c}{Galactic anti-centre} & PTSB\\
Mag.&\multicolumn{2}{c}{North} & \multicolumn{2}{c}{South} &
\multicolumn{2}{c}{North} & 
\multicolumn{2}{c}{South} &\\ & {($|b|=90\degr$)}
& {slope}&
{$|b|=90\degr$} & {slope}&
{$|b|=90\degr$} & {slope}&
{$|b|=90\degr$} & {slope}& $\%$\\\hline
$-6. \le V < -5.$ &    1.59 & $\mathbf{ 0.01}$ &    1.75 &  $\mathbf{ -0.02}$ &    1.58 &  $\mathbf{  0.01}$ &    1.74 &  $\mathbf{ -0.02}$ &   0.00 \\
$-5. \le V < -4.$ &    3.22 & $\mathbf{ 0.03}$ &    3.63 &  $\mathbf{ -0.04}$ &    3.21 &  $\mathbf{  0.03}$ &    3.62 &  $\mathbf{ -0.05}$ &   0.00 \\
$-4. \le V < -3.$ &    6.42 & $\mathbf{ 0.08}$ &    7.42 &  $\mathbf{ -0.10}$ &    6.40 &  $\mathbf{  0.08}$ &    7.40 &  $\mathbf{ -0.10}$ &   0.00 \\
$-3. \le V < -2.$ &   12.38 & $\mathbf{ 0.19}$ &   14.58 &  $\mathbf{ -0.18}$ &   12.33 &  $\mathbf{  0.18}$ &   14.51 &  $\mathbf{ -0.19}$ &   0.00 \\
$-2. \le V < -1.$ &   22.03 & $\mathbf{ 0.45}$ &   26.39 &  $\mathbf{ -0.18}$ &   21.92 &  $\mathbf{  0.43}$ &   26.23 &  $\mathbf{ -0.20}$ &   0.02 \\
$-1. \le V <  0.$ &   33.65 & $\mathbf{ 0.95}$ &   41.28 &  $\mathbf{  0.04}$ &   33.41 &  $\mathbf{  0.90}$ &   40.95 &  $\mathbf{ -0.02}$ &   0.07 \\
$ 0. \le V <  1.$ &   46.21 & $\mathbf{ 1.92}$ &   58.39 &  $\mathbf{  0.83}$ &   45.73 &  $\mathbf{  1.80}$ &   57.72 &  $\mathbf{  0.69}$ &   0.19 \\
$ 1. \le V <  2.$ &   60.78 & $\mathbf{ 3.81}$ &   79.89 &  $\mathbf{  2.68}$ &   59.90 &  $\mathbf{  3.53}$ &   78.65 &  $\mathbf{  2.33}$ &   0.44 \\
$ 2. \le V <  3.$ &   77.80 & $\mathbf{ 7.04}$ &  105.25 &  $\mathbf{  6.52}$ &   76.31 &  $\mathbf{  6.44}$ &  103.14 &  $\mathbf{  5.74}$ &   0.93 \\
$ 3. \le V <  4.$ &   97.42 & $\mathbf{12.20}$ &  133.22 &  $\mathbf{ 13.55}$ &   95.11 &  $\mathbf{ 10.97}$ &  129.97 &  $\mathbf{ 11.91}$ &   1.75 \\
$ 4. \le V <  5.$ &  114.73 & $\mathbf{19.73}$ &  156.69 &  $\mathbf{ 24.60}$ &  111.48 &  $\mathbf{ 17.43}$ &  152.15 &  $\mathbf{ 21.45}$ &   2.99 \\
$ 5. \le V <  6.$ &  126.08 & $\mathbf{29.80}$ &  170.50 &  $\mathbf{ 39.46}$ &  121.93 &  $\mathbf{ 25.81}$ &  164.76 &  $\mathbf{ 33.95}$ &   4.64 \\
$ 6. \le V <  7.$ &  129.27 & $\mathbf{41.86}$ &  171.86 &  $\mathbf{ 56.79}$ &  124.54 &  $\mathbf{ 35.43}$ &  165.45 &  $\mathbf{ 47.89}$ &   6.54 \\
$ 7. \le V <  8.$ &  121.18 & $\mathbf{54.42}$ &  157.62 &  $\mathbf{ 74.29}$ &  116.69 &  $\mathbf{ 44.79}$ &  151.73 &  $\mathbf{ 60.97}$ &   8.43 \\
$ 8. \le V <  9.$ &  103.83 & $\mathbf{65.20}$ &  131.20 &  $\mathbf{ 88.66}$ &  100.72 &  $\mathbf{ 51.80}$ &  127.52 &  $\mathbf{ 70.21}$ &   9.92 \\
$ 9. \le V < 10.$ &   81.34 & $\mathbf{71.83}$ &   98.88 &  $\mathbf{ 96.77}$ &   80.58 &  $\mathbf{ 54.68}$ &   98.81 &  $\mathbf{ 73.33}$ &  10.67 \\
$10. \le V < 11.$ &   57.95 & $\mathbf{73.32}$ &   67.15 &  $\mathbf{ 97.35}$ &   59.87 &  $\mathbf{ 53.14}$ &   70.98 &  $\mathbf{ 70.11}$ &  10.57 \\
$11. \le V < 12.$ &   36.40 & $\mathbf{70.18}$ &   39.66 &  $\mathbf{ 91.47}$ &   40.77 &  $\mathbf{ 48.12}$ &   46.80 &  $\mathbf{ 62.22}$ &   9.78 \\
$12. \le V < 13.$ &   17.82 & $\mathbf{63.85}$ &   17.30 &  $\mathbf{ 81.55}$ &   24.56 &  $\mathbf{ 40.99}$ &   27.23 &  $\mathbf{ 51.90}$ &   8.57 \\
$13. \le V < 14.$ &    3.10 & $\mathbf{55.52}$ &    0.36 &  $\mathbf{ 69.55}$ &   12.29 &  $\mathbf{ 32.83}$ &   12.94 &  $\mathbf{ 40.75}$ &   7.14 \\
$14. \le V < 15.$ &   -6.67 & $\mathbf{45.88}$ &  -10.36 &  $\mathbf{ 56.48}$ &    4.39 &  $\mathbf{ 24.57}$ &    4.05 &  $\mathbf{ 29.97}$ &   5.64 \\
$15. \le V < 16.$ &  -11.39 & $\mathbf{35.80}$ &  -15.04 &  $\mathbf{ 43.40}$ &    0.18 &  $\mathbf{ 17.16}$ &   -0.47 &  $\mathbf{ 20.61}$ &   4.21 \\
$16. \le V < 17.$ &  -12.28 & $\mathbf{26.38}$ &  -15.36 &  $\mathbf{ 31.54}$ &   -1.42 &  $\mathbf{ 11.17}$ &   -2.03 &  $\mathbf{ 13.23}$ &   2.96 \\
$17. \le V < 18.$ &  -10.73 & $\mathbf{18.32}$ &  -13.04 &  $\mathbf{ 21.64}$ &   -1.43 &  $\mathbf{  6.72}$ &   -1.85 &  $\mathbf{  7.86}$ &   1.96 \\
$18. \le V < 19.$ &   -7.94 & $\mathbf{11.88}$ &   -9.47 &  $\mathbf{ 13.88}$ &   -0.82 &  $\mathbf{  3.74}$ &   -1.04 &  $\mathbf{  4.33}$ &   1.22 \\
$19. \le V < 20.$ &   -5.07 & $\mathbf{ 7.21}$ &   -5.96 &  $\mathbf{  8.34}$ &   -0.34 &  $\mathbf{  2.00}$ &   -0.42 &  $\mathbf{  2.29}$ &   0.71 \\
$20. \le V < 21.$ &   -2.99 & $\mathbf{ 4.23}$ &   -3.47 &  $\mathbf{  4.84}$ &   -0.15 &  $\mathbf{  1.09}$ &   -0.18 &  $\mathbf{  1.24}$ &   0.41 \\
$21. \le V < 22.$ &   -1.82 & $\mathbf{ 2.51}$ &   -2.08 &  $\mathbf{  2.85}$ &   -0.12 &  $\mathbf{  0.62}$ &   -0.14 &  $\mathbf{  0.70}$ &   0.24 \\
$22. \le V < 23.$ &   -1.19 & $\mathbf{ 1.51}$ &   -1.34 &  $\mathbf{  1.71}$ &   -0.11 &  $\mathbf{  0.35}$ &   -0.13 &  $\mathbf{  0.39}$ &   0.14 \\
$23. \le V < 24.$ &   -0.77 & $\mathbf{ 0.89}$ &   -0.87 &  $\mathbf{  1.00}$ &   -0.09 &  $\mathbf{  0.19}$ &   -0.10 &  $\mathbf{  0.21}$ &   0.08 \\
$24. \le V < 25.$ &   -0.46 & $\mathbf{ 0.49}$ &   -0.52 &  $\mathbf{  0.55}$ &   -0.05 &  $\mathbf{  0.09}$ &   -0.06 &  $\mathbf{  0.10}$ &   0.04 \\
$25. \le V < 26.$ &   -0.25 & $\mathbf{ 0.24}$ &   -0.28 &  $\mathbf{  0.27}$ &   -0.03 &  $\mathbf{  0.04}$ &   -0.03 &  $\mathbf{  0.05}$ &   0.02 \\
$26. \le V < 27.$ &   -0.12 & $\mathbf{ 0.11}$ &   -0.13 &  $\mathbf{  0.12}$ &   -0.01 &  $\mathbf{  0.02}$ &   -0.01 &  $\mathbf{  0.02}$ &   0.01 \\
$27. \le V < 28.$ &   -0.05 & $\mathbf{ 0.04}$ &   -0.05 &  $\mathbf{  0.05}$ &    0.00 &  $\mathbf{  0.01}$ &   -0.01 &  $\mathbf{  0.01}$ &   0.00 \\
$28. \le V < 29.$ &   -0.02 & $\mathbf{ 0.02}$ &   -0.02 &  $\mathbf{  0.02}$ &    0.00 &  $\mathbf{  0.00}$ &    0.00 &  $\mathbf{  0.00}$ &   0.00 \\
$29. \le V < 30.$ &   -0.01 & $\mathbf{ 0.01}$ &   -0.01 &  $\mathbf{  0.01}$ &    0.00 &  $\mathbf{  0.00}$ &    0.00 &  $\mathbf{  0.00}$ &   0.00 \\
\noalign{\smallskip} \hline
\multicolumn{10}{c}{}\\
\multicolumn{10}{c}{Direct sum}\\ \hline
\multicolumn{10}{c}{}\\
$ V < 0$       &      79.28 & $\mathbf{   1.71}$ &      95.05 &    $\mathbf{  -0.48}$ &      78.83 &   $\mathbf{  1.63}$ &      94.46 & $\mathbf{ -0.57}$ \\
$0\le V < 7$   &     652.30 & $\mathbf{ 116.36}$ &     875.80 &    $\mathbf{ 144.44}$ &     635.00 &   $\mathbf{101.41}$ &     851.85 & $\mathbf{123.96}$ \\
$7\le V < 11$  &     364.30 & $\mathbf{ 264.78}$ &     454.84 &    $\mathbf{ 357.07}$ &     357.86 &   $\mathbf{204.42}$ &     449.04 & $\mathbf{274.62}$ \\
$11\le V < 20$ &       3.24 & $\mathbf{ 335.03}$ &     -11.90 &    $\mathbf{ 417.84}$ &      78.18 &   $\mathbf{187.29}$ &      85.21 & $\mathbf{233.17}$ \\
$20\le V < 30$ &      -7.67 & $\mathbf{  10.04}$ &      -8.77 &    $\mathbf{  11.42}$ &      -0.57 &   $\mathbf{  2.41}$ &      -0.66 & $\mathbf{  2.72}$ \\\noalign{\smallskip} \hline
Total          &    1091.45 & $\mathbf{ 727.92}$ &    1405.03 &    $\mathbf{ 930.29}$ &    1149.30 &   $\mathbf{497.15}$ &    1479.90 & $\mathbf{633.90}$ \\\noalign{\smallskip} \hline
\multicolumn{10}{c}{}\\	      
\multicolumn{10}{c}{\bf Total value: $5126$ ($|b|=90\degr$),
$2789$ (slope)}\\ \hline  
\end{tabular}
\end{center}
\end{table*}

\begin{table*}
\caption{Contributions from TYCHO-2 and USNO-A2 stars to the B surface
brightness, given in $10^{-9}\, \rm W\,m^{-2}$. We give the
values fitted to TYCHO-2 and USNO-A2 B data (see Figs.\
\ref{fig:usnob4} and \ref{fig:tychob4})  for
each sky area and magnitude bin. We also provide the corresponding
values obtained with simulations.}
\label{tab:usnob4}
\begin{center}
\begin{tabular}{rrrrrrrrr}
\hline \hline
&\multicolumn{4}{c}{Galactic Centre} &\multicolumn{4}{c}{Galactic Anti-Centre}\\ 
B mag.&\multicolumn{2}{c}{North} & \multicolumn{2}{c}{South} & \multicolumn{2}{c}{North} & \multicolumn{2}{c}{South}\\ 
\multicolumn{9}{r}{}\\ 
Test & {$|b|=90\degr$} & {\bf slope} & {$|b|=90\degr$} & {\bf slope} &  {$|b|=90\degr$} & {\bf slope} & {$|b|=90\degr$} & {\bf slope} \\\hline
\multicolumn{9}{c}{TYCHO-2 data}\\ 
$6\le B < 7$   &$ 117.26$ & $  \mathbf{42.60}$ & $ 127.84$ & $  \mathbf{44.63}$ & $ 156.98$ & $  \mathbf{18.52}$ & $ 148.84$ & $  \mathbf{52.60}$\\
$7\le B < 8$   &$ 144.37$ & $  \mathbf{36.62}$ & $ 143.04$ & $  \mathbf{45.85}$ & $ 154.85$ & $  \mathbf{34.38}$ & $ 157.57$ & $  \mathbf{53.65}$\\
$8\le B < 9$   &$ 134.97$ & $  \mathbf{46.45}$ & $ 150.55$ & $  \mathbf{55.16}$ & $ 166.30$ & $  \mathbf{41.65}$ & $ 156.32$ & $  \mathbf{54.84}$\\
$9\le B < 10$  &$ 147.35$ & $  \mathbf{47.23}$ & $ 151.01$ & $  \mathbf{64.20}$ & $ 164.94$ & $  \mathbf{50.75}$ & $ 158.00$ & $  \mathbf{56.83}$\\
$10\le B < 11$ &$ 143.05$ & $  \mathbf{53.38}$ & $ 154.52$ & $  \mathbf{69.20}$ & $ 156.43$ & $  \mathbf{56.14}$ & $ 151.58$ & $  \mathbf{53.71}$\\\hline
\multicolumn{9}{c}{USNO-A2 data}\\ 								      
$12\le B < 13$ & $ 98.71$ & $  \mathbf{71.33}$ & $   172.82$ & $ \mathbf{ 60.27}$ & $   105.69$ & $  \mathbf{56.87}$ & $   129.03$ & $ \mathbf{ 46.31}$ \\       
$13\le B < 14$ & $ 66.24$ & $  \mathbf{64.01}$ & $    88.08$ & $ \mathbf{ 80.64}$ & $    85.80$ & $  \mathbf{49.79}$ & $    93.22$ & $ \mathbf{ 41.05}$ \\       
$14\le B < 15$ & $ 44.11$ & $  \mathbf{61.04}$ & $    36.34$ & $ \mathbf{ 95.43}$ & $    60.89$ & $  \mathbf{45.45}$ & $    60.29$ & $ \mathbf{ 40.86}$ \\       
$15\le B < 16$ & $ 24.81$ & $  \mathbf{59.49}$ & $     2.90$ & $ \mathbf{102.94}$ & $    43.99$ & $  \mathbf{40.07}$ & $    34.36$ & $ \mathbf{ 39.84}$ \\       
$16\le B < 17$ & $ 10.97$ & $  \mathbf{54.41}$ & $   -12.62$ & $ \mathbf{ 97.11}$ & $    23.10$ & $  \mathbf{33.13}$ & $    16.33$ & $ \mathbf{ 33.32}$ \\       
$17\le B < 18$ & $ -2.58$ & $  \mathbf{49.07}$ & $   -15.81$ & $ \mathbf{ 79.41}$ & $     7.92$ & $  \mathbf{24.74}$ & $     6.19$ & $ \mathbf{ 24.73}$ \\       
$18\le B < 19$ & $ -5.50$ & $  \mathbf{36.95}$ & $   -11.52$ & $ \mathbf{ 58.80}$ & $     3.29$ & $  \mathbf{13.60}$ & $     0.43$ & $ \mathbf{ 15.43}$ \\\hline 
\multicolumn{9}{c}{Simulations}\\ 
$ 6. \le B <  7.$ &  134.14 &  $\mathbf{ 36.53}$ &  $177.34$ &  $\mathbf{ 49.10}$ &  $129.41$ &  $\mathbf{ 30.81}$ & $ 170.95$ &  $\mathbf{ 41.20}$ \\
$ 7. \le B <  8.$ &  132.53 &  $\mathbf{ 46.41}$ &  $171.57$ &  $\mathbf{ 62.97}$ &  $127.44$ &  $\mathbf{ 38.26}$ & $ 164.90$ &  $\mathbf{ 51.72}$ \\
$ 8. \le B <  9.$ &  122.53 &  $\mathbf{ 55.15}$ &  $154.82$ &  $\mathbf{ 74.66}$ &  $117.63$ &  $\mathbf{ 44.19}$ & $ 148.74$ &  $\mathbf{ 59.60}$ \\
$ 9. \le B < 10.$ &  105.65 &  $\mathbf{ 61.44}$ &  $129.94$ &  $\mathbf{ 82.38}$ &  $101.49$ &  $\mathbf{ 47.55}$ & $ 125.28$ &  $\mathbf{ 63.46}$ \\
$10. \le B < 11.$ &   83.91 &  $\mathbf{ 64.51}$ &  $100.27$ &  $\mathbf{ 85.24}$ &  $ 80.90$ &  $\mathbf{ 47.96}$ & $  97.53$ &  $\mathbf{ 62.99}$ \\\hline
$11. \le B < 12.$ &   59.73 &  $\mathbf{ 64.27}$ &  $ 69.13$ &  $\mathbf{ 83.45}$ &  $ 58.41$ &  $\mathbf{ 45.54}$ & $  68.77$ &  $\mathbf{ 58.72}$ \\\hline
$12. \le B < 13.$ &   35.89 &  $\mathbf{ 61.02}$ &  $ 39.71$ &  $\mathbf{ 77.81}$ &  $ 37.20$ &  $\mathbf{ 40.65}$ & $  42.65$ &  $\mathbf{ 51.44}$ \\
$13. \le B < 14.$ &   15.42 &  $\mathbf{ 55.09}$ &  $ 15.26$ &  $\mathbf{ 69.05}$ &  $ 20.32$ &  $\mathbf{ 33.80}$ & $  22.46$ &  $\mathbf{ 42.04}$ \\
$14. \le B < 15.$ &    0.95 &  $\mathbf{ 46.83}$ &  $ -1.46$ &  $\mathbf{ 57.79}$ &  $  9.22$ &  $\mathbf{ 25.94}$ & $   9.60$ &  $\mathbf{ 31.74}$ \\
$15. \le B < 16.$ &   -6.86 &  $\mathbf{ 37.20}$ &  $ -9.94$ &  $\mathbf{ 45.23}$ &  $  3.09$ &  $\mathbf{ 18.39}$ & $   2.81$ &  $\mathbf{ 22.17}$ \\
$16. \le B < 17.$ &   -9.61 &  $\mathbf{ 27.73}$ &  $-12.42$ &  $\mathbf{ 33.25}$ &  $  0.16$ &  $\mathbf{ 12.20}$ & $  -0.27$ &  $\mathbf{ 14.49}$ \\
$17. \le B < 18.$ &   -9.52 &  $\mathbf{ 19.62}$ &  $-11.73$ &  $\mathbf{ 23.22}$ &  $ -0.95$ &  $\mathbf{  7.62}$ & $  -1.33$ &  $\mathbf{  8.94}$ \\
$18. \le B < 19.$ &   -8.03 &  $\mathbf{ 13.22}$ &  $ -9.62$ &  $\mathbf{ 15.47}$ &  $ -1.05$ &  $\mathbf{  4.46}$ & $  -1.31$ &  $\mathbf{  5.17}$ \\\hline
\multicolumn{9}{c}{Data Sums}\\ \hline
$6\le B < 11$  & $ 687.00$ & $\mathbf{226.28}$ & $772.31$ & $\mathbf{271.63}$ & $799.50$ & $\mathbf{201.44}$ & $726.96$ & $\mathbf{279.04}$ \\
$12\le B < 19$ & $ 236.76$ & $ \mathbf{396.30}$ & $   260.19$ & $ \mathbf{574.60}$ & $   330.68$ & $ \mathbf{263.65}$ & $   339.85$ & $ \mathbf{241.54}$\\\hline\hline 
\multicolumn{9}{c}{Simulated Sums}\\ \hline
$ 6\le B < 11$& $578.76$ & $\mathbf{264.03}$ & $733.94$ & $\mathbf{354.35}$ & $556.88$ &$\mathbf{208.77}$ & $707.41$ & $\mathbf{278.97}$\\
$12\le B < 19$ & $18.24$ & $\mathbf{260.71}$ & $9.79$ & $\mathbf{321.82}$ & $67.99$ & $\mathbf{143.07}$ & $74.60$ & $\mathbf{175.99}$\\ \hline
\multicolumn{9}{c}{Total contributions}\\ \hline
&\multicolumn{4}{c|}{Data}&\multicolumn{4}{c}{Simulations}\\ \hline
&\multicolumn{2}{c}{$|b|=90\degr$}&\multicolumn{2}{c|}{\bf slope}&\multicolumn{2}{c}{$|b|=90\degr$}&\multicolumn{2}{c}{\bf slope}\\ \hline
$ -6\le B < 6$  & \multicolumn{2}{c}{} &  \multicolumn{2}{c|}{}& \multicolumn{2}{c}{$2841.74$} &  \multicolumn{2}{c}{$\mathbf{295.43}$}\\
$ 6\le B < 11$  & \multicolumn{2}{c}{$2985.77$} &  \multicolumn{2}{c|}{$\mathbf{978.39}$}& \multicolumn{2}{c}{$2577.02$} &  \multicolumn{2}{c}{$\mathbf{1106.12}$}\\
$11\le B < 12$  & \multicolumn{2}{c}{\em ($\mathit{98.44}$)} & \multicolumn{2}{c|}{\em ($\mathit {225.75}$)} & \multicolumn{2}{c}{$256.04$} &  \multicolumn{2}{c}{$\mathbf{251.98}$}\\
$12\le B < 19$  & \multicolumn{2}{c}{$1167.48$ } & \multicolumn{2}{c|}{$\mathbf{1476.09}$} & \multicolumn{2}{c}{$170.62$} &  \multicolumn{2}{c}{$\mathbf{901.59}$}\\\hline
$6 \le B < 19$   & \multicolumn{2}{c}{$4163.$} & \multicolumn{2}{c|}{$\mathbf{2680.}$} & \multicolumn{2}{c}{$5845.42.$} &  \multicolumn{2}{c}{$\mathbf{2555.12}$}\\\hline\hline
$ 6\le B < 11$ and $12\le B < 19$ & \multicolumn{2}{c}{$4153.$} & \multicolumn{2}{c|}{$\mathbf{2454.}$} & \multicolumn{2}{c}{$5845.42.$} &  \multicolumn{2}{c}{$\mathbf{2555.12}$}\\
$6 \le B < 19$   & \multicolumn{2}{c}{($7251.$)} & \multicolumn{2}{c|}{($\mathbf{3004.}$)} & \multicolumn{2}{c}{$5845.42.$} &  \multicolumn{2}{c}{$\mathbf{2555.12}$}\\\hline\hline
\end{tabular}
\end{center}
\end{table*}

\begin{table*}
\caption{USNO-A2 R fluxes, given in  \protect $10^{-9}\, \rm
W\,m^{-2}$. We give the values fitted to USNO-A2 R fluxes
(see Fig.~\protect\ref{fig:usnor4}) for \protect $1/\sin (b) <6$.}
\label{tab:usnor4}
\begin{center}
\begin{tabular}{rrrrrrrrr}
 \hline  \hline
&\multicolumn{4}{c}{Galactic Centre}
&\multicolumn{4}{c}{Galactic Anti-Centre}\\ 
R mag.&\multicolumn{2}{c}{North} &
\multicolumn{2}{c}{South} &
\multicolumn{2}{c}{North} &
\multicolumn{2}{c}{South}\\ &
{$|b|=90\degr$} & {\bf slope}&
{$|b|=90\degr$} & {\bf slope}&
{$|b|=90\degr$} & {\bf slope}&
{$|b|=90\degr$} & {\bf slope} \\ \hline
\multicolumn{9}{c}{USNO-A2 data}\\ 
$12\le R < 13$ &$ 79.69$ & $ \mathbf{145.06}$ & $   100.21$ & $ \mathbf{202.54}$ & $    92.32$ & $  \mathbf{95.03}$ & $    94.15$ & $ \mathbf{ 84.65}$\\      
$13\le R < 14$ &$ 37.06$ & $ \mathbf{150.61}$ & $     4.46$ & $ \mathbf{244.76}$ & $    64.51$ & $  \mathbf{89.38}$ & $    86.42$ & $ \mathbf{ 74.08}$\\      
$14\le R < 15$ &$ 13.02$ & $ \mathbf{129.08}$ & $   -38.66$ & $ \mathbf{212.66}$ & $    48.15$ & $  \mathbf{71.91}$ & $    41.65$ & $ \mathbf{ 68.61}$\\      
$15\le R < 16$ &$  1.80$ & $ \mathbf{110.32}$ & $   -53.84$ & $ \mathbf{182.94}$ & $    26.44$ & $  \mathbf{61.73}$ & $    22.37$ & $ \mathbf{ 56.58}$\\      
$16\le R < 17$ &$ -8.92$ & $ \mathbf{ 88.87}$ & $   -48.69$ & $ \mathbf{143.69}$ & $     8.12$ & $  \mathbf{43.45}$ & $     7.76$ & $ \mathbf{ 40.04}$\\      
$17\le R < 18$ &$-14.86$ & $ \mathbf{ 77.51}$ & $   -47.94$ & $ \mathbf{121.75}$ & $    -1.23$ & $  \mathbf{30.52}$ & $     1.66$ & $ \mathbf{ 29.04}$\\\hline
\multicolumn{9}{c}{}\\ 
\multicolumn{9}{c}{Simulations}\\ 
$-6.\le R < -5.$ & $   2.24$ &  $\mathbf{  0.02}$ & $   2.47$ & $ \mathbf{ -0.02}$ & $   2.24$ &  $\mathbf{  0.02}$ & $   2.47$ &  $\mathbf{ -0.03}$\\
$-5.\le R < -4.$ & $   4.19$ &  $\mathbf{  0.04}$ & $   4.62$ & $ \mathbf{ -0.04}$ & $   4.18$ &  $\mathbf{  0.03}$ & $   4.61$ &  $\mathbf{ -0.05}$\\
$-4.\le R < -3.$ & $   8.03$ &  $\mathbf{  0.08}$ & $   9.06$ & $ \mathbf{ -0.11}$ & $   8.01$ &  $\mathbf{  0.08}$ & $   9.04$ &  $\mathbf{ -0.11}$\\
$-3.\le R < -2.$ & $  16.20$ &  $\mathbf{  0.22}$ & $  18.77$ & $ \mathbf{ -0.25}$ & $  16.15$ &  $\mathbf{  0.21}$ & $  18.70$ &  $\mathbf{ -0.25}$\\
$-2.\le R < -1.$ & $  25.31$ &  $\mathbf{  0.43}$ & $  30.31$ & $ \mathbf{ -0.39}$ & $  25.19$ &  $\mathbf{  0.42}$ & $  30.15$ &  $\mathbf{ -0.40}$\\
$-1.\le R <  0.$ & $  37.62$ &  $\mathbf{  1.02}$ & $  46.91$ & $ \mathbf{ -0.24}$ & $  37.36$ &  $\mathbf{  0.98}$ & $  46.55$ &  $\mathbf{ -0.28}$\\
$ 0.\le R <  1.$ & $  51.95$ &  $\mathbf{  2.17}$ & $  67.20$ & $ \mathbf{  0.69}$ & $  51.40$ &  $\mathbf{  2.05}$ & $  66.43$ &  $\mathbf{  0.56}$\\
$ 1.\le R <  2.$ & $  68.12$ &  $\mathbf{  4.62}$ & $  91.55$ & $ \mathbf{  3.37}$ & $  67.05$ &  $\mathbf{  4.32}$ & $  90.05$ &  $\mathbf{  3.00}$\\
$ 2.\le R <  3.$ & $  88.53$ &  $\mathbf{  8.91}$ & $ 121.29$ & $ \mathbf{  8.90}$ & $  86.63$ &  $\mathbf{  8.22}$ & $ 118.61$ &  $\mathbf{  8.00}$\\
$ 3.\le R <  4.$ & $ 107.01$ &  $\mathbf{ 16.05}$ & $ 147.62$ & $ \mathbf{ 19.07}$ & $ 104.05$ &  $\mathbf{ 14.53}$ & $ 143.44$ &  $\mathbf{ 17.01}$\\
$ 4.\le R <  5.$ & $ 120.52$ &  $\mathbf{ 26.40}$ & $ 165.66$ & $ \mathbf{ 34.26}$ & $ 116.51$ &  $\mathbf{ 23.33}$ & $ 160.06$ &  $\mathbf{ 30.03}$\\
$ 5.\le R <  6.$ & $ 124.63$ &  $\mathbf{ 39.66}$ & $ 169.40$ & $ \mathbf{ 53.60}$ & $ 120.13$ &  $\mathbf{ 34.08}$ & $ 163.17$ &  $\mathbf{ 45.86}$\\
$ 6.\le R <  7.$ & $ 117.68$ &  $\mathbf{ 54.20}$ & $ 157.03$ & $ \mathbf{ 74.33}$ & $ 113.76$ &  $\mathbf{ 45.12}$ & $ 151.73$ &  $\mathbf{ 61.71}$\\
$ 7.\le R <  8.$ & $  99.92$ &  $\mathbf{ 67.51}$ & $ 129.94$ & $ \mathbf{ 92.73}$ & $  97.98$ &  $\mathbf{ 54.22}$ & $ 127.56$ &  $\mathbf{ 74.30}$\\
$ 8.\le R <  9.$ & $  76.71$ &  $\mathbf{ 76.42}$ & $  95.67$ & $ \mathbf{104.57}$ & $  77.90$ &  $\mathbf{ 58.91}$ & $  97.90$ &  $\mathbf{ 80.35}$\\
$ 9.\le R < 10.$ & $  53.41$ &  $\mathbf{ 79.19}$ & $  62.59$ & $ \mathbf{107.30}$ & $  58.05$ &  $\mathbf{ 58.28}$ & $  69.87$ &  $\mathbf{ 78.60}$\\
$10.\le R < 11.$ & $  33.95$ &  $\mathbf{ 75.96}$ & $  36.46$ & $ \mathbf{101.35}$ & $  41.17$ &  $\mathbf{ 53.22}$ & $  47.42$ &  $\mathbf{ 70.51}$\\
$11.\le R < 12.$ & $  18.59$ &  $\mathbf{ 68.85}$ & $  17.45$ & $ \mathbf{ 90.00}$ & $  27.17$ &  $\mathbf{ 45.78}$ & $  30.11$ &  $\mathbf{ 59.31}$\\\hline
$12.\le R < 13.$ & $   6.01$ &  $\mathbf{ 60.16}$ & $   3.06$ & $ \mathbf{ 76.90}$ & $  15.59$ &  $\mathbf{ 37.60}$ & $  16.55$ &  $\mathbf{ 47.60}$\\
$13.\le R < 14.$ & $  -3.99$ &  $\mathbf{ 50.96}$ & $  -7.85$ & $ \mathbf{ 63.78}$ & $   6.85$ &  $\mathbf{ 29.40}$ & $   6.65$ &  $\mathbf{ 36.45}$\\
$14.\le R < 15.$ & $ -10.34$ &  $\mathbf{ 41.37}$ & $ -14.40$ & $ \mathbf{ 50.85}$ & $   1.42$ &  $\mathbf{ 21.62}$ & $   0.72$ &  $\mathbf{ 26.33}$\\
$15.\le R < 16.$ & $ -12.74$ &  $\mathbf{ 31.82}$ & $ -16.40$ & $ \mathbf{ 38.51}$ & $  -1.06$ &  $\mathbf{ 14.82}$ & $  -1.81$ &  $\mathbf{ 17.77}$\\
$16.\le R < 17.$ & $ -12.02$ &  $\mathbf{ 23.04}$ & $ -14.90$ & $ \mathbf{ 27.50}$ & $  -1.50$ &  $\mathbf{  9.41}$ & $  -2.06$ &  $\mathbf{ 11.13}$\\
$17.\le R < 18.$ & $  -9.47$ &  $\mathbf{ 15.63}$ & $ -11.47$ & $ \mathbf{ 18.43}$ & $  -0.99$ &  $\mathbf{  5.55}$ & $  -1.30$ &  $\mathbf{  6.48}$\\\hline
$18.\le R < 19.$ & $  -6.46$ &  $\mathbf{  9.99}$ & $  -7.69$ & $ \mathbf{ 11.64}$ & $  -0.42$ &  $\mathbf{  3.13}$ & $  -0.55$ &  $\mathbf{  3.61}$\\\hline
$19.\le R < 20.$ & $  -4.04$ &  $\mathbf{  6.16}$ & $  -4.73$ & $ \mathbf{  7.11}$ & $  -0.18$ &  $\mathbf{  1.78}$ & $  -0.22$ &  $\mathbf{  2.03}$\\
$20.\le R < 21.$ & $  -2.58$ &  $\mathbf{  3.83}$ & $  -2.98$ & $ \mathbf{  4.37}$ & $  -0.17$ &  $\mathbf{  1.05}$ & $  -0.20$ &  $\mathbf{  1.19}$\\
$21.\le R < 22.$ & $  -1.81$ &  $\mathbf{  2.45}$ & $  -2.06$ & $ \mathbf{  2.77}$ & $  -0.19$ &  $\mathbf{  0.62}$ & $  -0.22$ &  $\mathbf{  0.71}$\\\hline
\multicolumn{9}{c}{}\\														       
  \multicolumn{9}{c}{Data sums }\\ \hline
$12\le R < 18$ & $107.79$ & $ \mathbf{701.45}$ & $   -84.46$ & $\mathbf{1108.34}$ & $ 238.31$ & $ \mathbf{392.02}$ & $   254.01$ & $ \mathbf{353.00}$ \\ \hline
\multicolumn{9}{c}{Simulated sums }\\ \hline
$-6\le R < 6 $ &  $ 654.35$ & $\mathbf{  99.62}$ & $874.86$ & $\mathbf{ 118.84}$ & $638.89$ &   $\mathbf{ 88.26}$ & $853.29$ & $\mathbf{103.34}$\\
$ 6\le R < 12$ &  $ 400.26$ & $\mathbf{ 422.13}$ & $499.16$ & $\mathbf{ 570.30}$ & $416.02$ &   $\mathbf{315.53}$ & $524.58$ & $\mathbf{424.79}$\\
$12\le R < 18$ &  $ -42.56$ & $\mathbf{ 222.99}$ & $-61.97$ & $\mathbf{ 275.96}$ & $ 20.31$ &   $\mathbf{118.39}$ & $ 18.74$ & $\mathbf{145.75}$\\
$18\le R < 22$ &  $ -14.88$ & $\mathbf{  22.43}$ & $-17.46$ & $\mathbf{  25.90}$ & $ -0.96$ &   $\mathbf{  6.58}$ & $ -1.19$ & $\mathbf{ 7.54}$ \\ 
\hline
\multicolumn{9}{c}{}\\ 
\multicolumn{9}{c}{Total contributions}\\ \hline
&\multicolumn{4}{c|}{Data}&\multicolumn{4}{c}{Simulations}\\ \hline
&\multicolumn{2}{c}{$|b|=90\degr$}&\multicolumn{2}{c|}{\bf slope}&\multicolumn{2}{c}{$|b|=90\degr$}&\multicolumn{2}{c}{\bf slope}\\ \hline
$-6\le R < 6$   & \multicolumn{4}{c|}{} &\multicolumn{2}{c}{$3021.39$} &  \multicolumn{2}{c}{$\mathbf{ 410.06}$}\\
$6\le R < 12$   & \multicolumn{4}{c|}{} &\multicolumn{2}{c}{$1840.02$} &  \multicolumn{2}{c}{$\mathbf{1732.75}$}\\
$12\le R < 18$  & \multicolumn{2}{c}{$515.65$} &\multicolumn{2}{c|}{$\mathbf{2554.81}$} & \multicolumn{2}{c}{$ -65.48$} &  \multicolumn{2}{c}{$\mathbf{763.09}$}\\
$18\le R < 22$  & \multicolumn{4}{c|}{} &\multicolumn{2}{c}{$-34.49$} &  \multicolumn{2}{c}{$\mathbf{ 62.45}$}\\\hline
$6\le R < 22$   & \multicolumn{4}{c|}{} &\multicolumn{2}{c}{$4761.44$} &  \multicolumn{2}{c}{$\mathbf{ 2968.55}$}\\\hline\hline
\end{tabular}
\end{center}
\end{table*}

\begin{table*}
\caption{Values of the surface brightness computed with the COBE/DIRBE
data, given in $10^{-9}\, \rm W\,m^{-2}$. We give the values fitted to
the COBE/DIRBE data for $1/\sin|b|<6$. The influence of the point
sources is removed with a 5$\sigma$ clipping. The slopes are computed
as explained for the optical (for $1/\sin|b|<6$), except for the 12
and 25 $\mu$m data. As explained in the text (\S\ \ref{sec:isb}),
these data exhibit a ``non-Galactic behaviour'' and the corresponding
fits are performed for $2<1/\sin|b|<6$.}
\label{tab:cobe}
\begin{center}
\begin{tabular}{rrrrrrrrr}
\hline\hline
\multicolumn{9}{c}{}\\
&\multicolumn{4}{c}{Galactic Centre}
&\multicolumn{4}{c}{Galactic Anti-Centre}\\
\multicolumn{1}{c}{$\lambda$}&\multicolumn{2}{c}{North} & \multicolumn{2}{c}{South} & \multicolumn{2}{c}{North} &
\multicolumn{2}{c}{South}\\ & {$|b|=90\degr$}
& {slope}&
{$|b|=90\degr$} & {slope}&
{$|b|=90\degr$} & {slope}&
{$|b|=90\degr$} & {slope}\\ \hline
$1.25 \mu$m & $2050.$ & $ \mathbf{ 2240.}$ & $1720.$ & $ \mathbf{2790.}$  & $2180.$ & $ \mathbf{1130.}$& $2320.$ & $ \mathbf{1330.}$\\
$2.20 \mu$m & $ 964.$ & $ \mathbf{ 1130.}$ & $ 787.$ & $ \mathbf{1390.}$  & $1020.$ & $ \mathbf{541.}$& $1050.$ & $ \mathbf{ 676.}$\\
$3.50 \mu$m & $ 397.$ & $ \mathbf{  393.}$ & $ 333.$ & $ \mathbf{ 472.}$  & $ 408.$ & $ \mathbf{191.}$& $ 417.$ & $ \mathbf{ 252.}$\\
$4.90 \mu$m & $ 324.$ & $ \mathbf{  155.}$ & $ 305.$ & $ \mathbf{ 171.}$  & $ 317.$ & $ \mathbf{77.8}$& $ 345.$ & $ \mathbf{ 97.3}$\\
$12 \mu$m   & $1630.$ & $ \mathbf{  121.}$ & $1500.$ & $ \mathbf{ 86.7}$  & $ 979.$ & $ \mathbf{181.}$& $1440.$ & $ \mathbf{ 139.}$\\
$25 \mu$m   & $1260.$ & $ \mathbf{  57.5}$ & $1260.$ & $ \mathbf{ 28.5}$  & $1020.$ & $ \mathbf{69.1}$& $1160.$ & $ \mathbf{ 59.0}$\\
$60 \mu$m   & $ 288.$ & $ \mathbf{  239.}$ & $ 136.$ & $ \mathbf{ 203.}$  & $ 148.$ & $ \mathbf{133.}$& $ 272.$ & $ \mathbf{ 201.}$\\
$100 \mu$m  & $ 665.$ & $ \mathbf{  549.}$ & $ 406.$ & $ \mathbf{ 439.}$  & $ 235.$ & $ \mathbf{385.}$& $ 563.$ & $ \mathbf{ 511.}$\\
$140\mu$m   & $ 764.$ & $ \mathbf{  800.}$ & $ 553.$ & $ \mathbf{ 567.}$  & $ 185.$ & $ \mathbf{638.}$& $ 660.$ & $ \mathbf{ 802.}$\\
$240\mu$m   & $ 315.$ & $ \mathbf{  351.}$ & $ 267.$ & $ \mathbf{ 238.}$  & $ 45.8$ & $ \mathbf{332.}$& $ 302.$ & $ \mathbf{ 382.}$\\\hline
\multicolumn{9}{c}{}\\
\multicolumn{9}{c}{Contribution from the Galactic Centre (left) and
Anti-Centre (right)}\\
\multicolumn{9}{c}{}\\\hline
$1.25 \mu$m &   \multicolumn{4}{c}{ 3782. ($|b|=90\degr$) 5040. (slope)}&  \multicolumn{4}{c}{  4500.  ($|b|=90\degr$) 2462. (slope)}\\  
$2.20 \mu$m &   \multicolumn{4}{c}{ 1752. ($|b|=90\degr$) 2520. (slope)}&  \multicolumn{4}{c}{  2066.  ($|b|=90\degr$) 1216. (slope)}\\  
$3.50 \mu$m &   \multicolumn{4}{c}{  730. ($|b|=90\degr$)  864. (slope)}&  \multicolumn{4}{c}{   826.  ($|b|=90\degr$)  442. (slope)}\\  
$4.90 \mu$m &   \multicolumn{4}{c}{  630. ($|b|=90\degr$)  324. (slope)}&  \multicolumn{4}{c}{   662.  ($|b|=90\degr$)  174. (slope)}\\  
$12 \mu$m   &   \multicolumn{4}{c}{ 3126. ($|b|=90\degr$)  208. (slope)}&  \multicolumn{4}{c}{  2418.  ($|b|=90\degr$)  320. (slope)}\\  
$25 \mu$m   &   \multicolumn{4}{c}{ 2522. ($|b|=90\degr$)   86. (slope)}&  \multicolumn{4}{c}{  2182.  ($|b|=90\degr$)  128. (slope)}\\  
$60 \mu$m   &   \multicolumn{4}{c}{  424. ($|b|=90\degr$)  442. (slope)}&  \multicolumn{4}{c}{   420.  ($|b|=90\degr$)  332. (slope)}\\  
$100 \mu$m  &   \multicolumn{4}{c}{ 1070. ($|b|=90\degr$)  988. (slope)}&  \multicolumn{4}{c}{   800.  ($|b|=90\degr$)  894. (slope)}\\      
$140\mu$m   &   \multicolumn{4}{c}{ 1318. ($|b|=90\degr$) 1366. (slope)}&  \multicolumn{4}{c}{   844.  ($|b|=90\degr$)  1440.(slope)}\\      
$240\mu$m   &   \multicolumn{4}{c}{  584. ($|b|=90\degr$)  590. (slope)}&  \multicolumn{4}{c}{   246.  ($|b|=90\degr$)  714. (slope)}\\ \hline
\multicolumn{9}{c}{}\\
\multicolumn{9}{c}{Surface brightness computed over the whole sky}\\ 
\multicolumn{9}{c}{}\\\hline
$1.25 \mu$m &   \multicolumn{8}{c}{8282.($|b|=90\degr$) 7502. (slope)}\\
$2.20 \mu$m &   \multicolumn{8}{c}{3818.($|b|=90\degr$) 3736. (slope)}\\
$3.50 \mu$m &   \multicolumn{8}{c}{1556.($|b|=90\degr$) 1306. (slope)}\\
$4.90 \mu$m &   \multicolumn{8}{c}{1292.($|b|=90\degr$)  500. (slope)}\\
$12 \mu$m   &   \multicolumn{8}{c}{5544.($|b|=90\degr$)  528. (slope)}\\
$25 \mu$m   &   \multicolumn{8}{c}{4704.($|b|=90\degr$)  214. (slope)}\\
$60 \mu$m   &   \multicolumn{8}{c}{ 844.($|b|=90\degr$)  774. (slope)}\\
$100 \mu$m  &   \multicolumn{8}{c}{1870.($|b|=90\degr$) 1882. (slope)}\\
$140\mu$m   &   \multicolumn{8}{c}{2162.($|b|=90\degr$) 2806. (slope)}\\
$240\mu$m   &   \multicolumn{8}{c}{ 930.($|b|=90\degr$) 1304. (slope)}\\\hline
\end{tabular}
\end{center}
\end{table*}

\clearpage

\section{Surface brightness definition}
\label{app:sb}
Following the usual conventions, the surface brightness $\nu S_\nu$ is
defined as:
\begin{equation}
\nu S_\nu = \frac{L_\lambda}{\pi R^2}
\end{equation}
where $R$ is the radius (in m) of the area probed and $L_\lambda$ its
luminosity in W at the wavelength $\lambda$. $\nu S_\nu$ can be
directly transformed in $L_{\lambda\odot} \,pc^{-2}$. It can also be
transformed in a magnitude scale $\mu$ (mag.arcsec$^{-2}$) with:
\begin{equation}
\mu_\lambda = -2.5 \log_{10} [\nu S_\nu \times \frac{1}{c_\lambda} \times
\frac{x_{sr\rightarrow arcsec^2}}{4\pi}]
\label{eq:mu}
\end{equation}
where $x_{sr\rightarrow arcsec^2}=23.5\times 10^{-12} arcsec^2 / sr$
is the solid angle normalisation. $c_\lambda$ is the conversion factor
of the observed surface brightness into magnitude. This conversion
factor is for instance provided by
NED\footnote{http://nedwww.ipac.caltech.edu/help/photoband.lst} For B, V
and R bands, we thus have the following conversions:
\begin{equation}
\mu_B = -2.5 \log_{10} [\nu_B S_B \times 
\frac{1}{2.90\times 10^{-8}} \times \frac{23.5\times 10^{-12}}{4\pi}]
\end{equation}
\begin{equation}
\mu_V = -2.5 \log_{10} [\nu_V S_V \times 
\frac{1}{1.97\times 10^{-8}}\times \frac{23.5\times 10^{-12}}{4\pi}]
\end{equation}
\begin{equation}
\mu_R = -2.5 \log_{10} [\nu_R S_R \times 
\frac{1}{1.44\times 10^{-8}}\times \frac{23.5\times 10^{-12}}{4\pi}]
\end{equation}
with $\nu_B S_B$, $\nu_V S_V$ and $\nu_R S_R$ the surface brightness
in B, V and R in W\,m$^{-2}$, while $\mu_B$, $\mu_V$ and $\mu_R$ are
in B mag\,arcsec$^{-2}$, V mag\,arcsec$^{-2}$ and R
mag\,arcsec$^{-2}$.

\section{Normalisation factor of the cosecant integration method}
\label{sect:norm}
We check in the simple case of a uniform distribution ($n(r,b,l,M_V) =
1$, $f(A_V(r,b))=1$ and $\phi(M_V) L(M_V) = 1$) the normalisation of
the cosecant integration method with respect to the integration in a
cylinder.
\begin{eqnarray}
\label{eq:eqn1}
S^\prime_{cyl}= && [ \int_{R=0}^{R_{cyl}}\int_{z=-z_{max}}^{+z_{max}}
	\int_{l=0}^{2\pi}  dl dz R \,dR \, ]/ {\pi R_{cyl}^2} = 2 z_{max}
\end{eqnarray}
\begin{eqnarray}
\label{eq:eqn2}
S^\prime_{cosec}= && \int_{b=-\frac{\pi}{2}}^{\frac{\pi}{2}}
\int_{l=0}^{2\pi} \int_{r=0}^{r_{max}(b)}
\frac{ r^2 dr dl \cos{b} \sin{|b|} db  }{4\pi r^2}\nonumber\\ 
 = &&  z_{max}\\
\end{eqnarray}
$S^\prime_{cyl}$ is the simplified version of the surface brightness
computed as the integration in a cylinder, while $S^\prime_{cosec}$
corresponds to the integration with the $\sin{|b|}$ weighting used for
the cosecant method. They differ by a factor of $2$ understood as the
difference of the volumes probed. As the behaviour of the exponential
disc geometry is close to the plane parallel geometry (see
e.g. Fig. \ref{fig:ppexpo}), we apply this factor 2 to the
normalisation of all estimates of the surface brightness based on the
cosecant approximation, namely the integration in a cone
(Eq. \ref{eq:cyl}), the corrected direct estimate
(Eqs. \ref{eq:direct} and \ref{eq:corrdirect}) and the cosecant
compilations of the various datasets. We check that this factor does
not change more than 7$\%$ with an exponential geometry.

\section{Luminosity function (LF)}  
\label{sect:LF}
\begin{figure}
\resizebox{0.5\textwidth}{!}{\includegraphics{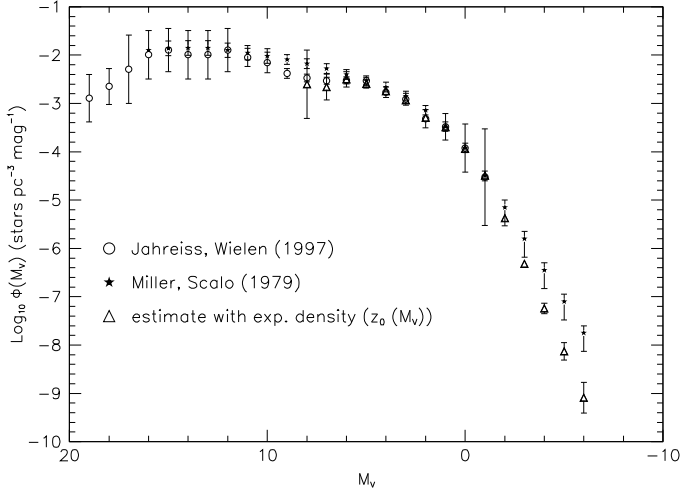}}
\caption{Luminosity function at the Solar Neighbourhood. The LF used
for the simulation is presented \protect
\citep{Miller:1979,Jahreiss:1997}. The LF used for the
direct/indirect estimation is also displayed with error bars
accounting for Poisson noise only. }
\label{fig:lf}
\end{figure}

We use two estimates of the LF to define $\phi(M_V)$: (1) the
\cite{Miller:1979} estimate for $M_V < 2$, (2) the
\cite{Jahreiss:1997} estimate based on Hipparcos data for $M_V \ge 2$.
They are displayed on Fig. \ref{fig:lf}. We find that this choice of
the LF combined with the variable scale height \protect $h_Z(M_V)$ of
\protect \cite{Miller:1979} is equivalent to a fixed scale height
\protect $h_Z=125pc$ 
(used to compare the plane parallel and exponential disc geometries).

\begin{figure}
\resizebox{0.5\textwidth}{!}{\includegraphics{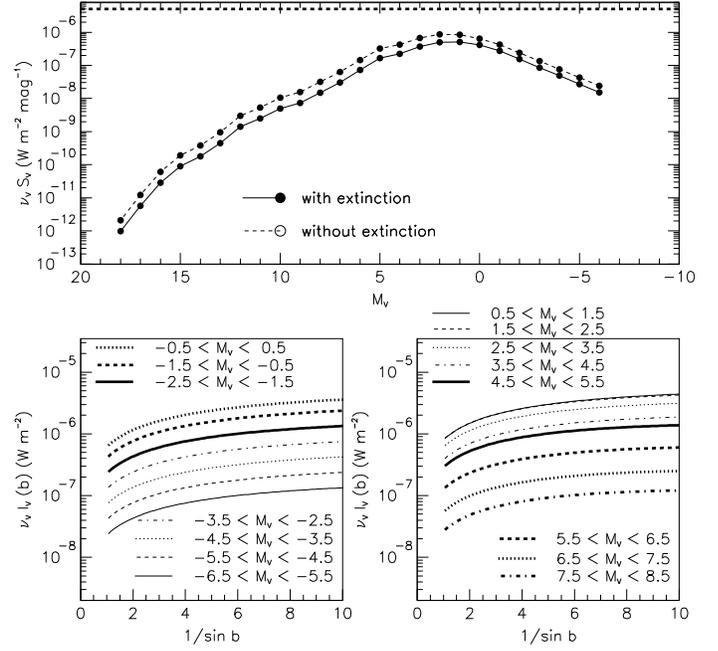}}
\caption{Contribution of each absolute magnitude bin to the $V$ surface
brightness (computed with $1/\sin |b| < 6$ from the Solar
Neighbourhood, with $h_Z(M_V)$). The upper panel displays the
contribution to $\nu_V S_V$ of each bin of absolute magnitude. The
horizontal dashed line corresponds to the cumulated value (the
estimated surface brightness). The lower panels exhibit the
corresponding decomposition of $\nu_V I_V(|b|)$ for each
absolute-magnitude bin (with extinction). These values have to be
increased by 65$\%$ (see Fig. \protect\ref{fig:ratio}) to correspond
to a face-on value estimate of the V surface
brightness.}\label{fig:newconsl2}
\end{figure}

\begin{figure}
\resizebox{0.5\textwidth}{!}{\includegraphics{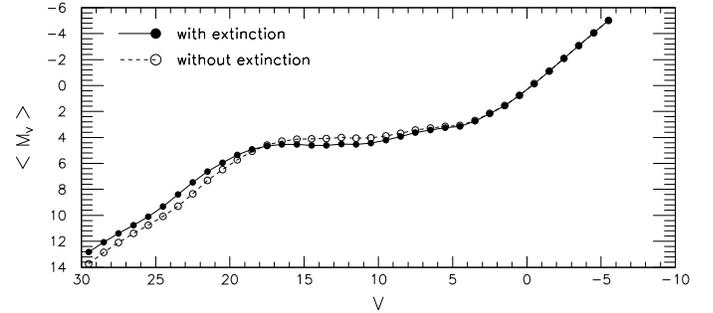}}
\caption{Mean absolute magnitude stars contribution to each apparent
magnitude bin.}\label{fig:meanmag1}  
\end{figure}

\subsection{Uncertainties}
\label{ssect:LF1}
Figure \ref{fig:newconsl2} shows the relative contribution to the
surface brightness from the Solar Neighbourhood of each
absolute-magnitude bin. This has been computed with the simulations
described in Sect. \ref{ssec:toy}.  Dim stars with $M_V>7$ do not
contribute significantly to the surface brightness ($1\%$), while the
magnitude interval $-6 \le M_V \le 6$ is the main contributor to the
surface brightness ($98\%$). It is important to note that the bright
end ($M_V<0.5$) of the LF constitutes a significant contribution ($34\%$ of
the surface brightness). Hence, theoretical estimations of the surface
brightness based on the LF depend on the uncertainties on the upper
tail. However, stars with $M_V<-1.5$ (resp. $M_V<-3.5$) correspond to
only $11\%$ (resp. $3\%$) of the surface brightness. This effect
disappears when one considers stellar catalogues, which provide
apparent magnitudes (see Fig. \ref{fig:newconsl3}). Figure
\ref{fig:newconsl2} displays the value of the absolute magnitude of
the ``mean star'' of each apparent magnitude bin, which gives a
qualitative behaviour of these related parameters.

\subsection{Computation of the LF for direct/indirect estimates}
\label{ssect:LF2}
\begin{figure}
\resizebox{0.5\textwidth}{!}{\includegraphics{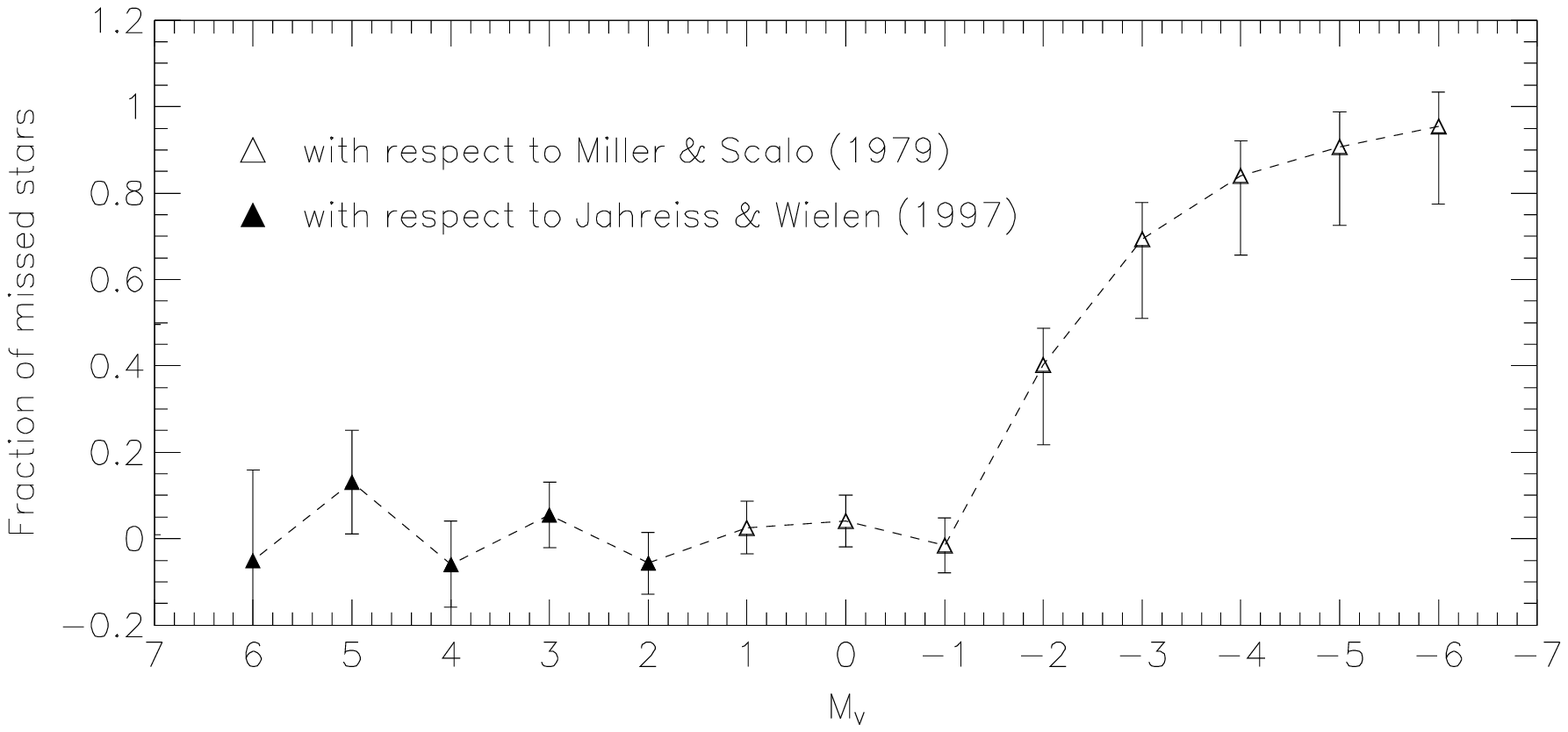}}
\caption{Fraction $f_i^{missed}$ of the stars missed with respect to 
\citet{Miller:1979} for each magnitude bin. The correction factor
$f_i^{LF} = 1/(1-f_i^{missed})$ is applied for the computation of the
corrected and indirect estimates in Sect. \ref{sec:synth} in order to
be compatible with the simulations. The error bars account for the
uncertainties on the
\citet{Jahreiss:1997,Miller:1979} LF used for the simulations and the
Poisson noise of our estimate.}\label{fig:lfcorr}
\end{figure}
We present here the estimate of the LF, based on the Hipparcos
catalogue, used to compute the direct and indirect estimates.  The
Hipparcos Catalogue \citep{Perryman:1997,vanLeeuwen:1997} is complete
for $V<7.3$, and parallaxes are available for most of the stars. We
consider that the number of stars in each magnitude bin
$M_V-\frac{1}{2} \le m < M_V+\frac{1}{2}$, where $m$ is the absolute
magnitude, is complete within a sphere of radius $R(M_V+\frac{1}{2})=
10^{(7.3-(M_V+\frac{1}{2}))/5.+1}\,\rm pc$ for $-6.5
\le M_V < 6.5$. We count stars in these volumes $vol(R)$ using,
\begin{equation} 
\phi(M_V) = \frac{N(M_V-\frac{1}{2} \le m < M_V+\frac{1}{2})} 
		{vol(R)}
\label{eq:lf} 
\end{equation} 
where $N(M_V-\frac{1}{2} \le m < M_V +\frac{1}{2})$ is the number of
stars found with an absolute magnitude $m$ in this bin, and the volume
is defined as: $vol(R)= \int_{r<R} \eta(r,b,l;h_Z(M_V)) r^2 d\sin b dl
dr$, and where $\eta(r,b,l;h_Z(M_V))=n(r,b,l,M_V)/n_\odot(M_V)$ (open
triangles in Fig.~\ref{fig:lf}).  This accounts for the exponential
profile of the Galaxy, with different vertical scales for different
magnitude ranges. We adopt the same scale heights as
\cite{Miller:1979} ($h_Z(M_V\le 0)=80$pc).   We also correct each star
for interstellar extinction according to our extinction modelling
described in Appendix \ref{sec:ext}.

Among the 20\,906 stars with $V<7.3$, we consider the 18\,381 stars
with a parallax significant at $3\sigma$. We further restrict this
sample by 29$\%$ to account the completion limit for each magnitude
bin. We then treat the $13\,374$ stars as follows:
\begin{itemize}
\item 9\,444 stars have a standard spectral type with a luminosity
class, that can be found in the \cite{Pickles:1998} library.
\item 
548 stars either have a standard spectral type with no luminosity
class available in \cite{Pickles:1998} and that has to be interpolated
between existing templates, or have two spectral types (the average is
then adopted).
\item 
3\,114 stars have a well defined spectral type, but no luminosity
class. We rely on $M_V$ to estimate it, and derive a template.
\item 1 Wolf-Rayet star and 20 C/N stars, which do not exceed
$3.06\times 10^{-8}$~W m$^{-2}$ sr$^{-1}$ (corrected direct) and
$1.40\times 10^{-9}$~W m$^{-2}$ sr$^{-1}$ (indirect) in V.
\item 247 stars, with no precise spectral type ($63\%$ Am/Ap, $18\%$
M), are neglected. We estimate that their contribution to the V
surface brightness is $6.80\times 10^{-8}$~W m$^{-2}$ sr$^{-1}$
(corrected direct) and $5.69\times 10^{-9}$~W m$^{-2}$ sr$^{-1}$
(indirect).
\end{itemize}

Figure \ref{fig:lf} summarises the LF thereby obtained.  The
discrepancy with respect to \citet{Miller:1979} is significant for
luminous stars. The corresponding stellar population is strongly
inhomogeneous \citep{deZeeuw:1999}, and this part of the LF certainly
suffers uncertainties larger than Poisson noise (see also
Appendix \ref{ssect:3D}). In order, to use the same LF as for the
simulations, we correct the LF as displayed in Fig. \ref{fig:lfcorr},
for stars brighter than magnitude $-1.5$.

\subsection{3D visualisation of the Hipparcos data}
\label{ssect:3D}
We present a 3D visualisation of the Hipparcos catalogue on
http://www.obspm.fr/sbhip06. We also position a schematic Gould's belt
in the Galactic plane following Guillout et al. (1998). Gould's belt
is significantly larger than the volume probed by the Hipparcos
catalogue.  The early-type stars are distributed in the Galactic Disk,
but do not exhibit a clear (20deg) inclination. More generally, the
Hipparcos stars distribution is clearly flatten towards the Galactic
Poles, but no clear correlation with Gould's belt is detected due to
the small volume sampled.  For the majority of the stars, the cut in
magnitude is clearly seen as a spheroidal distribution of
stars. However, the bright stars exhibit a more irregular
distribution. The apparent contours are roughly compatible with the
local cavity observed in the gas distribution
(e.g. \citealt{Lallement:2003}), but there is not enough information to
conclude about an extinction bias.

\section{Extinction Modelling}
\label{sec:ext}
\begin{figure}
\resizebox{0.5\textwidth}{!}{\includegraphics{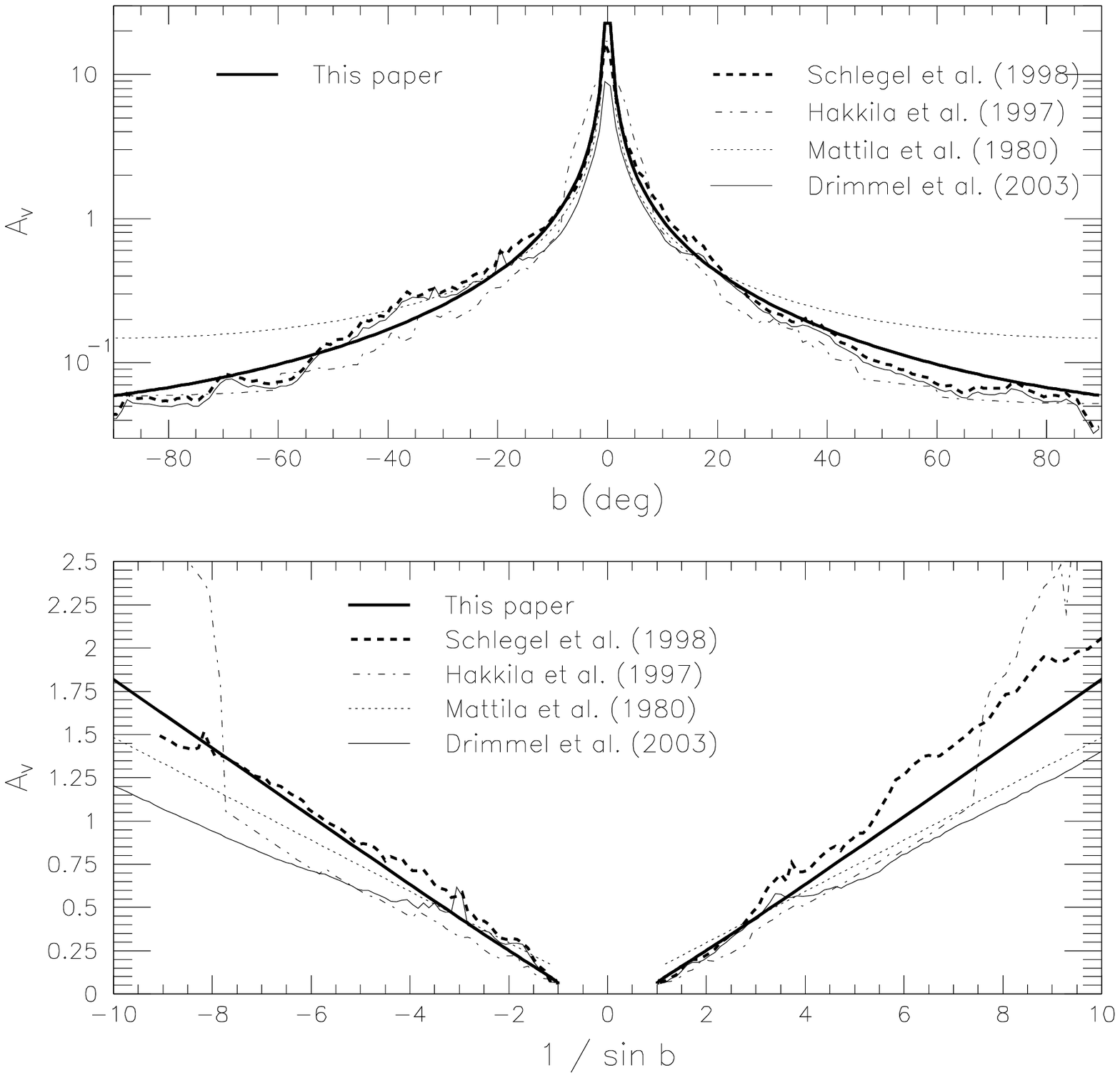}}
\caption{$A_V$ extinction dependence on Galactic latitude. The
extinction used for the simulations is compared with other models
\citep{Drimmel:2003,Hakkila:1997,Mattila:1980a} and the dust
maps \citep{Schlegel:1998}.}\label{fig:ext} 	 
\end{figure}
In Figure \ref{fig:ext}, we display the dependence of the $A_V$
extinction with the Galactic latitude obtained from dust maps by
\citet{Schlegel:1998}. We find systematic differences with
\citet{Hakkila:1997}, based on a compilation of various models. For the
purpose of this work, we wish to use a simple analytical model to
describe first order effects linked to extinction. We consider the
extinction model of \citet{Mattila:1980a}, based on a 2-components
model. Compared to the $A_V$ extinction derived from
\citet{Schlegel:1998}, it tends to overestimate the extinction towards
the pole (even though this excess remains within the error bars quoted by
\citealt{Hakkila:1997}), and to underestimate it at intermediate
latitudes. We further tune this type of law, and adopt the following empirical
model:
\begin{equation}
A_{V}(r,b) =  a_{v} \beta / \sin |b| \times (\frac{\pi}{b+\pi})^3  \times ( 1 - \exp(-(r{|\sin {b}|})/\beta)) 
\label{eq:av}
\end{equation}
with $a_{v} = 1.48$ mag/kpc and $\beta = 135$ pc. The inverse cube
power of $b$ has been defined empirically to reproduce the $b$
dependence on Fig. \ref{fig:ext}. This empirical law is also in good
agreement with the z-dependence of the extinction modelled by
\citet{Hakkila:1997}.
\subsection*{Numerical values}
\begin{table}
\caption{Values of the extinction correction applied to cosecant
measurements for the main filters used in this paper (see
Fig. \ref{fig:ratio} and Table \ref{tab:extval}). We consider
$R_{cyl}=1$\, kpc.}
\label{tab:extval}
\begin{center}
\begin{tabular}{rr|rr}
\hline
$\lambda$ ($\mu$m) & Correction & $\lambda$ ($\mu$m) & Correction
\\\hline 0.44 (B) &1.72 & 0.69 (R$_J$) & 1.36\\ 
0.55 (V) &1.52 & 1.25 (J) & 1.12\\
0.64 (R$_C$)& 1.41 & 2.20 (K) & 1.05\\
\hline
\end{tabular}
\end{center}
\end{table}
The extinction correction used to derive a surface brightness
measurement of the Galaxy at the Solar Neighbourhood is provided in
Table \ref{tab:extval}.

\section{Comparison of the USNO-A2 with UCAC-1 data}
\label{sec:compucac}
\begin{figure}[h]
\resizebox{0.5\textwidth}{!}{ \includegraphics{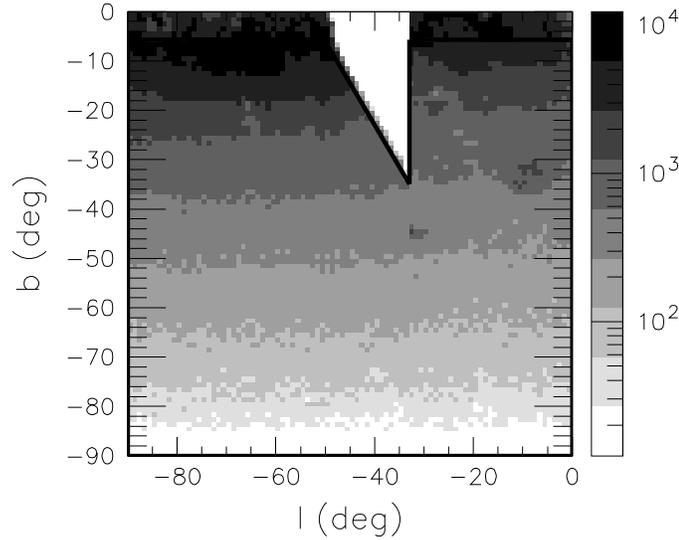}}
\caption{Galactic plane coverage of the  UCAC-1 catalogue \protect
\citep{Zacharias:2000}. We restrict our study to the area indicated
with thick lines, defined by $b<0$ and $-90\degr < l < 0\degr$, and we
exclude the area between $l=-33\degr$ and $b=-1.7 \times l
+522\degr$. The intensity corresponds to the number of stars with
$11\le R_U<16$.}
\label{fig:ucac}
\end{figure}
\begin{figure}[h]
\resizebox{0.5\textwidth}{!}{ \includegraphics{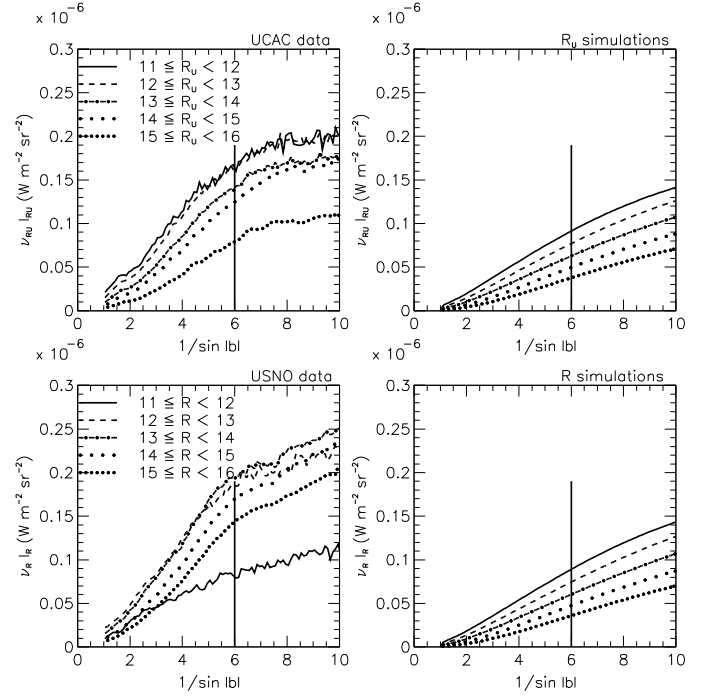}}
\caption{Integrated $R$ flux as a function of $1/\sin(|b|)$ for the
UCAC (top) and USNO-A2 (bottom) stars lying in the UCAC-1 area defined
in Fig.~\protect\ref{fig:ucac}.  These fluxes correspond to the
contribution of this area to the total flux received at the Sun. No
correction has been applied. The results of the fits for
$1/\sin(|b|)<6$ are provided in Table~\protect\ref{tab:slopeucac}.}
\label{fig:bucac0}
\end{figure}
\begin{table}
\caption{$R_U$ (UCAC-1) and $R$ (USNO-A2) fluxes, given in $10^{-9}\rm
W\,m^{-2}\,sr^{-1}$. We give the values (2) and (3) fitted to $R_U$
(UCAC-1) and $R$ (USNO-A2) fluxes for $1/\sin |b| <6$ (see
Fig.~E.1) over the area defined in Fig.~\ref{fig:ucac},
for each magnitude bin (1).  Last, we provide the ratio (4) between
the UCAC-1 and USNO-A2 slopes. No correction is applied.}
\label{tab:slopeucac}
\begin{center}
\begin{tabular}{rrrrr}
\hline\hline
Magnitude range & {$|b|=90\degr$} & {Slope} & {$|b|=90\degr$} & {Slope} \\
(1) & (2) & (3) & (4)& (5) \\
\hline
&\multicolumn{2}{l}{UCAC-1 data}& \multicolumn{2}{l}{$R_U$ Simulation}\\ 
$~8 \le R_U <~9$ & ${22.49}$ & $\mathbf{  8.07}$& $7.32$ &   $\mathbf{31.20} $ \\
$~9 \le R_U <10$ & ${31.98}$ & $\mathbf{ 16.38}$& $5.74$ &   $\mathbf{26.86} $ \\
$10 \le R_U <11$ & ${26.88}$ & $\mathbf{ 25.62}$& $4.22$ &   $\mathbf{22.22} $ \\
$11 \le R_U <12$ & ${17.56}$ & $\mathbf{ 31.55}$& $2.87$ &   $\mathbf{17.50} $ \\
$12 \le R_U <13$ & ${ 7.72}$ & $\mathbf{ 32.98}$& $1.80$ &   $\mathbf{12.99} $ \\
$13 \le R_U <14$ & ${ 0.03}$ & $\mathbf{ 28.88}$& $1.03$ &   $\mathbf{ 9.06} $ \\
$14 \le R_U <15$ & ${-3.95}$ & $\mathbf{ 25.61}$& $0.55$ &   $\mathbf{ 5.90} $ \\
$15 \le R_U <16$ & ${-4.87}$ & $\mathbf{ 16.36}$& $0.29$ &   $\mathbf{ 3.59} $ \\
&\multicolumn{2}{l}{USNO-A2 data}& \multicolumn{2}{l}{$R_C$ Simulation}\\ 
$11\le R <12$ & $13.51$ & $\mathbf{15.16}$&    $2.87$ &   $\mathbf{17.50}$\\
$12\le R <13$ & $14.94$ & $\mathbf{35.80}$&    $1.80$ &   $\mathbf{12.99}$\\
$13\le R <14$ & $ 4.00$ & $\mathbf{39.58}$&    $1.03$ &   $\mathbf{ 9.06}$\\
$14\le R <15$ & $-4.45$ & $\mathbf{34.63}$&    $0.55$ &   $\mathbf{ 5.90}$\\
$15\le R <16$ & $-7.39$ & $\mathbf{29.38}$&    $0.29$ &   $\mathbf{ 3.59}$\\
\hline
\end{tabular}
\end{center}
\end{table}
Figure \ref{fig:ucac} displays the area of the sky considered for this
comparison. Figure \ref{fig:bucac0} and Table \ref{tab:slopeucac} compare
the 2 data sets with simulations.

\clearpage

\section{Calibration of the USNO-A2 data}
\label{sec:calib}
\begin{table*}[hb]
\caption{Coefficients used for the calibration of the USNO
magnitudes. In blue (B), \protect $B_J = \alpha + \beta
(B_{\mathrm{USNO}}-15) + \gamma (B_{\mathrm{USNO}} -15)^2$.  In red
(R), \protect $R_C = \alpha + \beta (R_{\mathrm{USNO}}-15) + \gamma
(R_{\mathrm{USNO}}-15)^2$. The columns are defined as follows: (1)
the zone number, (2) the declination band covered, (3) the filter
used, (4) $\alpha$, (5) $\beta$, (6) $\gamma$, (7) the RMS error of
the polynomial fit, (8) the number of stars effectively used in the
fit. We thus used 104\,354 stars in blue and 154\,491 stars in red.}
\label{tab:calib}
\begin{center}
\begin{tabular}{lllllrlr}
\hline\hline
 {Zone} & {DEC} & {Filters} & {$\alpha$} & {$\beta$} & \multicolumn{1}{c}{$\gamma$} &
{Residuals} & {N$_{\mathrm{stars}}$}\\
 075 & -82.5$\degr$ to -75.0$\degr$ & B & 15.24 & 1.14 & 0.0 & 0.38 & 1111 \\
 & & R & 15.06 & 1.02 & $6.99\times 10^{-2}$ & 0.34 & 724\\
 150 & -75.0$\degr$ to -67.5$\degr$ & B & 15.59 & 0.95 & $2.32\times 10^{-2}$ & 0.45 & 2048\\
 & & R & 15.05 & 1.01 & $7.68\times 10^{-1}$ & 0.42 & 1320\\
 225 & -67.5$\degr$ to -60.0$\degr$ & B & 15.80 & 1.00 & $-1.67\times 10^{-1}$ & 0.67 & 2006\\
 & & R & 14.92 & 0.97 & $6.45\times 10^{-2}$ & 0.41 & 1653\\
 300 & -60.0$\degr$ to -52.5$\degr$ & B & 15.22 & 1.16 & 0.0 & 0.42 & 2509\\
 & & R & 15.04 & 0.95 & $8.09\times 10^{-2}$ & 0.36 & 2204\\
 375 & -52.5$\degr$ to -45.0$\degr$ & B & 15.48 & 1.09 & 0.0 & 0.40 & 3488\\
 & & R & 15.03 & 0.99 & $7.31\times 10^{-2}$ & 0.38 & 1949\\
 450 & -45.0$\degr$ to -37.5$\degr$ & B & 15.27 & 1.15 & 0.0 & 0.36 & 2206\\
 & & R & 15.05 & 0.98 & $5.13\times 10^{-2}$ & 0.33 & 1220\\
 525 & -37.5$\degr$ to -30.0$\degr$ & B & 15.47 & 1.03 & $9.07\times 10^{-3}$ & 0.50 & 7590\\
& & R & 14.90 & 0.88 & $8.26\times 10^{-3}$ & 0.47 & 5061\\
 600 & -30.0$\degr$ to -22.5$\degr$ & B & 15.63 & 1.00 & $9.59\times 10^{-3}$ &  0.48 & 8879\\
 & & R & 15.01 & 0.94 & $4.88\times 10^{-2}$ & 0.40 & 5771\\
 675 & -22.5$\degr$ to -15.0$\degr$ & B & 15.58 & 1.00 & $1.44\times 10^{-2}$ & 0.42 & 6065\\
 & & R & 14.99 & 0.95 & $3.96\times 10^{-2}$ & 0.42 & 4578\\
 750& -15.0$\degr$ to -7.5$\degr$ & B & 15.60 & 1.05 & 0.0 & 0.38 & 4196\\
 & & R & 15.16 & 0.92 & $1.21\times 10^{-2}$ & 0.41 & 6624\\
 825& -7.5$\degr$ to 0.0$\degr$ & B & 15.39 & 1.06 & 0.0 & 0.33 & 2470\\
 & & R & 15.07 & 0.97 & 0.0 & 0.34 & 3949\\
 900& 0.0$\degr$ to 7.5$\degr$ & B & 15.39 & 1.07 & 0.0 & 0.30 & 2171\\
 & & R & 15.09 & 0.99 & 0.0 & 0.33 & 5047\\
 975& 7.5$\degr$ to 15.0$\degr$ & B & 15.68 & 1.00 & 0.0 & 0.34 & 8428\\
 & & R & 15.13 & 0.97 & 0.0 & 0.33 & 14084\\
 1050& 15.0$\degr$ to 22.5$\degr$ & B & 15.47 & 0.94 & $2.69\times 10^{-2}$ & 0.38 & 5945\\
 & & R & 15.11 & 0.95 & $1.68\times 10^{-2}$ & 0.39 & 11035\\
 1125& 22.5$\degr$ to 30.0$\degr$ & B & 15.28 & 1.05 & $9.92\times 10^{-3}$ & 0.33 & 7549\\
 & & R & 15.11 & 0.96 & $1.47\times 10^{-2}$ & 0.33 & 12974\\
 1200& 30.0$\degr$ to 37.5$\degr$ & B & 15.21 & 1.13 & $-6.01\times 10^{-3}$ & 0.28 & 12011\\
 & & R & 15.11 & 0.96 & $8.08\times 10^{-3}$ & 0.33 & 17769\\
 1275& 37.5$\degr$ to 45.0$\degr$ & B & 15.40 & 0.99 & $1.68\times 10^{-2}$ & 0.33 & 12665\\
 & & R & 15.02 & 0.94 & $1.37\times 10^{-2}$ & 0.39 & 21022\\
 1350 & 45.0$\degr$ to 52.5$\degr$ & B & 15.34 & 1.08 & 0.0 & 0.32 & 9248\\
 & & R & 15.07 & 0.97 & $4.56\times 10^{-3}$ & 0.39 & 17967\\
 1425& 52.5$\degr$ to 60.0$\degr$ & B & 15.65 & 1.08 & $-3.69\times 10^{-2}$ & 0.76 & 3720\\
& & R & 15.12 & 0.93 & 0.0 & 0.55 & 8983\\
 1500& 60.0$\degr$ to 67.5$\degr$ & B & 15.23 & 1.15 & $-1.08\times 10^{-2}$ & 0.31 & 6468\\
& & R & 15.10 & 0.95 & 0.0 & 0.43 & 9996\\
 1575&67.5$\degr$ to 75.0$\degr$ & B & 15.65 & 1.16 & $-2.03\times 10^{-2}$ & 0.25 & 412\\
& & R & 14.61 & 1.00 & $2.90\times 10^{-2}$ & 0.56 & 1031\\\hline
\end{tabular}
\end{center}
\end{table*}
Given the dispersion of USNO-A2 magnitudes, accounting for colour
effects would introduce an additional source of noise for single
wavelength flux measurements. As we are interested in integrating the
total flux at given wavelengths, we deliberately neglect colour
effects and perform the calibration between single filters. The red
measurements correspond to the plate emulsions 103a-E (North) and
IIIa-F (south). The equivalent wavelengths are respectively 6450~\AA\
and 6400~\AA\, which are very close to the $R_C$ filter. The blue
measurements (103a-O (POSS-I O) for the North, and IIIa-J (POSS-I E)
for the south) differ more significantly with equivalent wavelengths
of 4050~\AA\ and 4680~\AA.  We estimate that a 10\% uncertainty is
introduced when aligning these measurements to the $B_J$ data, which
remains negligible for our purposes.

We cross-identify the 526\,280\,881 USNO-A2 stars \citep{Monet:1998}
with those of the GSPC2.1 catalogue \citep{Bucciarelli:2001} composed
of 305\,017 stars with CCD (B)VR photometry within a $5''\times5''$
window. In the USNO-A2 catalog, the sky is partitioned into 24 zones
of South Polar Distance (SPD), each of width $7^\circ\hskip-2pt
.5$. We thereby find 257\,352 stars spread over 21 zones. For each
zone, we plot the residuals in RA and DEC, and keep the 186\,568 stars
within 1$\sigma$ of the mean of each zone. We then compute the
relationships between different magnitude systems as shown in Figure
\ref{fig:calib}. The corresponding coefficients are provided in Table
\ref{tab:calib}. $B_{\mathrm{USNO}}$ and $R_{\mathrm{USNO}}$
magnitudes are thus converted into $B_J$ ($\lambda_o^B=$4400~\AA) and
$R_C$ ($\lambda_o^R=$6400~\AA) magnitudes, which are in turn converted
into fluxes with $F_{B,R} = C_{B,R} 10^{-0.4\times {B,R}}$
($C_B=4.26\times 10^{-23}$W m$^{-2}$ Hz$^{-1}$, $C_R=3.08\times
10^{-23}$W m$^{-2}$ Hz$^{-1}$).

Note that we do not detect any significant bias between northern and
southern hemispheres.  Systematic effects are observed for
$R_\mathrm{USNO}$ data for $R_\mathrm{USNO}>17.5$ and
$\delta<-14\degr$.  They will tend to overestimate the contribution of
faint stars and are most probably an effect of crowding.

\clearpage

\begin{figure*}
\includegraphics[width=0.33\textwidth]{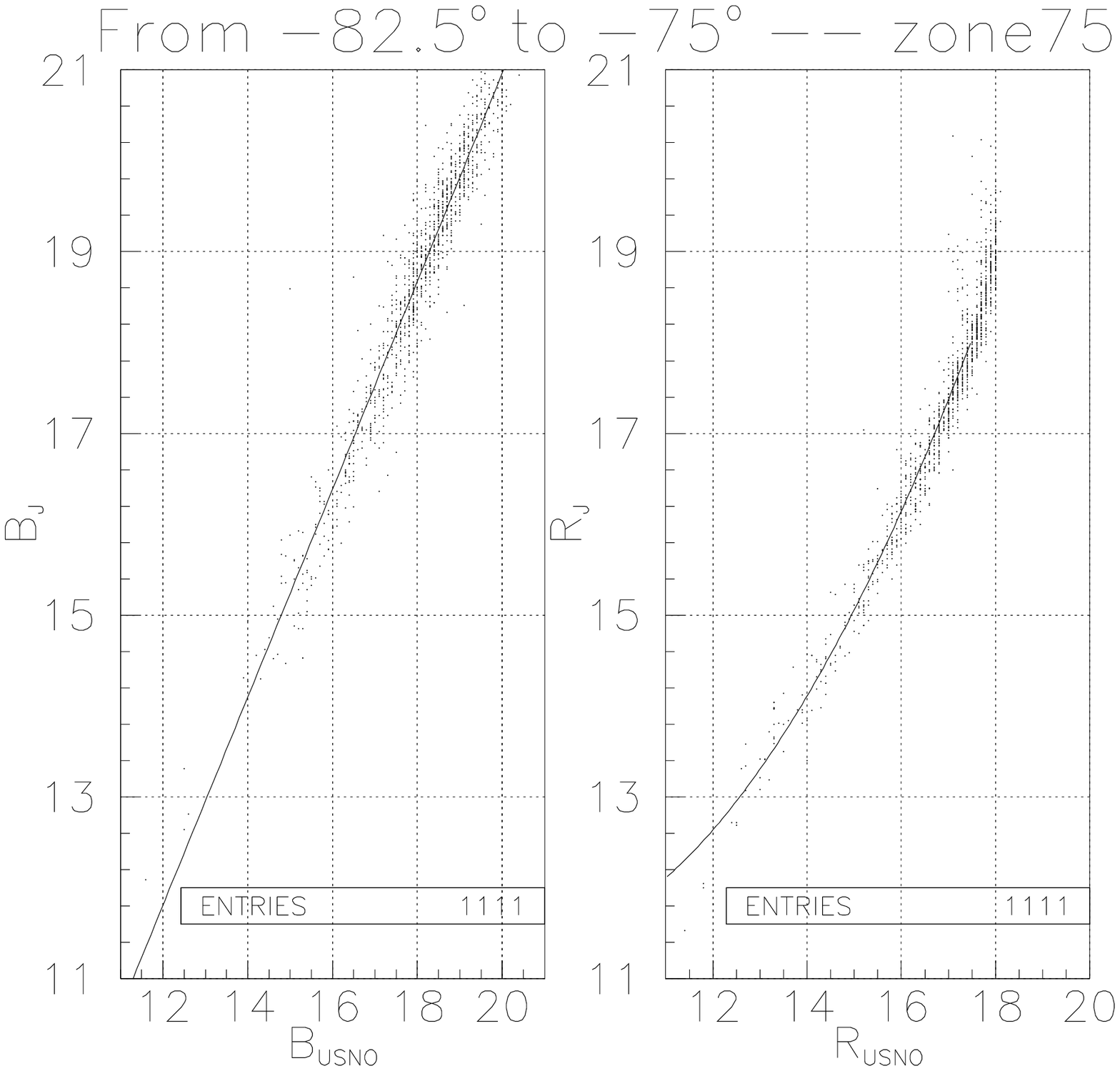} 
\includegraphics[width=0.33\textwidth]{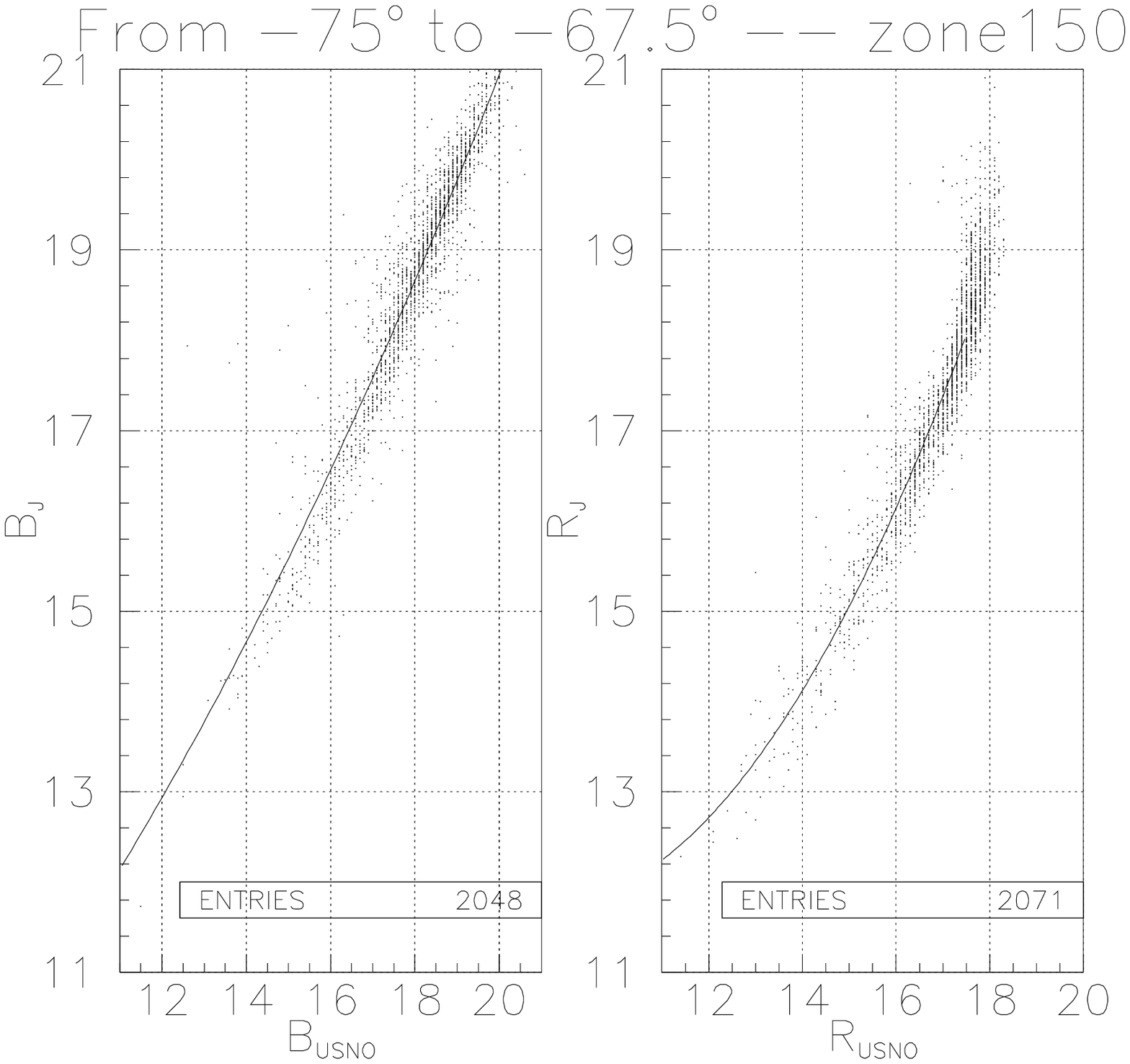} 
\includegraphics[width=0.33\textwidth]{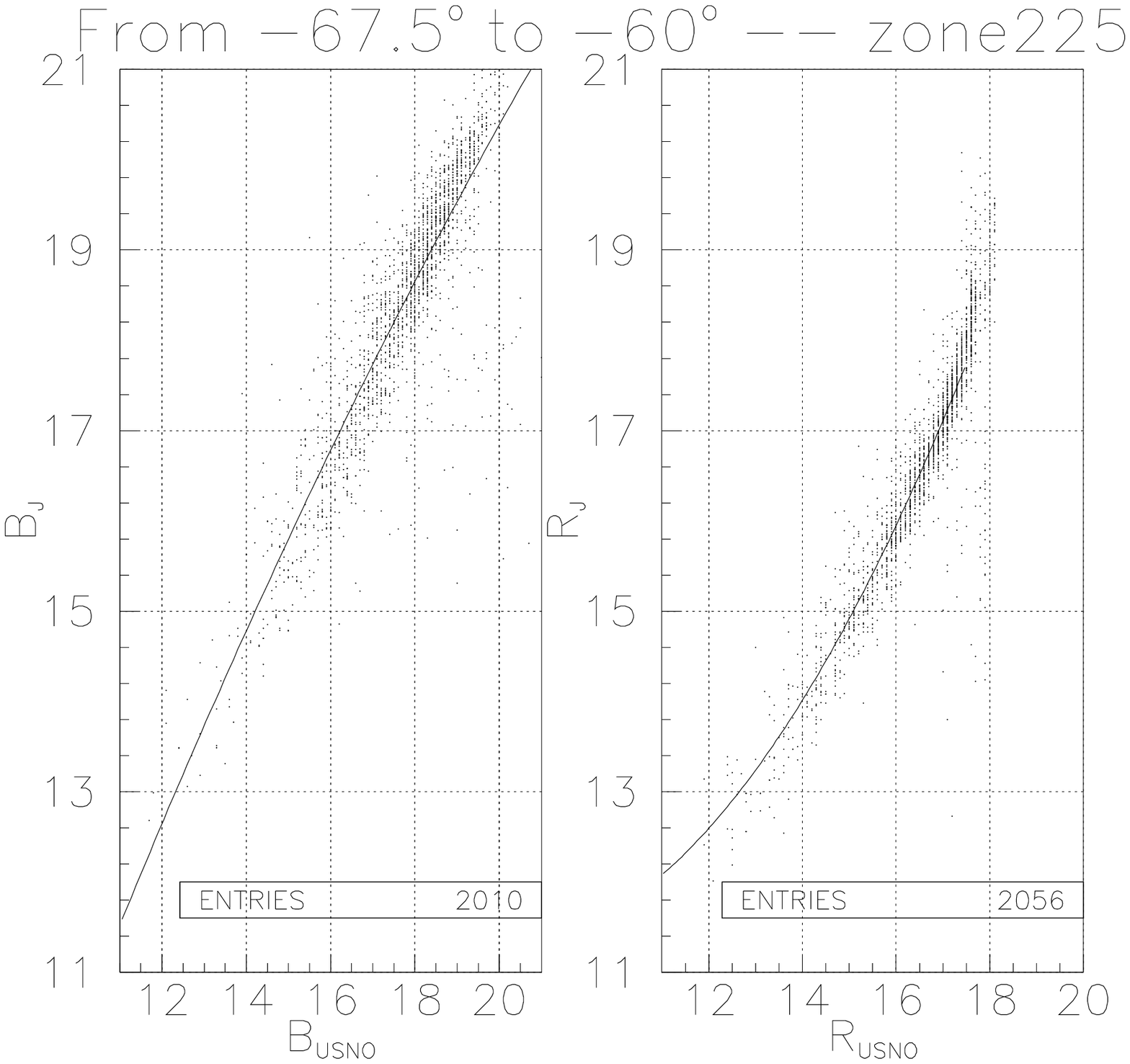} 
\includegraphics[width=0.33\textwidth]{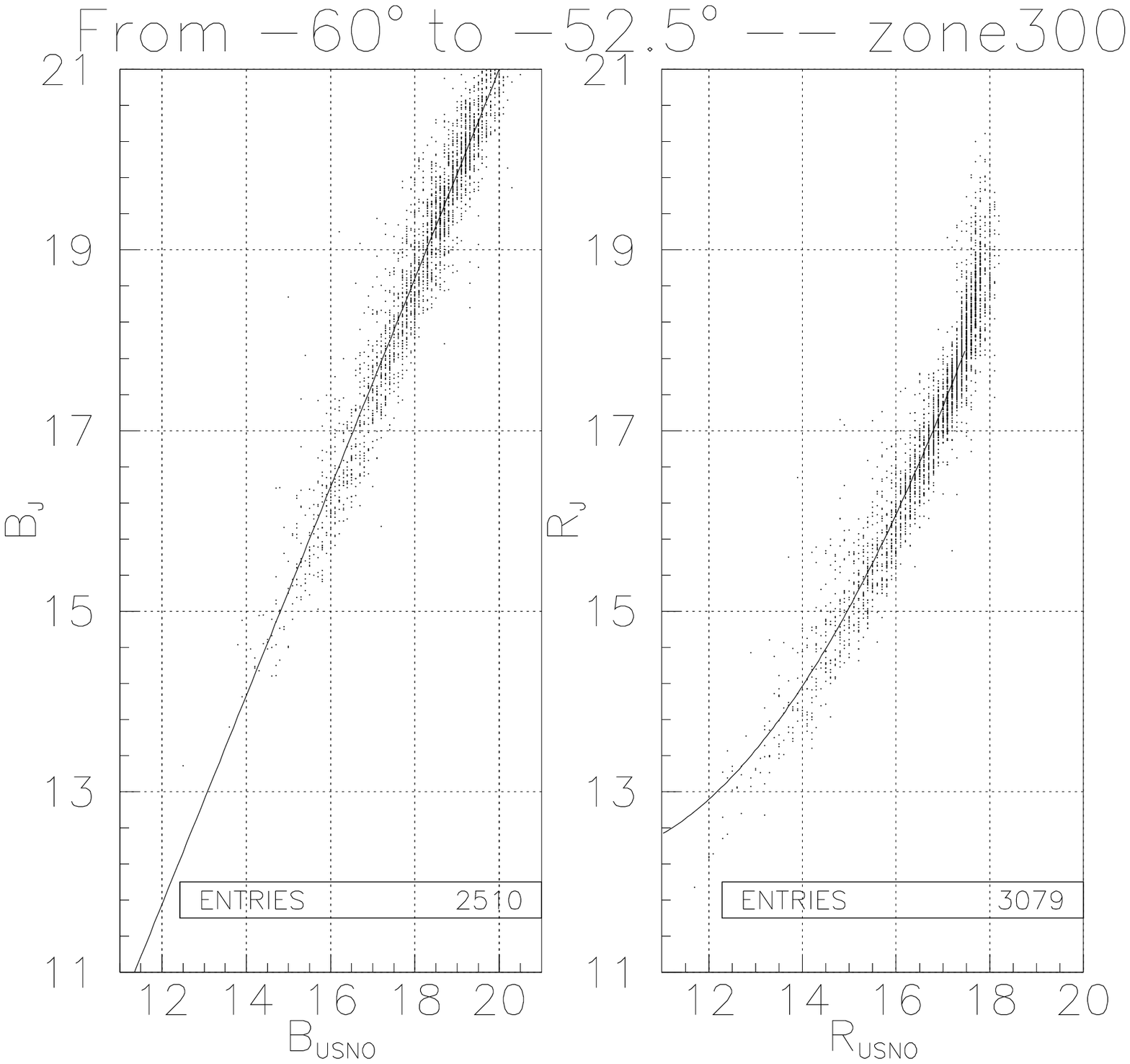} 
\includegraphics[width=0.33\textwidth]{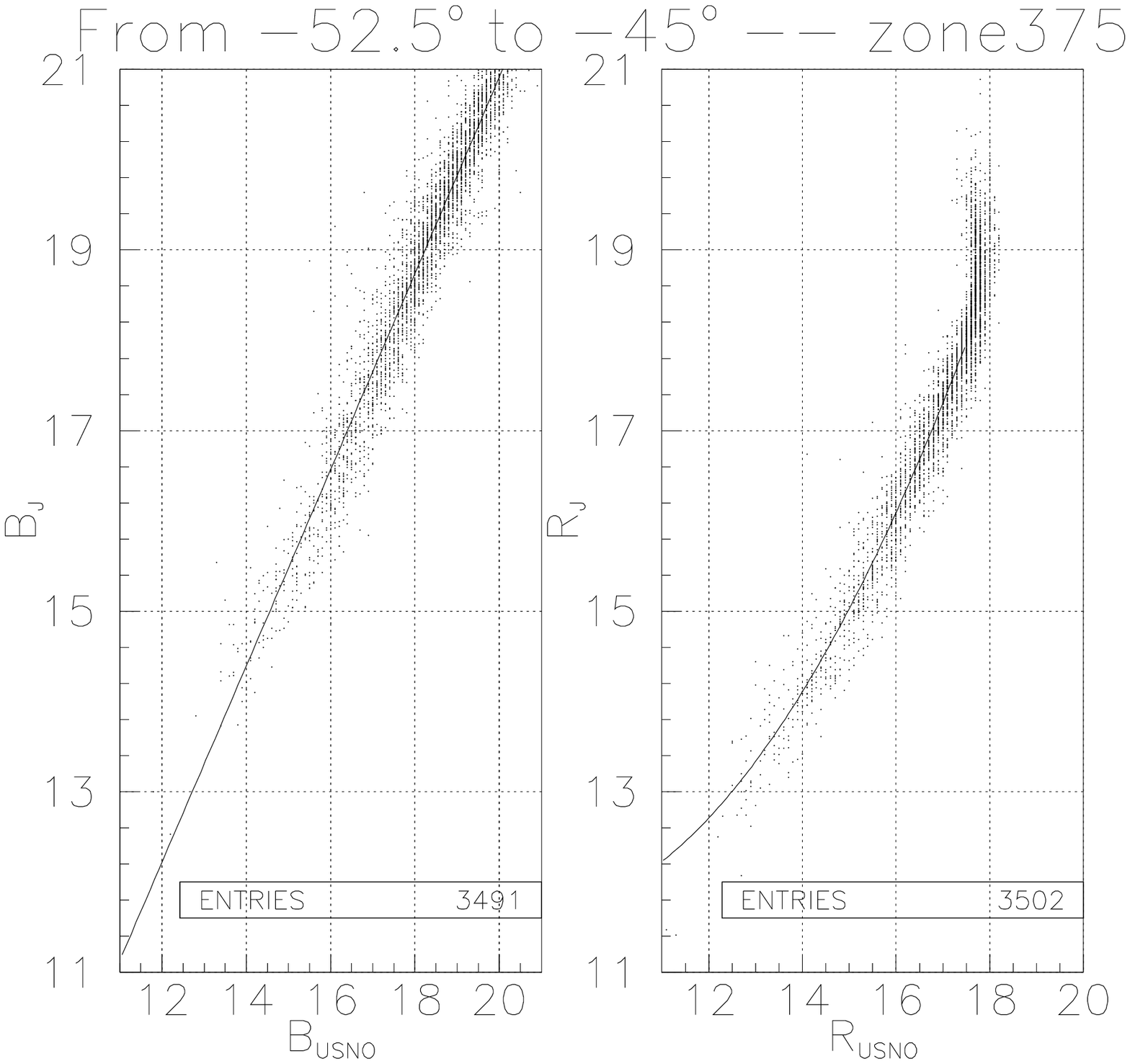} 
\includegraphics[width=0.33\textwidth]{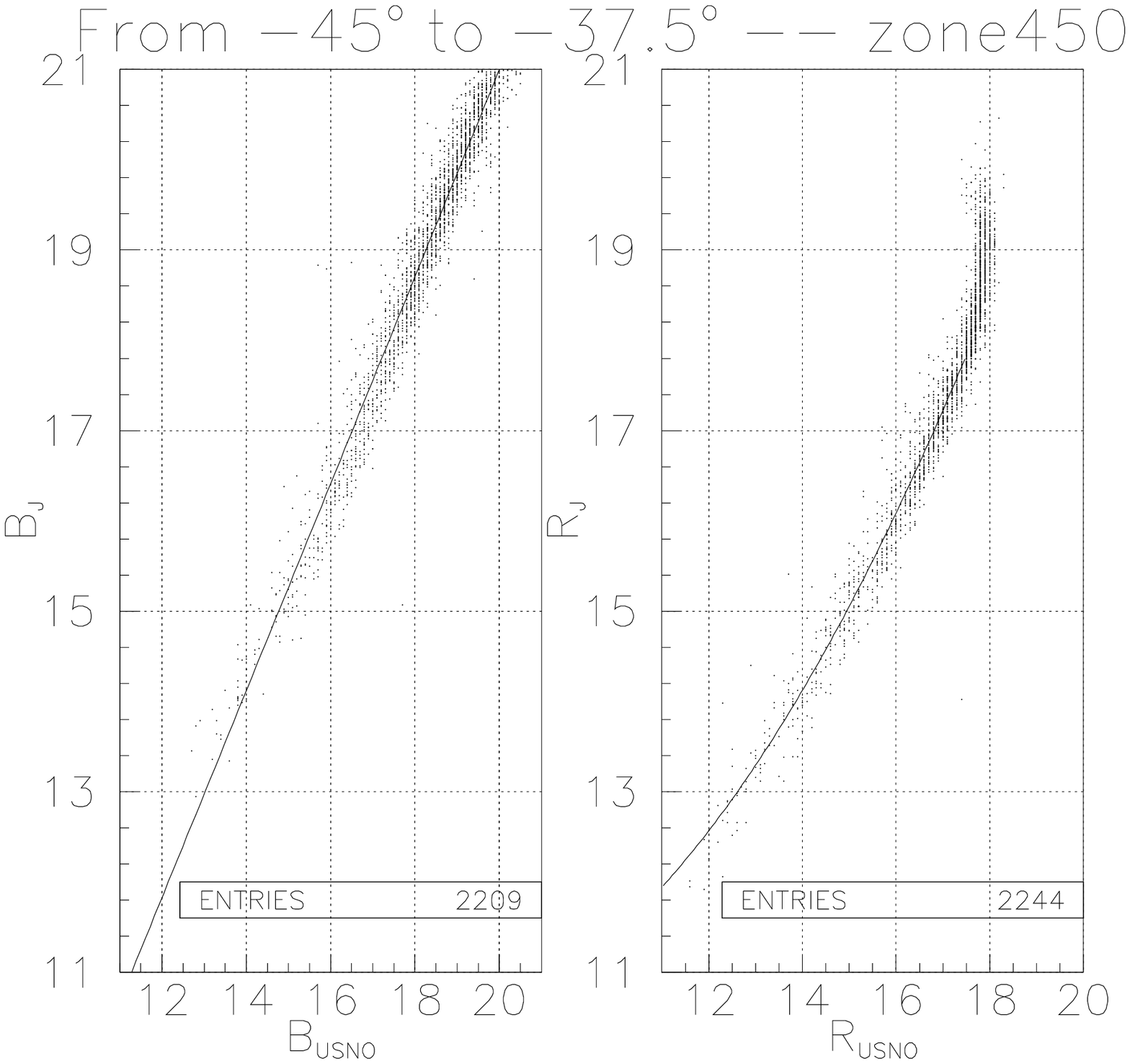} 
\includegraphics[width=0.33\textwidth]{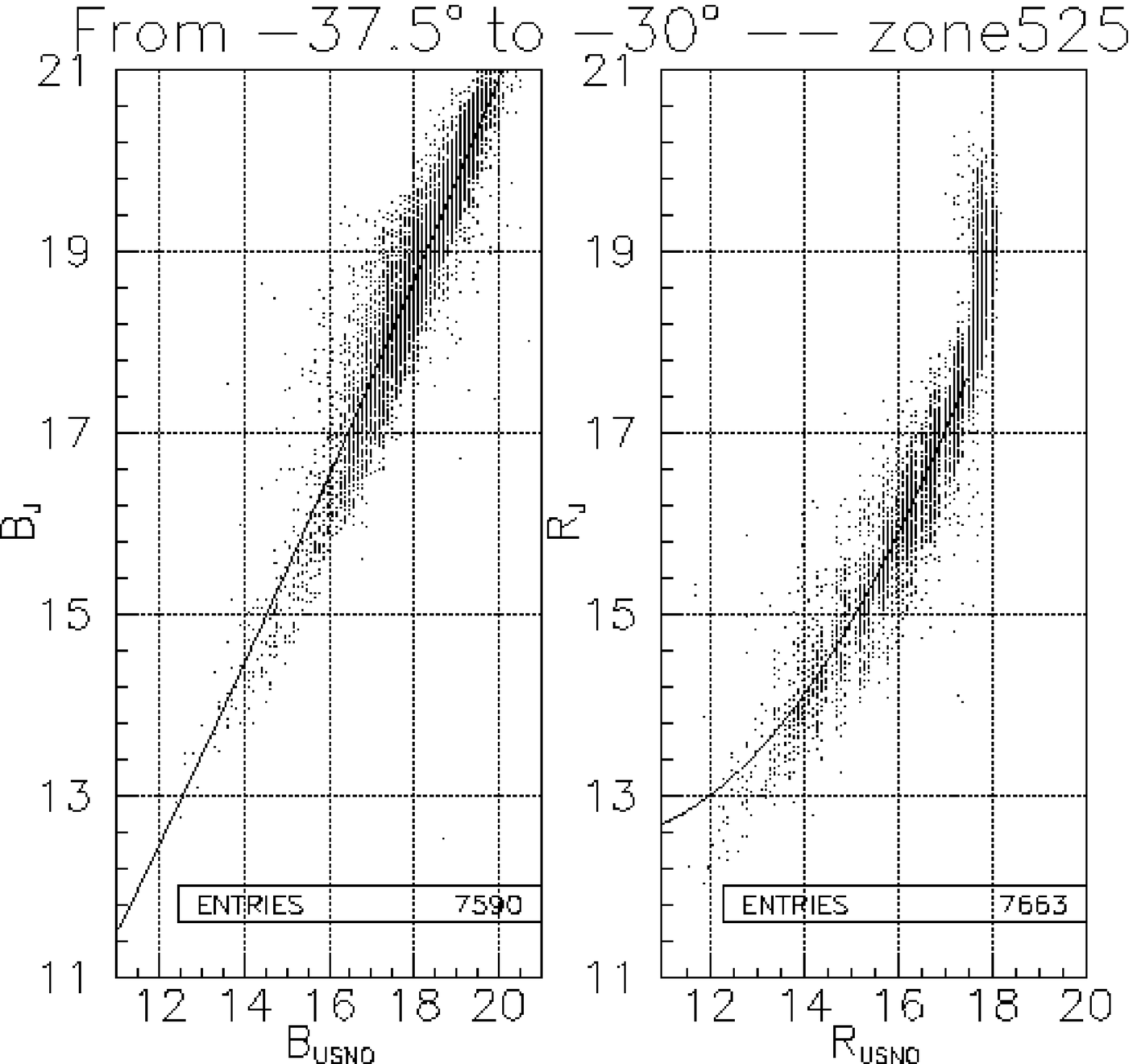} 
\includegraphics[width=0.33\textwidth]{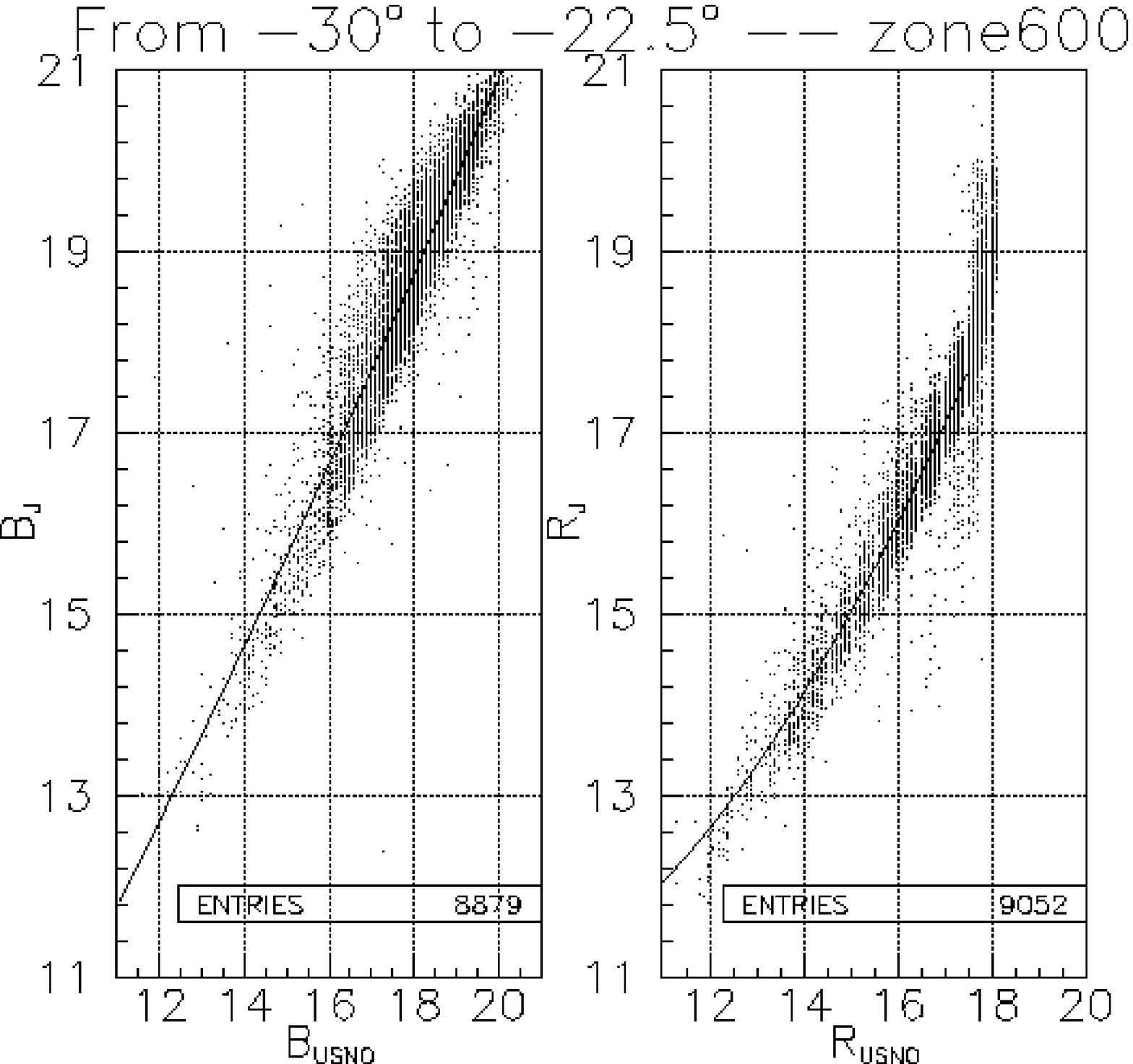} 
\includegraphics[width=0.33\textwidth]{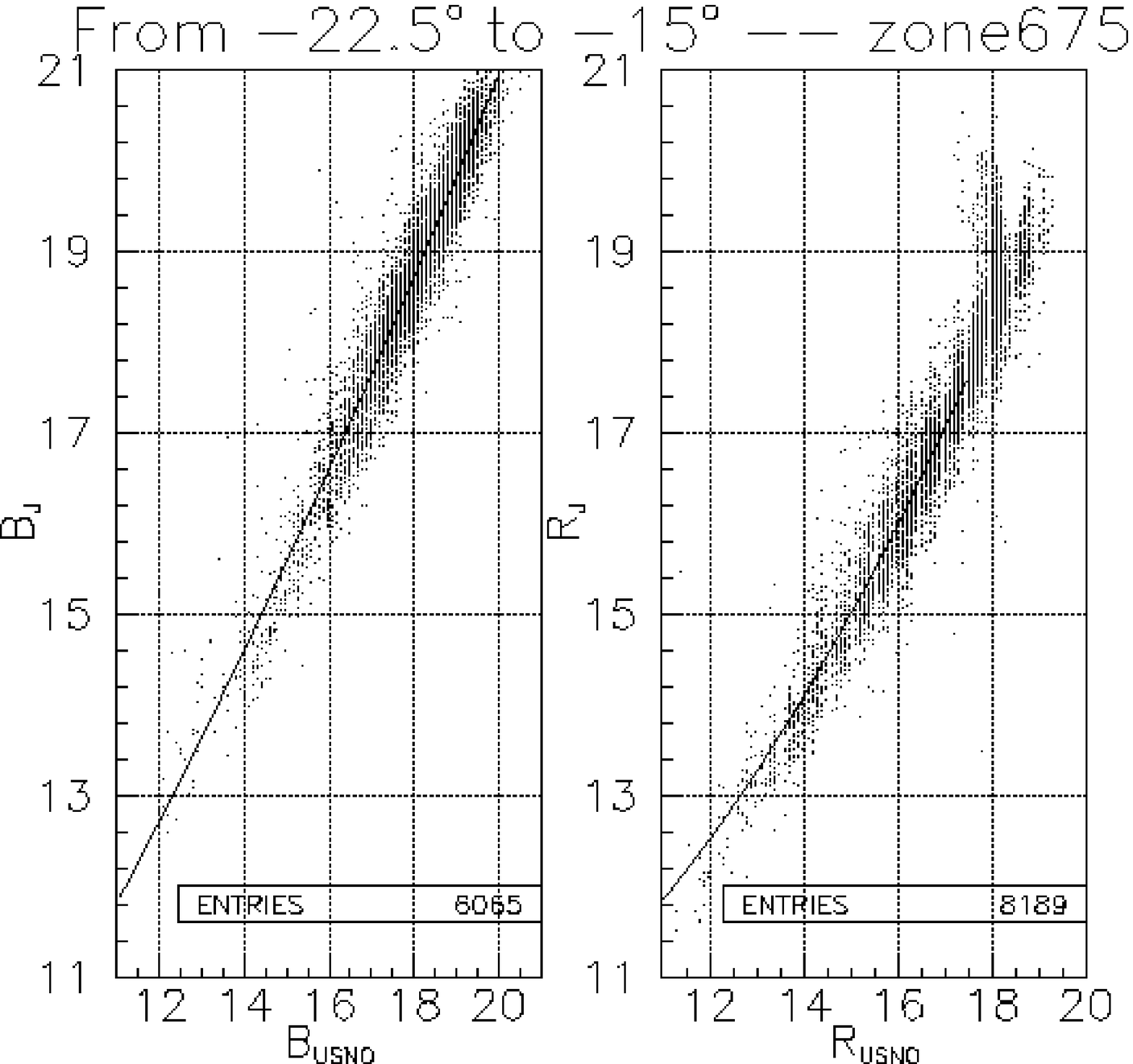} 
\includegraphics[width=0.33\textwidth]{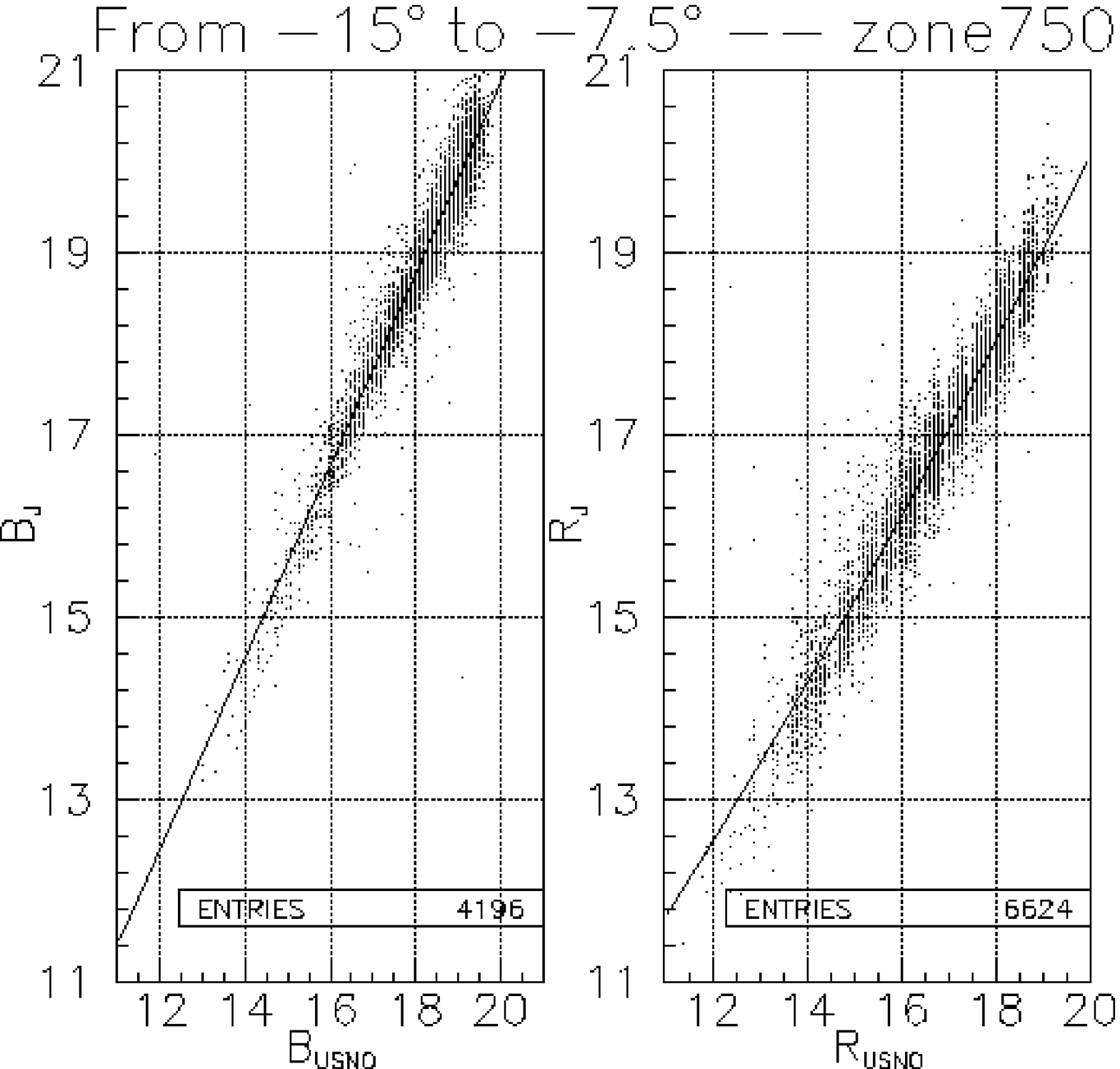} 
\includegraphics[width=0.33\textwidth]{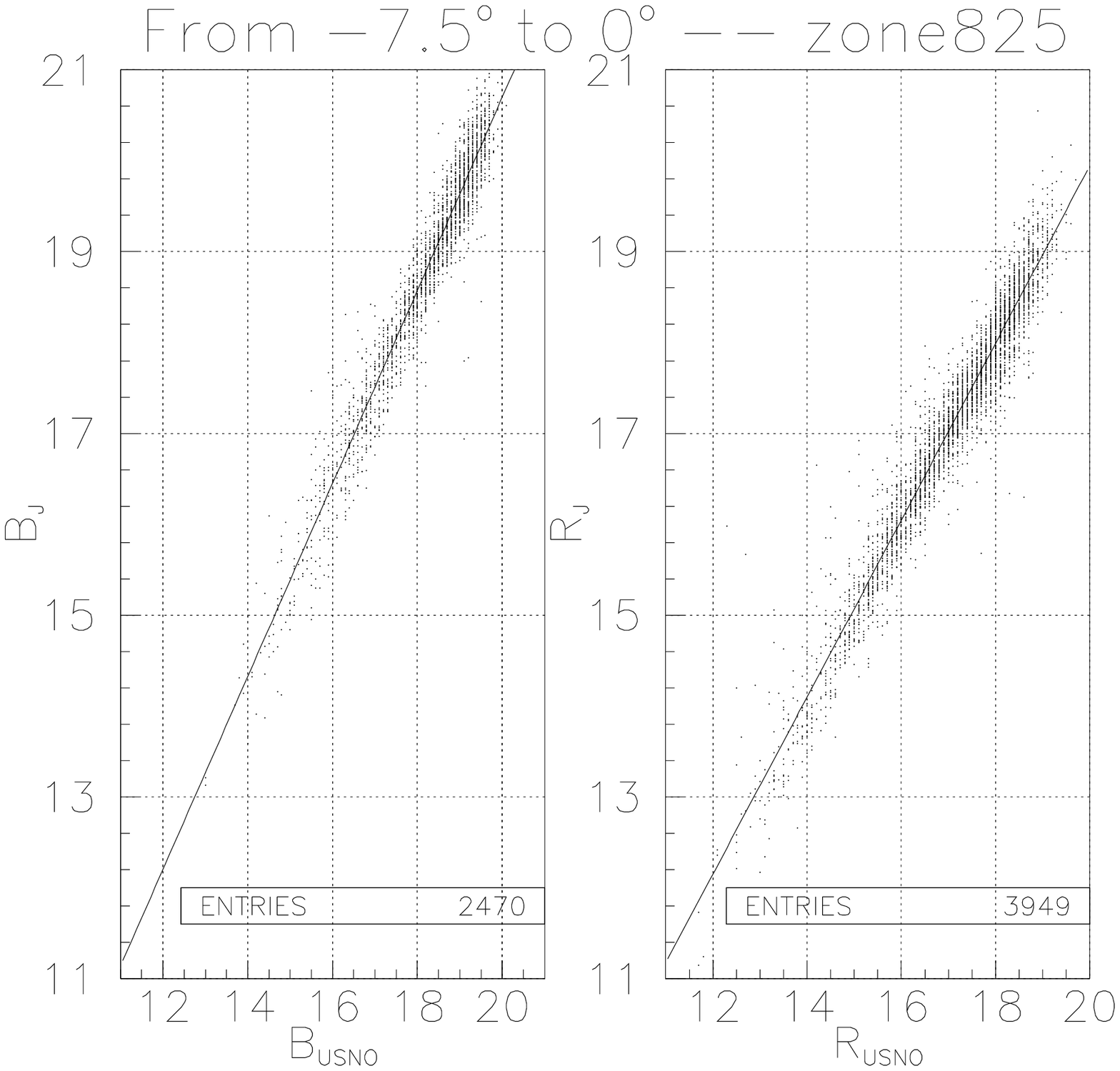} 
\includegraphics[width=0.33\textwidth]{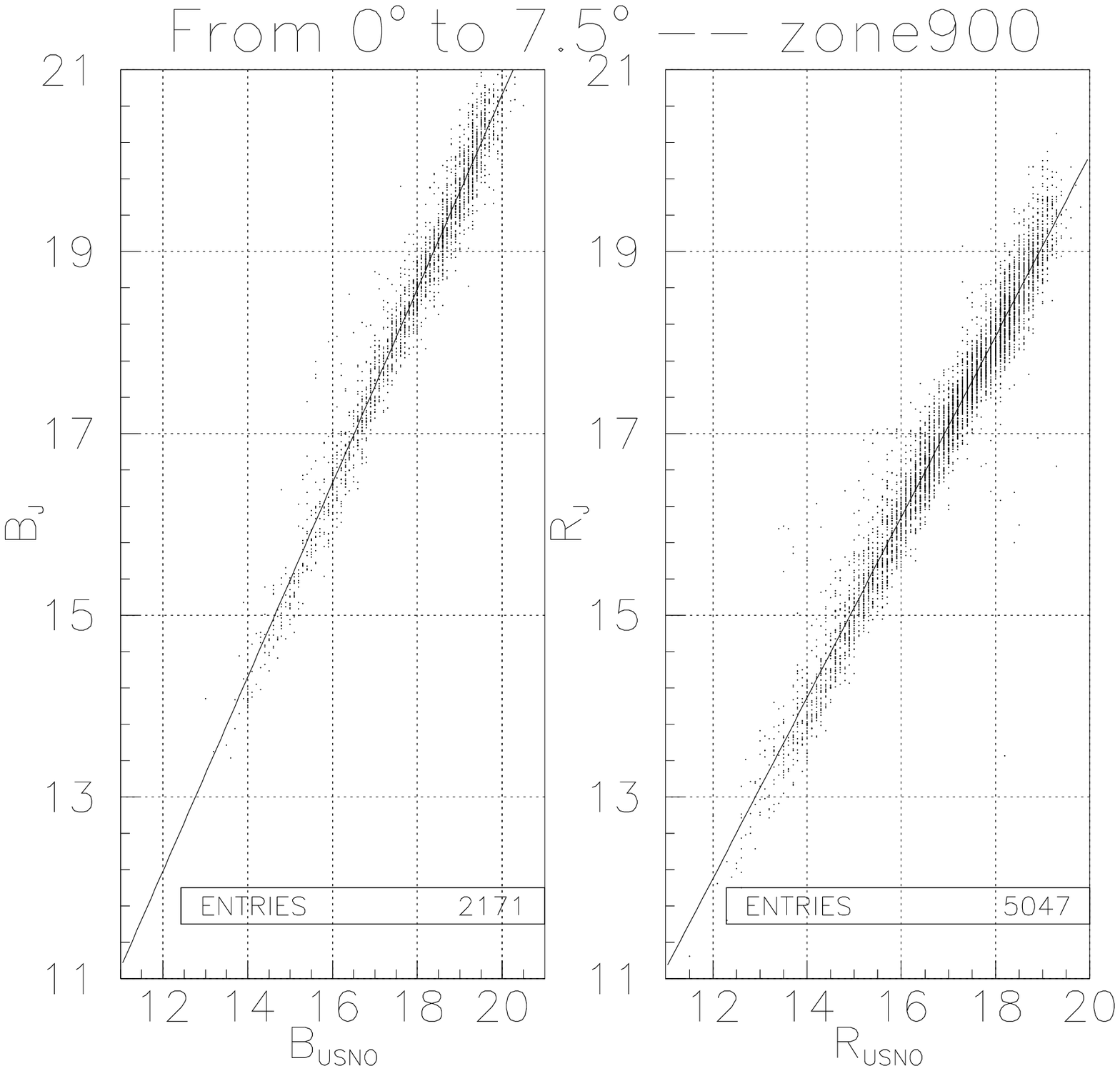} 
\caption{Each panel presents the $B_J$ versus B$_\mathrm{USNO}$ and
$R_C$ versus R$_\mathrm{USNO}$ relations for each South Polar Distance
(SPD) interval corresponding to a zone. The polynomial adjustments are
superimposed for each relation, and the corresponding coefficients are
provided in Table E.1.}  \label{fig:calib}
\end{figure*}
\addtocounter{figure}{-1}
\begin{figure*}[Ht]
\includegraphics[width=0.33\textwidth]{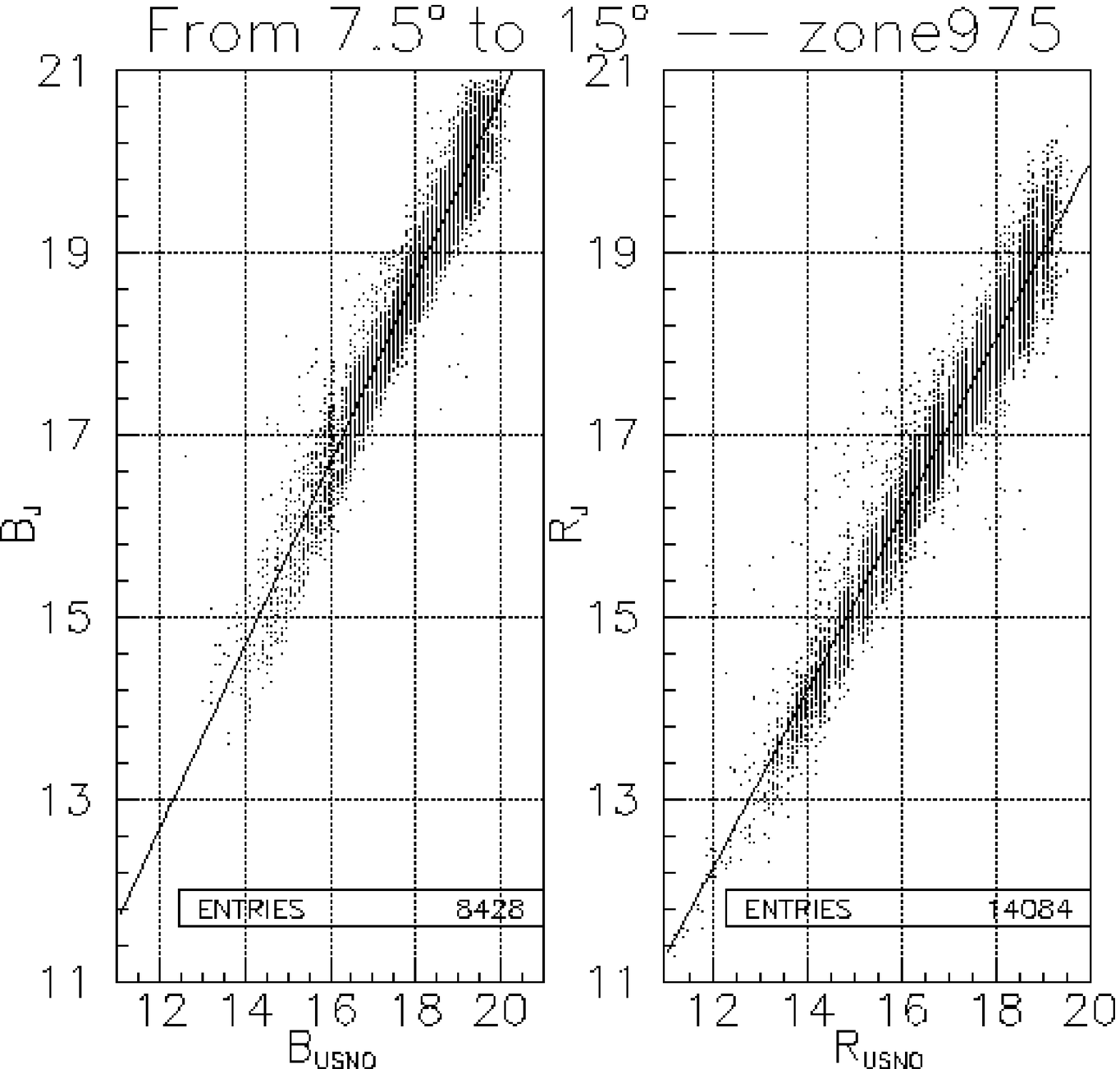} 
\includegraphics[width=0.33\textwidth]{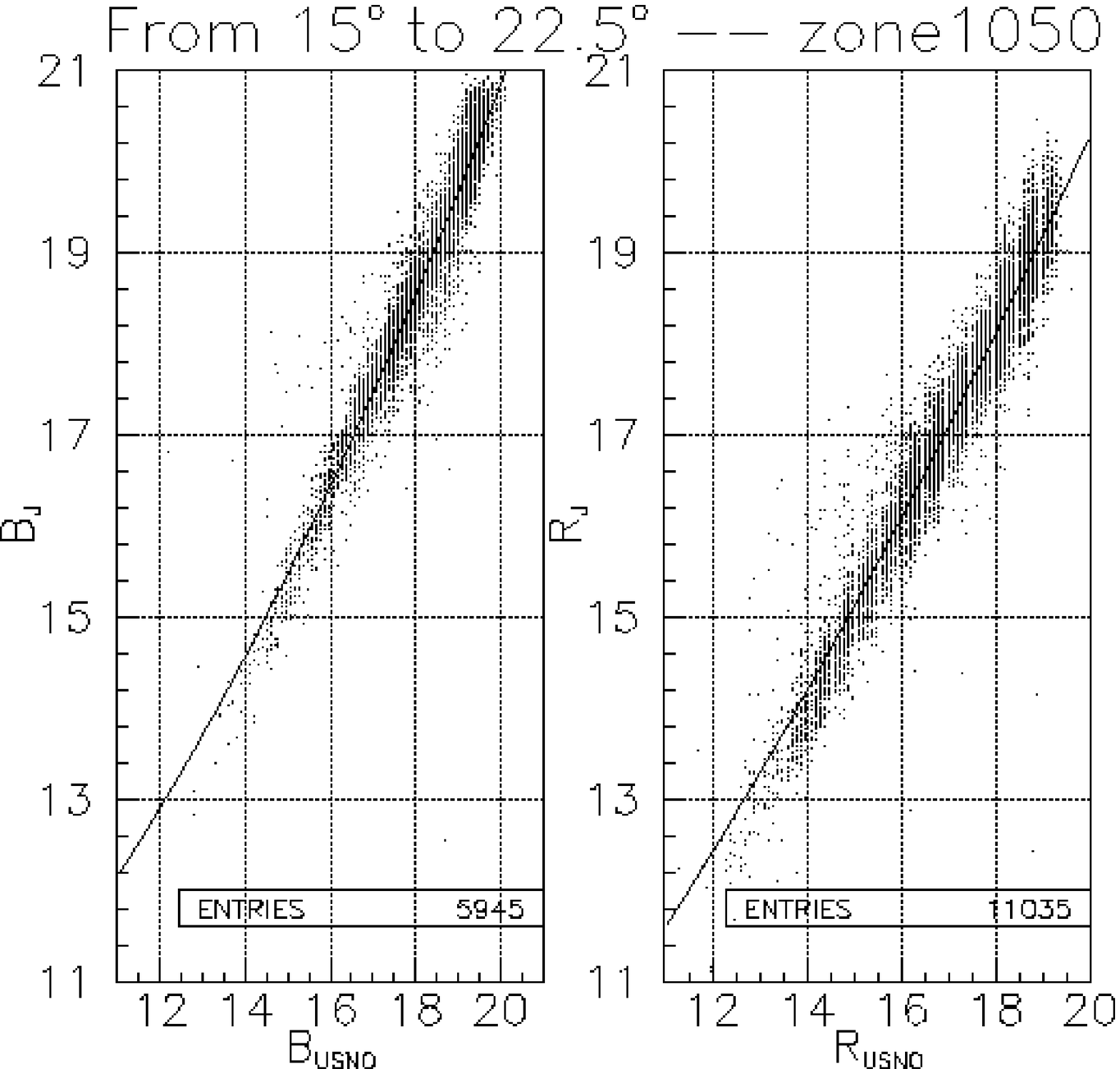} 
\includegraphics[width=0.33\textwidth]{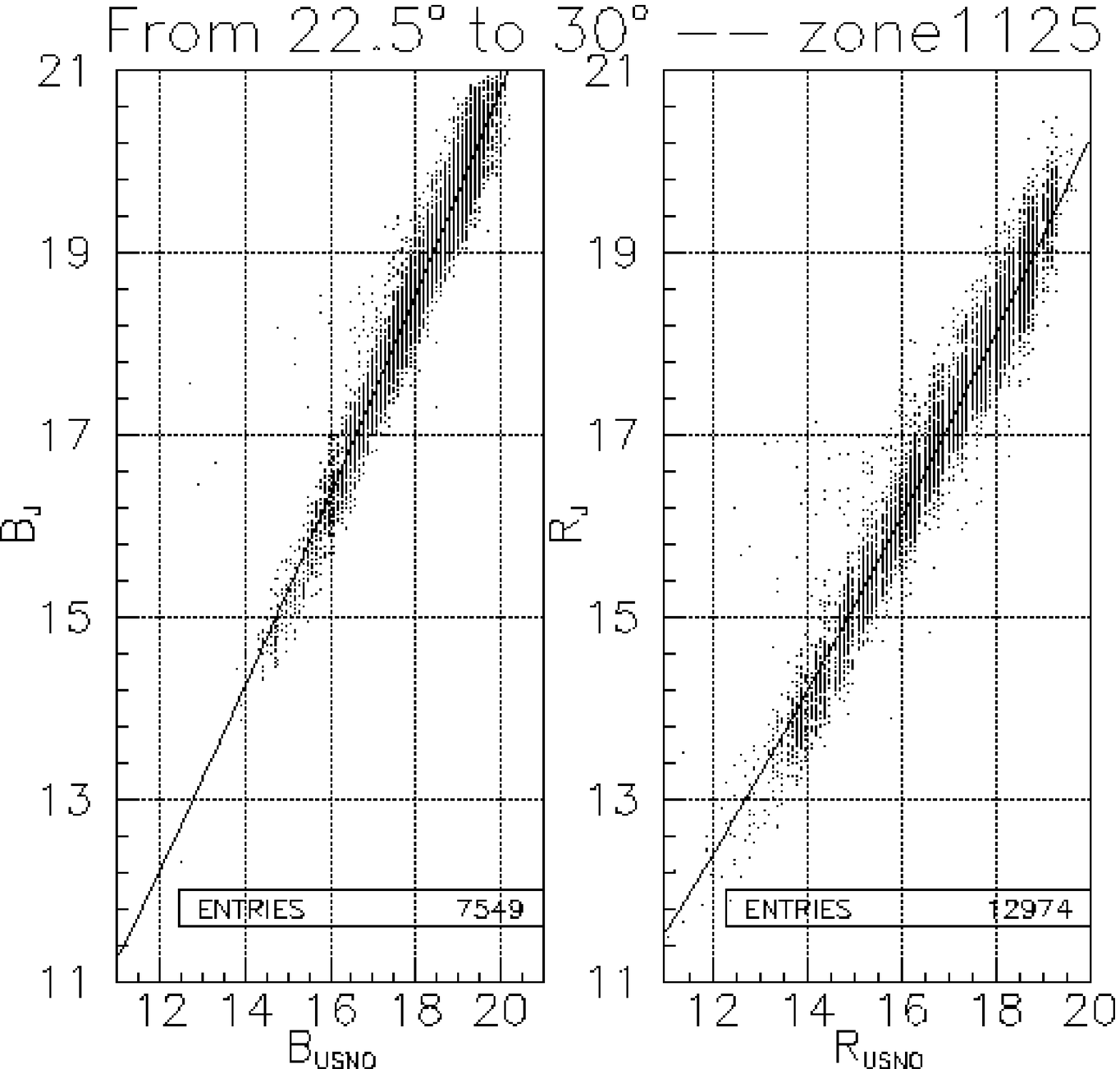} 
\includegraphics[width=0.33\textwidth]{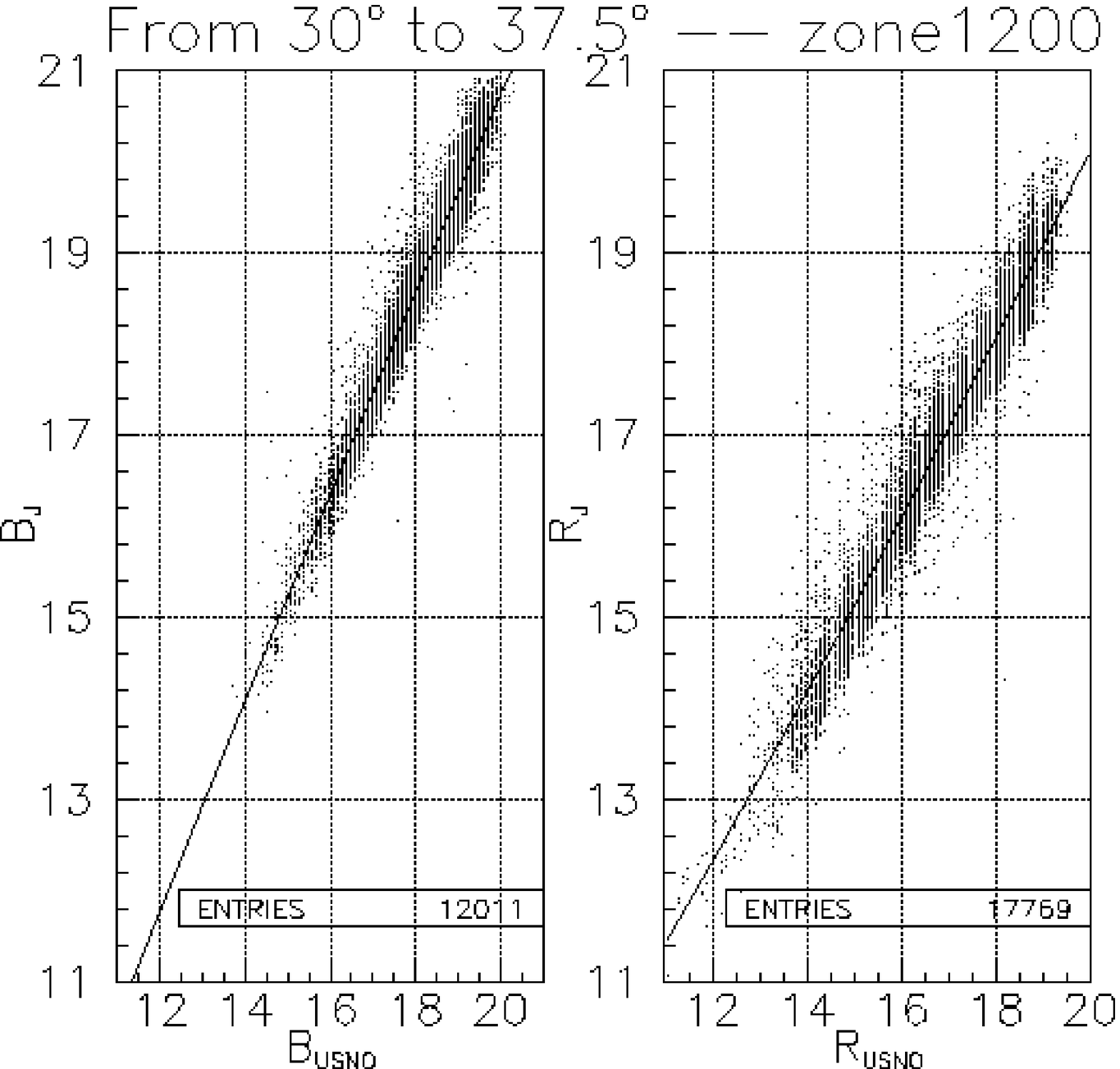} 
\includegraphics[width=0.33\textwidth]{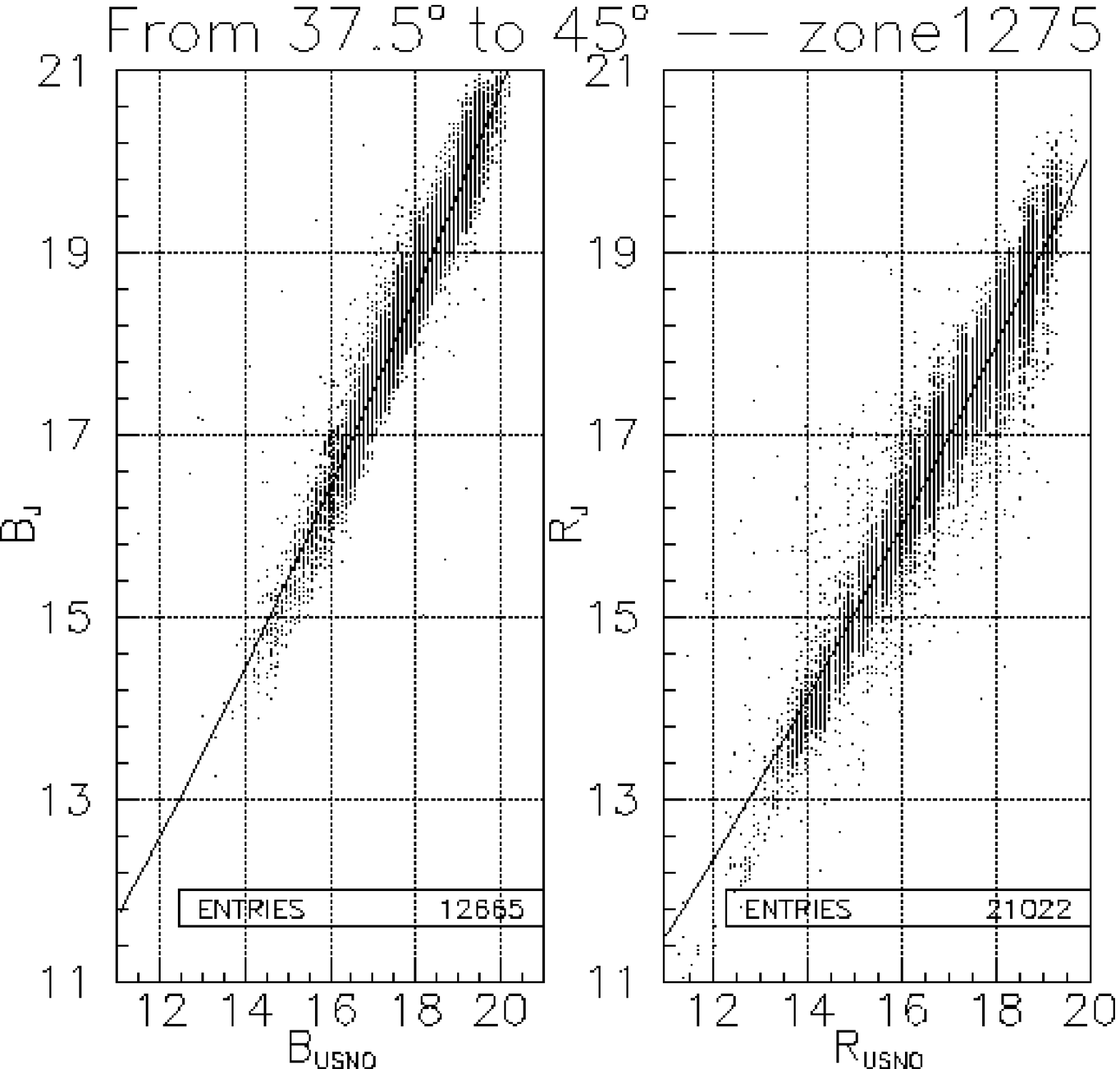} 
\includegraphics[width=0.33\textwidth]{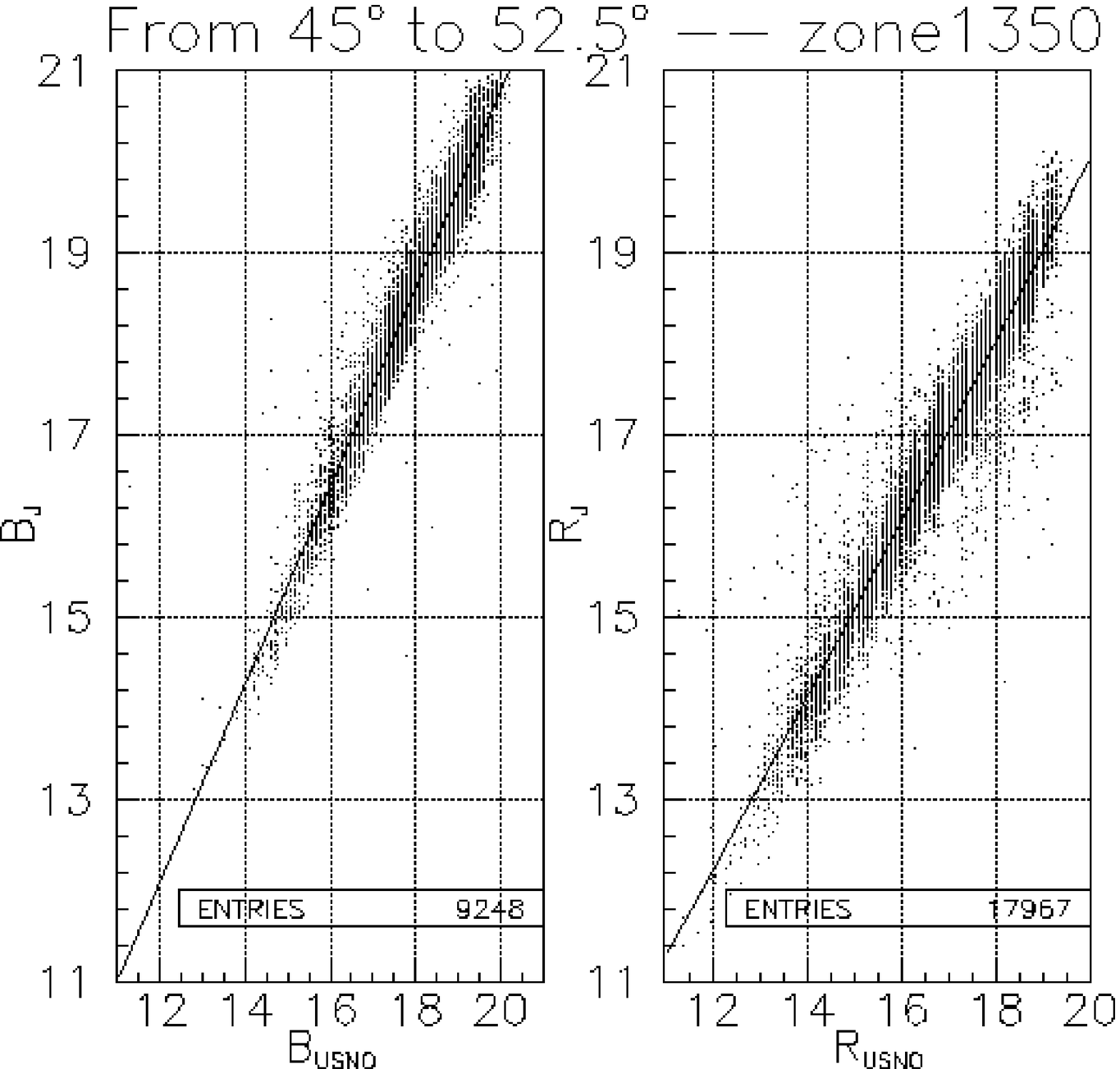} 
\includegraphics[width=0.33\textwidth]{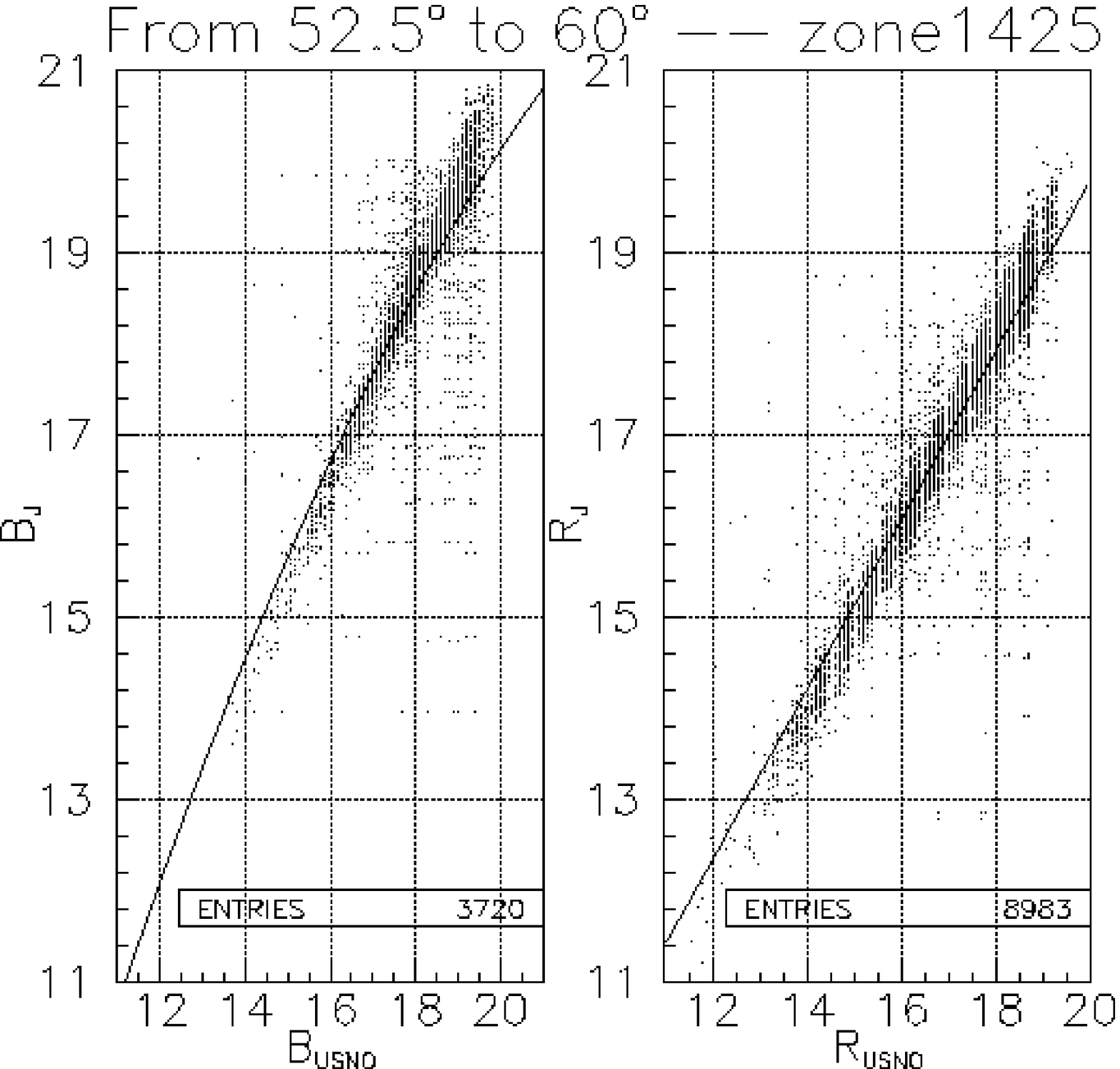} 
\includegraphics[width=0.33\textwidth]{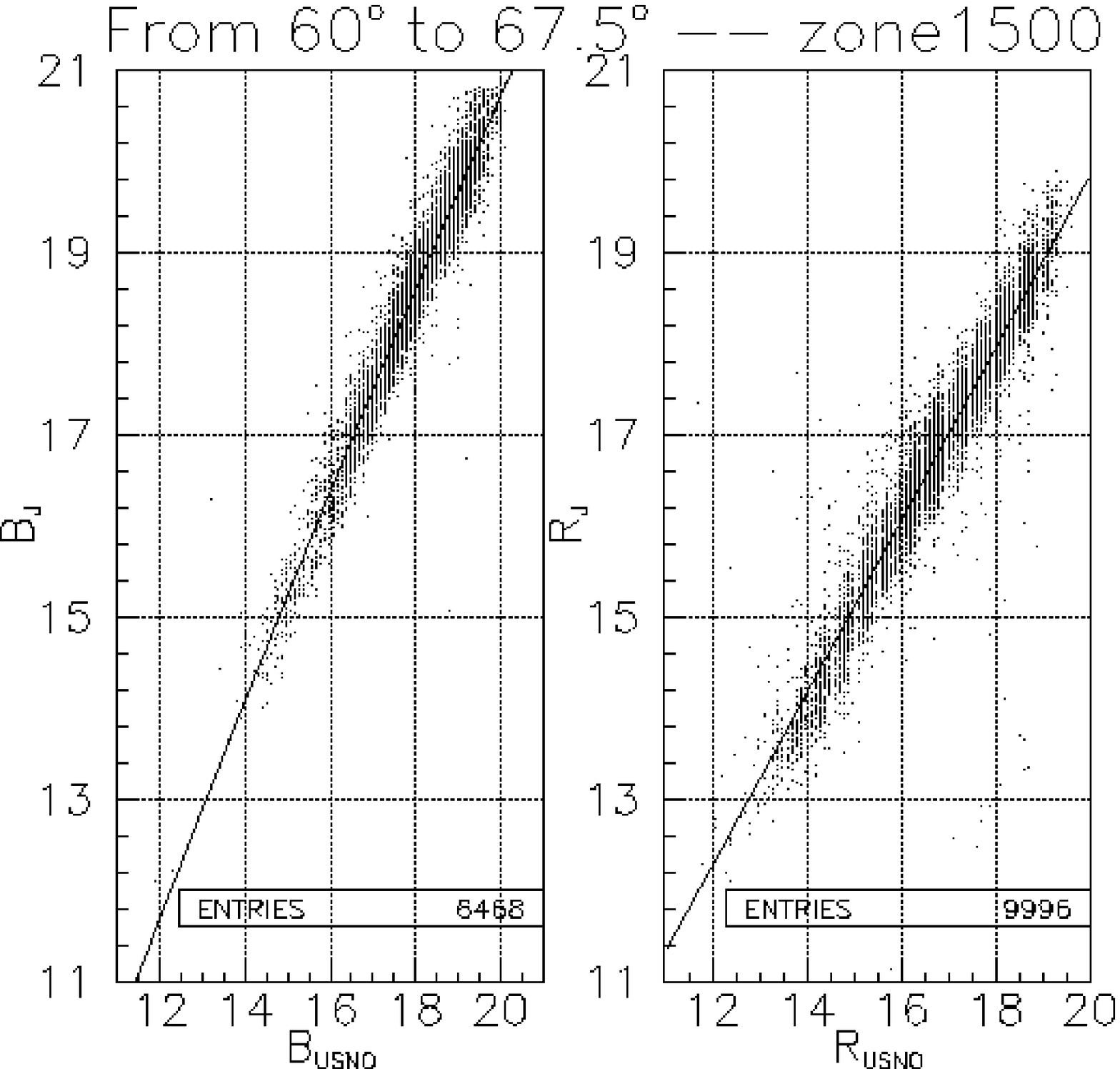} 
\includegraphics[width=0.33\textwidth]{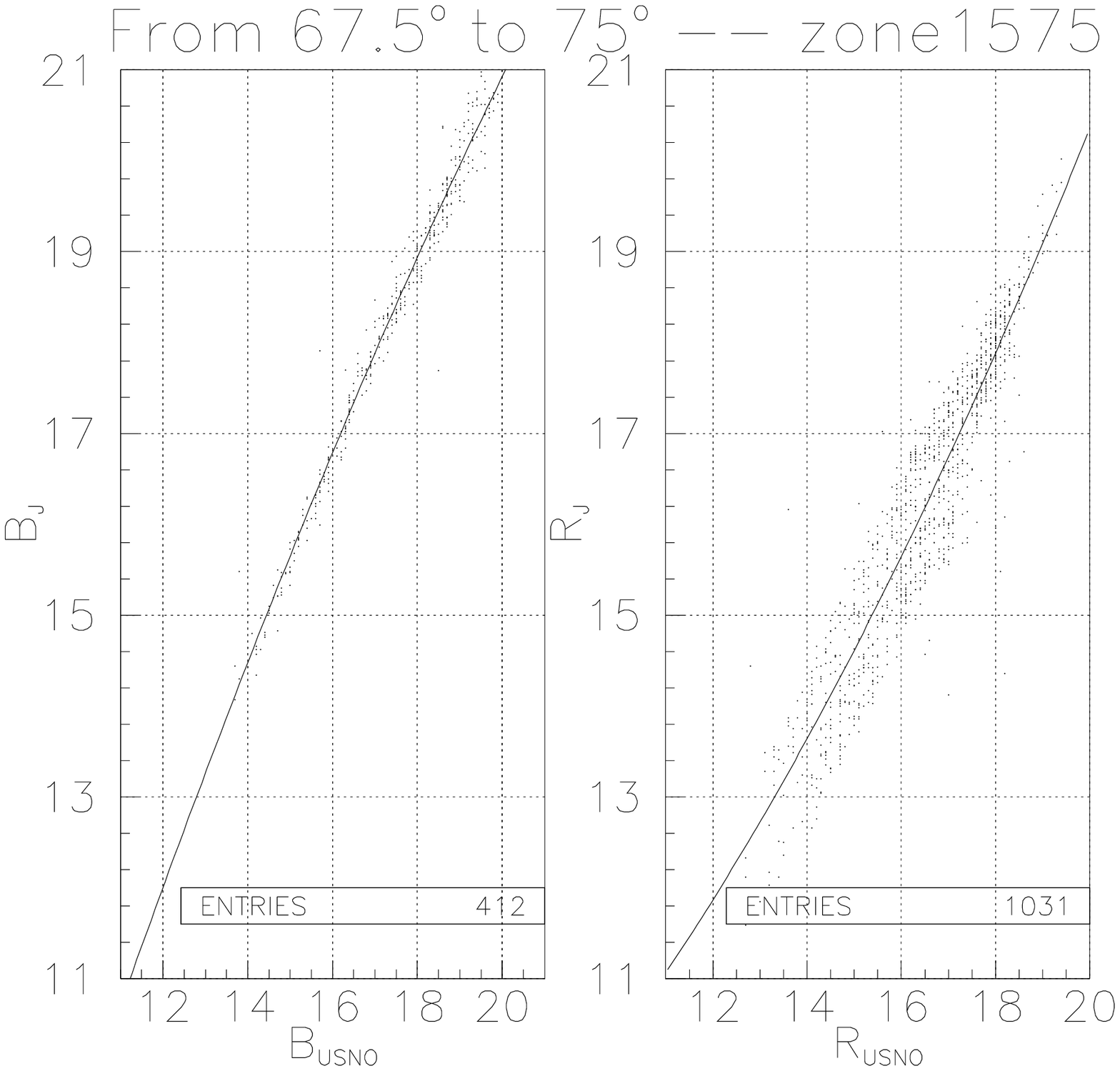} 
\caption{Continued}  
\end{figure*}
\stepcounter{figure}

\end{document}